\documentclass[twocolumn,showpacs,aps,superscriptaddress,amsmath,amssymb,preprintnumbers, floatfix]{revtex4} %, draft

\usepackage{graphicx}% Include figure files
\usepackage{dcolumn}% Align table columns on decimal point
\usepackage{bm}% bold math
\usepackage{amscd}

\usepackage[T1]{fontenc}

\usepackage{url}
\usepackage{color}
\ifx\color@rgb\@undefined\else
 \definecolor{purple}{rgb}{0.6,0,0.6}
\fi
\usepackage{colordvi}

\usepackage{citesort}
\usepackage{xspace}
\usepackage{german}
\usepackage[bookmarks=true,bookmarksnumbered=true,hypertexnames=false]{hyperref}

\renewcommand{\Re}{\operatorname{Re}}

\renewcommand{\vec}{\operatorname{vec}}
\newcommand{\unity}{\ensuremath{{\rm 1 \negthickspace l}{}}}
\newcommand{\Adr}{\operatorname{Ad}}
\newcommand{\adr}{\operatorname{ad}}
\newcommand{\tr}{\operatorname{tr}}
\newcommand{\diag}{\operatorname{diag}}

\newcommand{\grad}{\operatorname{grad}}

\newcommand{\Hess}{\mathcal{H}}
\newcommand{\Mat}{\operatorname{Mat}}

\newcommand{\R}[1]{\ensuremath{\mathbb R}^{\,#1}{}}
\newcommand{\C}[1]{\ensuremath{\mathbb C}^{\,#1}{}}
\newcommand{\Z}{\mathbb{Z}}

\newcommand{\bra}[1]{\ensuremath{\langle #1 |}{}}
\newcommand{\ket}[1]{\ensuremath{| #1 \rangle}{}}
\newcommand{\braket}[2]{\ensuremath{\langle #1 | #2 \rangle}{}}
\newcommand{\ketbra}[2]{\ensuremath{| #1 \rangle \langle #2 |}{}}

\newcommand{\Partial}[2]{\ensuremath \frac{\partial{#1}}{\partial{#2}}{}}
\newcommand{\tPartial}[2]{\ensuremath \tfrac{\partial{#1}}{\partial{#2}}{}}

\newcommand{\grape}{{\sc grape}\xspace}
\newcommand{\krotov}{{Krotov}\xspace}
\newcommand{\dynamo}{{\sc dynamo}\xspace}
\newcommand{\lbfgs}{{\sc l-bfgs}\xspace}
\newcommand{\bfgs}{{\sc bfgs}\xspace}

\newcommand{\matlab}{{\sc matlab}\xspace}
\newcommand{\cpu}{{\sc cpu}\xspace}

\newcommand{\subspace}{subspace\xspace}

\newcommand{\subspaces}{subspaces\xspace}
\newcommand{\Subspaces}{Subspaces\xspace}

\newcommand{\timeslice}{time slice\xspace}

\newcommand{\timeslices}{time slices\xspace}

\newcommand{\squishlist}{
 \begin{list}{$\bullet$}
  { \setlength{\itemsep}{0pt}
     \setlength{\parsep}{3pt}
     \setlength{\topsep}{3pt}
     \setlength{\partopsep}{0pt}
     \setlength{\leftmargin}{1.5em}
     \setlength{\labelwidth}{1em}
     \setlength{\labelsep}{0.5em} } }

\newcommand{\squishlisttwo}{
 \begin{list}{$\bullet$}
  { \setlength{\itemsep}{0pt}
     \setlength{\parsep}{0pt}
    \setlength{\topsep}{0pt}
    \setlength{\partopsep}{0pt}
    \setlength{\leftmargin}{2em}
    \setlength{\labelwidth}{1.5em}
    \setlength{\labelsep}{0.5em} } }

\newcommand{\squishend}{
  \end{list}  }
\begin{document}
%\bibliographystyle{prsty}

%\preprint{First contribution in a line of comparisons coordinated by M.~Murphy and T.~Calarco from the University of Ulm}

\title{Comparing, Optimising and Benchmarking Quantum Control Algorithms \\[1mm] in a Unifying Programming Framework}

\author{S.~Machnes}
	\affiliation{Quantum Group, Department of Physics, Tel-Aviv University, Tel Aviv 69978, Israel}
    \affiliation{Institute for Theoretical Physics, University of Ulm, D-89069 Ulm, Germany}
\author{U.~Sander}
\author{S.~J.~Glaser}
	\affiliation{Department of Chemistry, Technical University of Munich (TUM), D-85747 Garching, Germany}
\author{P.~de Fouqui{\`e}res}
	\affiliation{Department of Applied Mathematics and Theoretical Physics, University of Cambridge, CB3 0WA, UK}
\author{A.~Gruslys}
	\affiliation{Department of Applied Mathematics and Theoretical Physics, University of Cambridge, CB3 0WA, UK}
\author{S.~Schirmer}
	\affiliation{Department of Applied Mathematics and Theoretical Physics, University of Cambridge, CB3 0WA, UK}
\author{T.~Schulte-Herbr{\"u}ggen}\email{tosh@tum.de}
	\affiliation{Department of Chemistry, Technical University of Munich (TUM), D-85747 Garching, Germany}

\date{\today}% It is always \today, today,

% PACS, the Physics and Astronomy Classification Scheme.

\begin{abstract}
For paving the way to novel applications in quantum simulation, computation, and technology,
increasingly large quantum systems have to be steered with high precision.
It is a typical task amenable to numerical optimal control to turn the time course
of pulses, i.e. piecewise constant control amplitudes, iteratively into an optimised shape.
Here, we present the first comparative study of optimal control algorithms for a wide
range of finite-dimensional applications. We focus on the most commonly used algorithms: 
\grape methods which update all controls concurrently, and \krotov-type methods which do
so sequentially. Guidelines for their use are given and open research questions are pointed out. ---
Moreover we introduce a novel unifying algorithmic framework, \dynamo
({\em dynam}ic {\em o}ptimisation platform) designed to provide the quantum-technology community 
with a convenient \matlab-based toolset for optimal control. In addition, it gives researchers in 
optimal-control techniques a framework for benchmarking and comparing new proposed algorithms
to the state-of-the-art. It allows for a mix-and-match approach  with various types
of gradients, update and step-size methods as well as \subspace choices. 
Open-source code including examples is made available at \href{http://qlib.info}{http://qlib.info}.

\end{abstract}

\maketitle
%%%%%%%%%%%%%%%%%%%%%%%%%%%%%%%%%%%%%%%%%%%%%%%%%%%%%%%%%%%%%%%%%%%
%%%%%%%%%%%%%%%%%%%%%%%%%%%%%%%%%%%%%%%%%%%%%%%%%%%%%%%%%%%%%%%%%%%
%%%%%%%%%%%%%%%%%%%%%%%%%%%%%%%%%%%%%%%%%%%%%%%%%%%%%%%%%%%%%%%%%%%

%%%

%\noindent
%\fbox{
%\RED{PLEASE OBSERVE: BRITISH ENGLISH \& MACROS}}
%optimise, realise, synthesise, centre, colour, neighbour;
%{\verb|\grape|},
%{\verb|\krotov|},
%{\verb|\matlab|},
%{\verb|\lbfgs|},
%{\verb|\bfgs|},
%{\verb|\dynamo|}

\section{Introduction}
%%%%%%%%%%%%%%%%%%%%%%
%%%%%%%%%%%%%%%%%%%%%%
For unlocking the inherent quantum treasures of future quantum technology,
it is essential to steer experimental quantum dynamical systems in a fast, accurate, and
robust way \cite{DowMil03, WisMil09}.
While the accuracy demands in quantum computation (the \/`error-correction threshold\/')
may seem daunting at the moment, quantum simulation is far less sensitive.

In practice, using coherent superpositions as a resource is often tantamount to protecting
quantum systems against relaxation without compromising accuracy. In order to tackle these
challenging quantum engineering tasks, optimal control algorithms are establishing themselves
as indispensable tools. They have matured from principles \cite{Sam90+} and early
implementations \cite{TR85,Rabitz87,Rabitz90} via spectroscopic applications \cite{KLG,XYOFR04,Poetz06}
to advanced numerical algorithms \cite{Krotov,GRAPE} for state-to-state transfer and quantum-gate
synthesis \cite{PRA05} alike.

In engineering high-end quantum experiments, 
progress has been made in many areas including cold atoms in optical lattice potentials \cite{GMEHB02,BDZ08},
trapped ions \cite{LBM+03, GZC03, GZC05, DCZ05, BW08, Wunder09, Wunder10}, 
and superconducting qubits \cite{Mart09,CCG09} to name just a few.
To back these advances, optimal control among numerical tools have become increasingly important, 
see, e.g., \cite{SK-RMP10} for a recent review.
For instance, near time-optimal control may take pioneering realisations of solid-state qubits
being promising candidates for a computation platform \cite{CW_Nat08},
from their fidelity-limit to the decoherence-limit \cite{PRA07}.
More recently, open systems governed by a Markovian master
equation have been addressed \cite{PRL_decoh}, and even smaller non-Markovian subsystems can be tackled, if they can be
embedded into a larger system that in turn interacts in a Markovian way with its environment \cite{PRL_decoh2}.
Taking the concept of decoherence-free \subspaces \cite{ZanRas97,KBL+01} to more realistic scenarios,
avoiding decoherence in encoded \subspaces \cite{NS09} complements recent approaches
of dynamic error correction \cite{KV08,KV09a}.--- Along these lines, quantum control is anticipated
to contribute significantly to bridging the gap between quantum principles demonstrated in pioneering
experiments and high-end quantum engineering \cite{DowMil03,WisMil09}.

\subsection*{Scope and Focus}

The schemes by which to locate the optimal control sequence within the space of
possible sequences are varied. The values taken by the system controls over time may
be parameterised by piece-wise constant control amplitudes in the time domain, or
in frequency space \cite{CRAB}, by splines 
or other methods. For specific aspects of the toolbox of
quantum control, see {e.g.} \cite{Lloyd00, PK02, PK03, GZC03, OTR04, GRAPE, PRA05, Tarn05,
ST04+06, MSZ+06, PRA07,  Rabitz07, SP09, NS09, Schi09, PRL_decoh2, DRAG}, while a recent
review can be found in \cite{dAll08}.
Here, we concentrate on piece-wise constant controls in the time domain.
For this parametrisation of the control space, there are two well-established
optimal control approaches: \krotov-type methods \cite{KK83,KK99,PK02,PK03} which update all
controls within a single {\timeslice} once before proceeding on to the next {\timeslice}
(cycling back to the first slice when done), and \grape-type methods \cite{GRAPE}
which update all controls in all \timeslices concurrently. Here we refer to the former as
\emph{sequential-update schemes} and to the latter as \emph{concurrent-update schemes}.

Sequential methods have mainly been applied to provide control fields in 
(infinite-dimensional) systems of atomic and molecular optics 
characterised by energy potentials \cite{PK02,PK03,KK10,K10}, 
while concurrent methods have mostly been applied
to (finite-dimensional) qubit systems of spin nature \cite{GRAPE,PRA05},
or to Josephson elements \cite{PRA07,PRL_decoh2}, ion traps \cite{TGW08,NHR09},
or 2D-cavity grids in quantum electrodynamics \cite{PRB10}.
Here we compare sequential vs.~concurrent algorithms in finite-dimensional systems.

Both of the methods require a mechanism to control the selection of
the next point to sample. For sequential-update methods, which perform a single or
few iterations per parameter \subspace choice, first-order methods are most often used;
yet for algorithms repeatedly modifying the same wide segment of parameter space at
every iteration, second-order methods, such as the well-established one by 
Broyden-Fletcher-Goldfarb-Shanno (\bfgs) \cite{NocWri06}, seem better-suited. These
choices, however, are by no means the final word and are subject of on-going research.

Controlling quantum systems via algorithms on classical computers naturally comes 
with unfavourable scaling. Thus it is essential to optimise the code by minimising
the number of operations on matrices which scale with the system size,
and by parallelising computation on high-performance clusters. 
While elements of the latter have been accomplished \cite{EP06}, here we focus on the former. 

{To this end, we present a new unifying programming framework, the \dynamo platform,
allowing to combine different methods of \subspace selection, gradient calculation,
update controls, step-size controls, etc. The framework allows for benchmarking the
various methods on a wide range of problems in common usage, allowing future research
to quickly compare proposed methods to the current state-of-the-art. It also makes
significant strides towards minimising the number of matrix operations required for serial,
concurrent, and generalised hybrid schemes. Full \matlab code of the platform is provided
to the community alongside this manuscript at \href{http://qlib.info}{http://qlib.info}.}
 --- 
We benchmark \krotov-type algorithms and \grape algorithms over a selection
of scenarios, giving the user of control techniques guidelines
as to which algorithm is appropriate for which problem.

\medskip
The paper is organised as follows: 
In Sec.~II we provide a generalised algorithmic framework
embracing the established algorithms \grape and \krotov as limiting cases.
Sec.~III shows how the formal treatment applies to concrete standard settings of 
optimising state transfer and gate synthesis in closed and open quantum systems.
In Sec.~IV we compare the computational performance of concurrent vs.~sequential
update algorithms for a number of typical test problems of synthesising gates or
cluster states. Computational performance is discussed in terms of
costly multiplications and exponentials of matrices.

Sec.~V provides the reader with an outlook
on emerging guidelines  as to which type of problem asks for which flavour of
algorithm in order not to waste computation time. --- Finally, we point at a list of
open research questions, in the persuit of which \dynamo is anticipated to prove useful.

%%%%%%%%%%%%%%

%%%%%%%%%%%%%%%%%%
\section{Algorithmic Settings}
%%%%%%%%%%%%%%%%%%
%%%%%%%%%%%%%%%%%%
Most of the quantum control problems boil down to a single general form, namely
steering a dynamic system following an internal drift under additional external controls,
such as to maximise a given figure of merit.
Because the underlying equation of motion is taken to be linear both in the drift as well as in
the control terms, dynamic systems of this form are known as {\em bilinear control systems} $(\Sigma)$
\begin{equation}\label{eqn:bilinear_contr2}
         \dot X(t) = -\big(A + \sum_{j=1}^m u_j(t) B_j\big) \; X(t)
\end{equation}
with \/`state\/' $X(t)\in\C N$, drift $A\in \Mat_N(\C{})$, controls $B_j\in \Mat_N(\C{})$,
and control amplitudes $u_j(t)\in\R{}$. Defining the $A_u(t):= A + \sum_{j=1}^m u_j(t) B_j$
as generators, the formal solution reads
\begin{equation}
X(t) = \mathbb T\, \exp\big\{-\int\limits_0^t d\tau\, A_u(\tau)\big\}\;X(0)\quad,
\end{equation}
where $\mathbb T$ denotes Dyson's time ordering operator. ---
In this work, the pattern of a bilinear control system will turn out to
serve as a convenient unifying frame for applications in closed and open
quantum systems, which thus can be looked upon as a variation of a theme.

\subsection{Closed Quantum Systems}

Throughout 
this work we study systems that are 
{\em fully controllable} \cite{SJ72JS,RaRa95,Lloyd96,TOSH-Diss,SchiFuSol01,SchiPuSol01,AA03},
i.e.~those in which---neglecting relaxation---every unitary gate can be
realised. Finally, unless specified otherwise, we allow for unbounded control amplitudes.

Closed quantum systems are defined by
the system Hamiltonian $H_d$ as the only {\em drift term}\/, while the
\/`switchable\/' {\em control Hamiltonians} $H_j$ express external manipulations in
terms of the quantum system itself, where each control Hamiltonian can be
steered in time by its (here piece-wise constant) {\em control amplitudes} $u_j(t)$.
Thus one obtains a bilinear control system in terms of the controlled Schr{\"o}dinger equations
\begin{eqnarray}
         \ket{\dot\psi(t)} &=& -i\big(H_d + \sum_{j=1}^m u_j(t) H_j\big) \;\ket{\psi(t)}
         \label{eqn:schroed_contr1}\\
         {\dot U(t)} &=& -i\big(H_d + \sum_{j=1}^m u_j(t) H_j\big) \;{U(t)} \quad,
         \label{eqn:schroed_contr2}
\end{eqnarray}
where the second identity can be envisaged as lifting the first one to an operator equation.
For brevity we henceforth concatenate all Hamiltonian components and write
\begin{equation}\label{eqn:Ham-c}
H_u(t):= H_d + \sum_{j=1}^m u_j(t) H_j\quad.
\end{equation}
Usually one wishes to absorb unobservable global phases by taking
density-operator representations of states $\rho(t)$. Their
time evolution is brought about by unitary conjugation
$\widehat{U}(\cdot) := U(\cdot)U^\dagger\equiv \Adr_U(\cdot)$ generated by commutation
with the Hamiltonian 
${\widehat{H}}_u(\cdot):= [H_u, (\cdot)]\equiv\adr_{H_u}(\cdot)$.
So in the projective representation in Liouville space, Eqns.~\eqref{eqn:schroed_contr1} and
\eqref{eqn:schroed_contr2} take the form
\begin{eqnarray}
\dot \rho(t) &=&  -i \widehat{H}_u\; \rho(t)\\[0mm]
\tfrac{d}{dt} {\widehat{U}}(t) &=& -i \widehat{H}_u\; \widehat{U}(t)\quad.
\end{eqnarray}
It is now easy to accommodate dissipation to this setting.

\subsection{Open Quantum Systems}
Markovian relaxation can readily be introduced on the level of the equation
of motion by the operator $\Gamma$, which may, e.g., take the GKS-Lindblad form.
Then the respective controlled master equations for state transfer
and its lift for gate synthesis read
\begin{eqnarray}%\label{eqn:master}
\dot{\rho}(t) &=& -(i \widehat{H}_u \,+\,\Gamma)\; \rho(t) \label{eqn:master}\\[0mm]
\dot F(t)  &=& -(i \widehat{H}_u\,+\,\Gamma)\; F(t) \quad.\label{eqn:super_master}
\end{eqnarray}
Here $F$ denotes a {\em quantum map} in ${\rm GL(N^2)}$ as linear image over all basis
states of the Liouville space representing the open system, where
henceforth $N:=2^n$ for an $n$"~qubit system.
Note that only in the case of $[\widehat{H}_u \,,\,\Gamma\,]=0$ the map $F(t)$ boils down to a
mere contraction of the unitary conjugation $\widehat{U}(t)$.
In the generic case, it is the intricate interplay of the respective coherent ($i\widehat{H}_u$)
and incoherent ($\Gamma$) part of the time evolution \cite{DHKS08}
that ultimately entails the need for relaxation-optimised control based on the full knowledge
of the master Eqn.~\eqref{eqn:super_master}.

%%%%%%%%%%%%%%%%%%
\subsection{Figures of Merit}
%%%%%%%%%%%%%%%%%%

In this work, we treat quality functions only depending on the final state $X(T)$ of
the system without taking into account running costs, which, however, is no principal limitation
\footnote{
  Depending on the state of the system $X(t)$ over a time interval $[0,T]$
  and on the %corresponding 
  control amplitudes $u(t)$, 
  a general quality function may be formulated to take the form\\
\mbox{
  $
  \qquad f := f_T\big(X(T),T\big) + \int\limits_{0}^{T} f_0\big(X(t),t,u(t)\big)\, dt\;,
  $
}\\
  where $f_T\big(X(T),T\big)$ is the component solely depending on the final
  state of the system $X(T)$ and independent of the control amplitudes, 
  while $f_0\big(X(t),t,u(t)\big)$ collects the {\em running costs} usually 
  depending on the amplitudes $u(t)$. In optimal control and variational calculus,
  the general case ($f_T\neq 0, f_0\neq 0$) is known as {\em problem of {\sc Bolza}}, 
  while the special case of zero running costs ($f_T\neq 0, f_0=0$) is termed 
  {\em problem of {\sc Mayer}}, whereas ($f_T=0, f_0\neq 0$) defines the {\em problem of {\sc Lagrange}}.
  --- 
  Henceforth, we will not take into account any running costs 
  (thereby also allowing for unbounded control amplitudes). Thus here all our problems take
  the form of {\sc Mayer}. On the other hand, many applications of {\sc Krotov}-type algorithms have
  included explicit running costs \cite{PK02,PK03,KK10,K10} to solve problems of {\sc Bolza} form,
  which are also amenable to \grape (as has been shown in \cite{GRAPE}). 
  --- 
  Yet, though well known, it should be pointed out again that 
  a problem of {\sc Bolza} can always be transformed 
  into a problem of {\sc Mayer}, and ultimately all the three types of problems above
  are equivalent \cite{Berkovitz76}, which can even be traced back to the pre-control
  era in the calculus of variations \cite{Bliss46}. 
  The implications for convergence of the respective algorithms are treated in detail
  in \cite{SF11}.
}.

No matter whether the $X(t)$ in Eqn.~\eqref{eqn:bilinear_contr2} denote states or gates, a common
natural figure of merit is the projection onto the target in terms of the overlap %expressed by
\begin{equation}
 g = \tfrac{1}{{\| X_{\rm target}\|}_2} \tr\{X^\dagger_{\rm target} X(T)\}\quad.
\end{equation}
Depending on the setting of interest, one may choose as
the actual figure of merit $f_{\rm SU}:= \Re g$ or $f_{\rm PSU}:=|g|$.

More precisely, observe there are two scenarios
for realising quantum gates or modules $U(T)\in {\rm SU(N)}$ with maximum trace fidelities:
Let
\begin{equation}\label{eq:overlap}
g:= \tfrac{1}{N} \tr \{U_{\rm target}^\dagger U(T)\}
\end{equation}
define the normalised overlap of the generated gate $U(T)$ with the target.
Then the quality function 
\begin{equation}
f_{{\rm SU}} := \tfrac{1}{N}\, \Re \tr \{U_{\rm target}^\dagger U(T)\} \;=\; \Re g
\end{equation}
covers the case where overall global phases shall be
respected, whereas if a global phase is immaterial \cite{PRA05},
another quality function $f_{\rm PSU}$ applies, whose square reads
\begin{equation}
f^2_{{\rm PSU}} := \tfrac{1}{N^2}\, \Re \tr \{\widehat{U}_{\rm target}^\dagger\, \widehat{U}(T)\}
 = \big| g \big|^2\quad.
\end{equation}
The latter identity is most easily seen \cite{PRA05} in the so-called
$vec$-representation \cite{HJ2} of $\rho$, where
$\widehat{U} = \bar{U}\otimes U\in {\rm PSU(N)}$
(with $\bar U$ as the complex conjugate)
%and the commutator superoperator $\widehat{H} = \unity\otimes H - H^t\otimes\unity$):
recalling the projective unitary group is 
${\rm PSU(N)}=\tfrac{\rm U(N)}{\rm U(1)} = \tfrac{\rm SU(N)}{\Z_N}$.
Now observe that
$\tr\{(\bar U \otimes U)(\bar V \otimes V)\}=
\tr\{\bar U\bar V \otimes UV\}=|\tr\{UV\}|^2$.

\bigskip
%%%%%%%%%%%%%%%%
%%%%%%%%%%%%%%%%
%%%%%%%%%%
\begin{table}[Ht!]
\begin{center}
\caption{Bilinear Quantum Control Systems}
\label{tab:glossary}
\begin{tabular}{ll|cccc}
\hline\hline\\[-1mm]
\multicolumn{2}{l}{Setting and Task }&\phantom{.}& Drift &\phantom{.}& Controls \\[0mm]
\multicolumn{2}{l}{$\dot X(t) = -\big(A + \sum_j u_j(t) B_j\big) X(t)$} &\phantom{.}& $A$ &\phantom{.}&  $B_j$ \\[1mm]
\hline\\[-3.7mm]
{} &          &&      &&  \\[-2mm]
{\em closed systems:} &          &&      &&  \\[0mm]
\; pure-state transfer & $X(t) = \ket{\psi(t)}$ \hspace{0mm} && $i H_0$  && $i H_j$  \\[0mm]
\; gate synthesis I & $X(t) = U(t)$ \hspace{0mm} && $i H_0$  && $i H_j$  \\[0mm]
\; state transfer & $X(t) = \rho(t)$ \hspace{0mm} && $i \widehat{H}_0$  && $i \widehat{H}_j$  \\[0mm]
\; gate synthesis II & $X(t) = \widehat{U}(t)$ \hspace{0mm} && $i \widehat{H}_0$  && $i \widehat{H}_j$  \\[2mm]
\hline%\\[-1mm]
{} &          &&      &&  \\[-2mm]
{\em open systems:}   &         &&              &&          \\[0mm]
\; state transfer & $X(t) = \rho(t)$ \hspace{0mm} && $i \widehat{H}_0 + \Gamma$ &&  $i \widehat{H}_j$  \\[0mm]
\; map synthesis & $X(t) = F(t)$ \hspace{0mm} && $i \widehat{H}_0 + \Gamma$ &&  $i \widehat{H}_j$  \\[2mm]
\hline\hline
\end{tabular}\hspace{15mm}
\end{center}
\end{table}
%%%%%%%%%%
%%%%%%%%%%%%%%%%%
%%%%%%%%%%%%%%%%%

\subsection{Core of the Numerical Algorithms:\\ Concurrent and Sequential}\label{sec:Alg-core}
%Having seen that 
Since the equations of motion for closed and open quantum systems
as well as the natural overlap-based quality functions are of common form,
we adopt the unified frame for the numerical algorithms
to find optimal steerings $\{u_j(t)\}$.
To this end, we describe first-order and second-order methods to
iteratively update the set of control amplitudes in a unified way
for bilinear control problems.
%%%%%%%%%%%%%%%%%%%%%%%%%%%%%%%%%%%%%%%%%%
\begin{table}[Ht!]
\begin{center}
\caption{List of Symbols}
\label{tab:nomenclature}
\begin{tabular}{ll}
\hline\hline\\[-1mm]
\; Symbol & Meaning \\[2mm]
\hline\\[-1mm]
\; $j$ & control Hamiltonian index ($1{\ldots}m$) \\
\; $k$ & \timeslice index ($1{\ldots}M$) \\
\; $u_j(t_k)$ & control amplitude to Hamiltonian $j$ \\
		& in {\timeslice} $k$ (more lables below)\\
\; $A$ & non-switchable drift term (see Tab.~\ref{tab:glossary})  \\
\; $B_j$ & switchable control terms (see Tab.~\ref{tab:glossary})  \\
\; $X_0$ & initial condition (see Tab.~\ref{tab:glossary-ini}) \\
\; $X_{\rm target}\equiv X_{M+1}$ & final condition (see Tab.~\ref{tab:glossary-ini}) \\
\; $X_k$ & propagator from time $t_{k-1}$ to $t_k$ \\
\; $X_{k:0}$ & {\em forward} propagation of initial state\\[-.5mm] 
		&up to time $t_k$, i.e.\ $X_k X_{k-1}\cdots X_1 X_0$\\
\; $X_{M+1:k+1}$ & {\em backward} propagation of target state\\[-.5mm]
		&up to time $t_k$, i.e.\ $X^\dagger_{\rm target} X_M\cdots X_{k+1}$ \\[2mm] %\\[.5mm]
\hline\\[-2mm]
\; $q$ & \subspace selection counter (outer loop) \\
\; $s$ & step-within-\subspace counter (inner loop) \\
\; $r$ & global counter (overall number of steps) \\
\; $\mathcal{T}^{(q)}\subseteq\{1{\ldots}M\}$ & set of \timeslices belonging to \subspace $q$ \\  % \subseteq\{1{\ldots}M\}
\; $M^{(q)}$ & number of \timeslices in $\mathcal{T}^{(q)}$ \\
\; $t^{(q)}_k$ & tag members of $\mathcal{T}^{(q)}$ with $k\in\{1{\ldots}M^{(q)}\}$ \\
\; $u_{j}^{(r)}(t_k^{(q)})$ & control amplitude to Hamiltonian $j$ for  \\[-.5mm]
\; & \subspace $q$, {\timeslice} $t^{(q)}_k$, iteration $r$ \\[1mm]
\; $f$ & figure(s) of merit \\[2mm]
\hline\hline
\end{tabular}\hspace{15mm}
\end{center}
\end{table}
%%%%%%%%%%%%%%%%%%%%%%%%%%%%%%%%%%%%%%%%%%%%

\subsubsection*{Discretising Time Evolution} 
For algorithmic purposes one discretises the time evolution. 
To this end, the {\em control terms} $B_j$ are switched
by piecewise constant {\em control amplitudes} $u_j(t_k)\in \mathcal U \subseteq \mathbb R$
with $t_k\in[0,T]$, where $T$ is a fixed final time and $ \mathcal U$
denotes some subset of admissible control amplitudes.
For simplicity, we henceforth assume equal discretised time spacing $\Delta t:= t_k-t_{k-1}$
for all \timeslices $k=1,2,\dots,M$. So $T=M\Delta t$.
Then the total generator (i.e.~Hamiltonian or Lindbladian)
governing the evolution in the time interval $(t_{k-1},t_k]$
shall be labelled by its final time $t_k$ as
\begin{equation}
 A_u(t_k) := A + \sum_j u_j(t_k) B_j
\end{equation}
generating the propagator
\begin{equation}
 X_k:= e^{-\Delta t A_u(t_k)}
\end{equation}
which governs the controlled time evolution in the \timeslice $(t_{k-1},t_k]$.
%%%%
%%%%
\begin{figure}[Ht!]
 (a) concurrent (\grape-type) $\hfill$\\[1mm] \includegraphics[width=.99\linewidth]{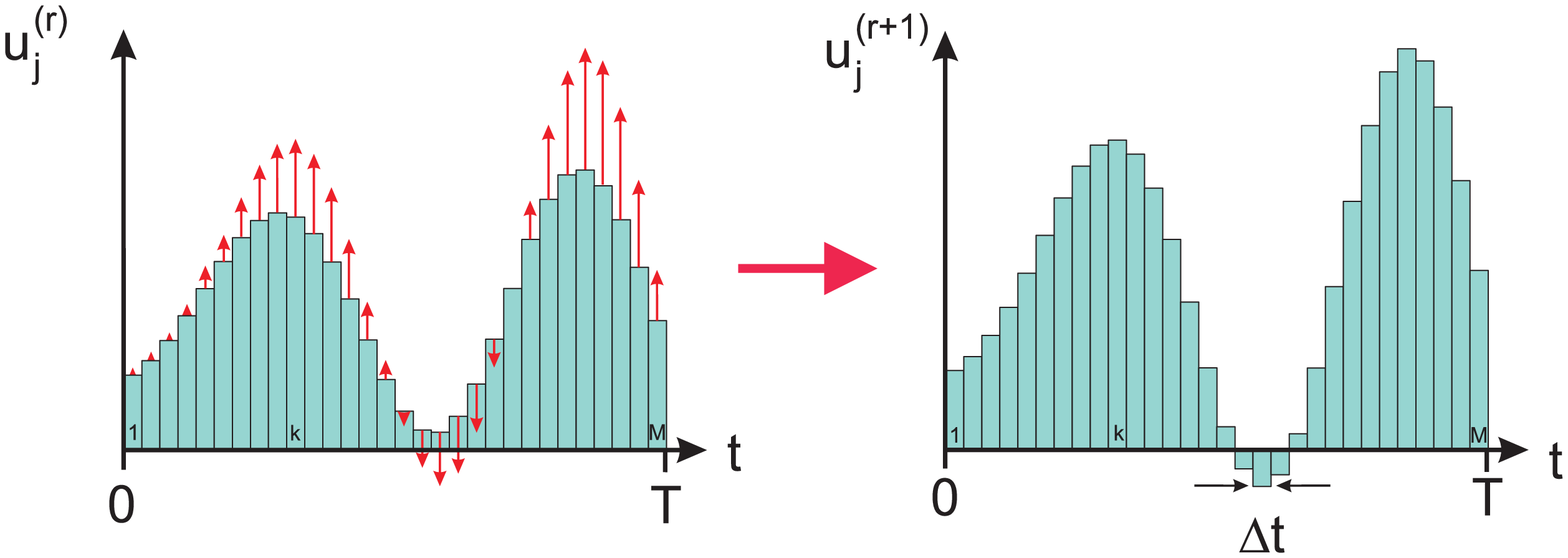}\\[2mm]
 (b) sequential (\krotov-type) $\hfill$\\[1mm] \includegraphics[width=.99\linewidth]{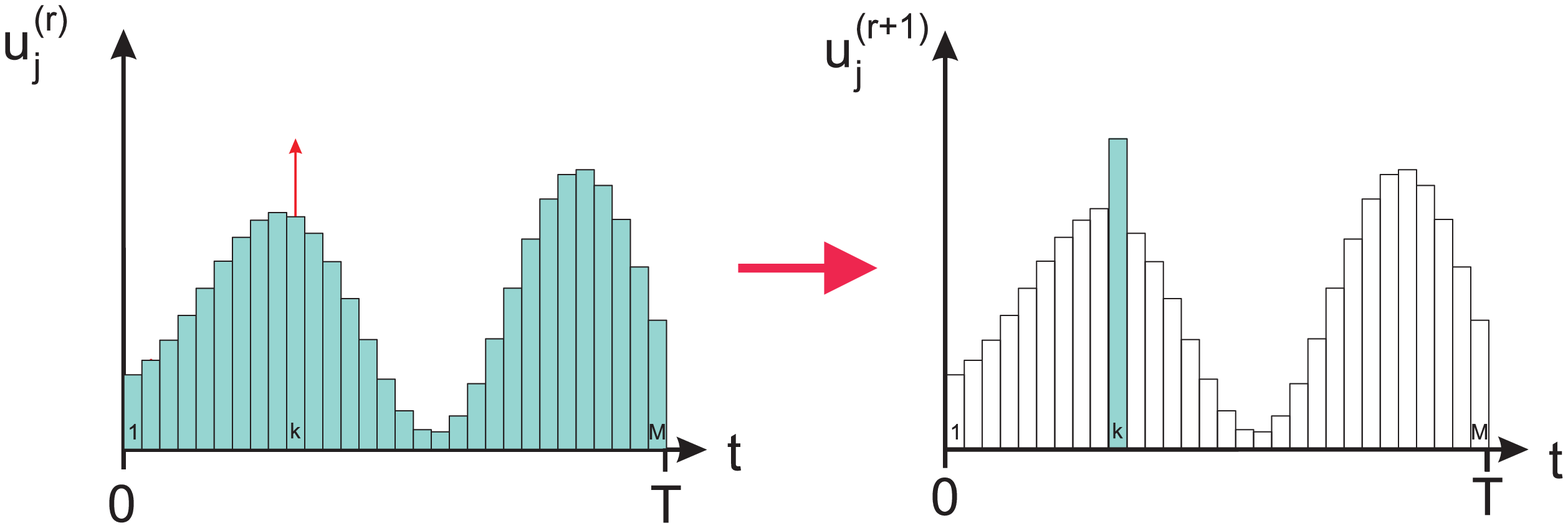}\\[2mm]
 (c) hybrid $\hfill$\\[1mm] \includegraphics[width=.99\linewidth]{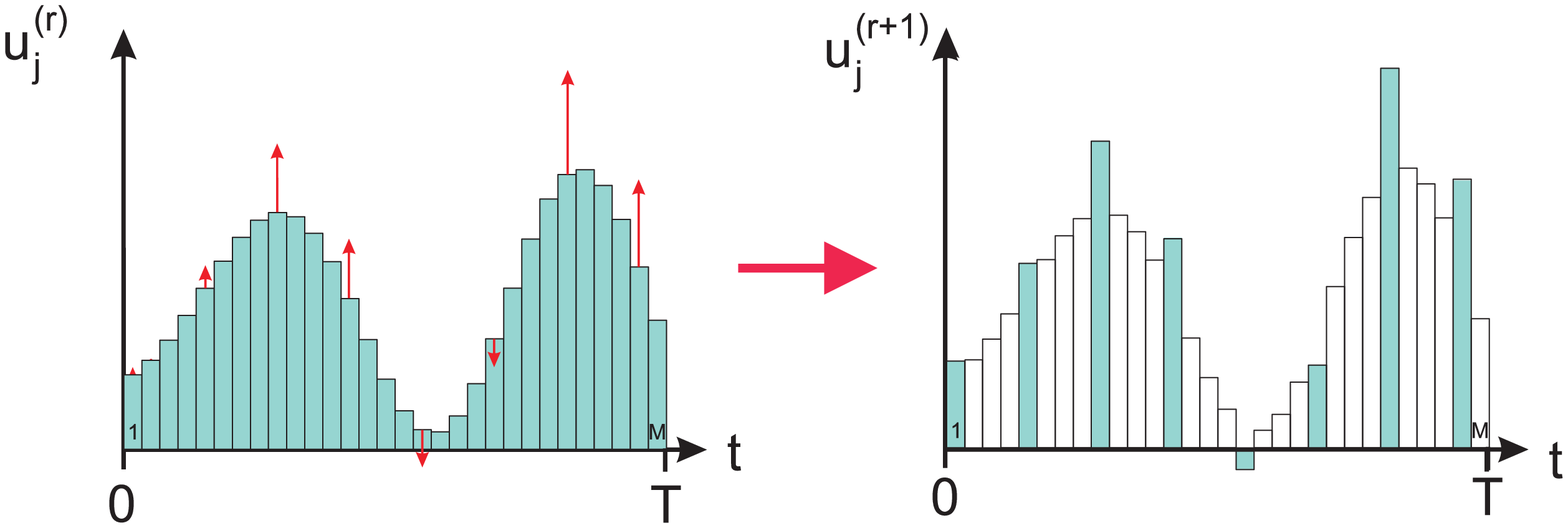}
 \caption{\label{fig:GKH-algs} (Colour online)
 Overview on the update schemes of gradient-based optimal control algorithms in terms of the
 set of \timeslices   $\mathcal T^{(q)}=\{k_1^{(q)}, k_2^{(q)}, \dots k_{M^{(q)}}^{(q)}\}$ for which the
 control amplitudes are concurrently updated in each iteration.
\Subspaces are enumerated by $q$, gradient-based steps within each \subspace by $s$, and $r$ is the global step counter.
 In \grape (a) all the $M$ piecewise constant control amplitudes are updated
 at every step, so $\mathcal T^{(1)}=\{1, 2,\dots M\}$ for the single iteration $q{\equiv}1$.
 Sequential update schemes (b) update a single {\timeslice} once, in the degenerate inner-loop $s{\equiv}1$,
 before moving to the subsequent {\timeslice} in the outer loop, $q$;
 therefore here $\mathcal T^{(q)}=\{q \mod M\}$.
 Hybrid versions (c) follow the same lines: for instance,
 they are devised such as to update a (sparse or block) subset of $p$ different \timeslices
 before moving to the next (disjoint) set of \timeslices.
 } % caption
 \end{figure}
%%%%
%%%%%%%%%%%%%%%%%%%%
Next, we define as boundary conditions $X(0):=X_0$ and $X_{M+1}:= {X}_{\rm target}$.
They specify the problem and are therefore discussed in more detail in Sec.~\ref{sec:appl}, Tab.~\ref{tab:glossary-ini}.
A typical problem is unitary gate synthesis, where $X_0\equiv\unity$ and $X_{\rm target} \equiv U_{\rm target}$,
whereas in pure-state transfer $X_0\equiv\ket{\psi_0}$ and $X_{\rm target} \equiv \ket{\psi}_{\rm target}$. ---
In any case, the state of the system is given by the discretised evolution
\begin{equation}
 X(t_k)=X_{k:0}:=X_k X_{k-1}\cdots X_1 X_0\quad.
\end{equation}
Likewise, the state of the adjoint system also known as {\em co-state} $\Lambda^\dagger(t_k)$ results
from the backward propagation of $X_{M+1}\equiv X_{\rm target}$
\begin{equation}
\begin{split}
 \Lambda^\dagger(t_k) &:=X^\dagger_{\rm target} X_M X_{M-1}\cdots X_{k+1}\\
                      &\phantom{:}= X^\dagger_{M+1}  X_M X_{M-1}\cdots X_{k+1}=: X_{M+1:k+1}
\end{split}
\end{equation}
which is needed to evaluate the figure of merit here taken to be
\begin{equation}
f:=\tfrac{1}{N} |\tr \{\Lambda^\dagger(t_k) X(t_k)\}| = |\tr \{X^\dagger_{\rm target}  X(T)\}|\,\,{\forall}k
\end{equation}
as the (normalised) projection of the final state under controlled discretised
time evolution upto time $T$ onto the target state.

\bigskip
\subsubsection*{Algorithmic Steps} 
\begin{figure}[Ht!]
\begin{center}
\includegraphics[width=0.45\textwidth]{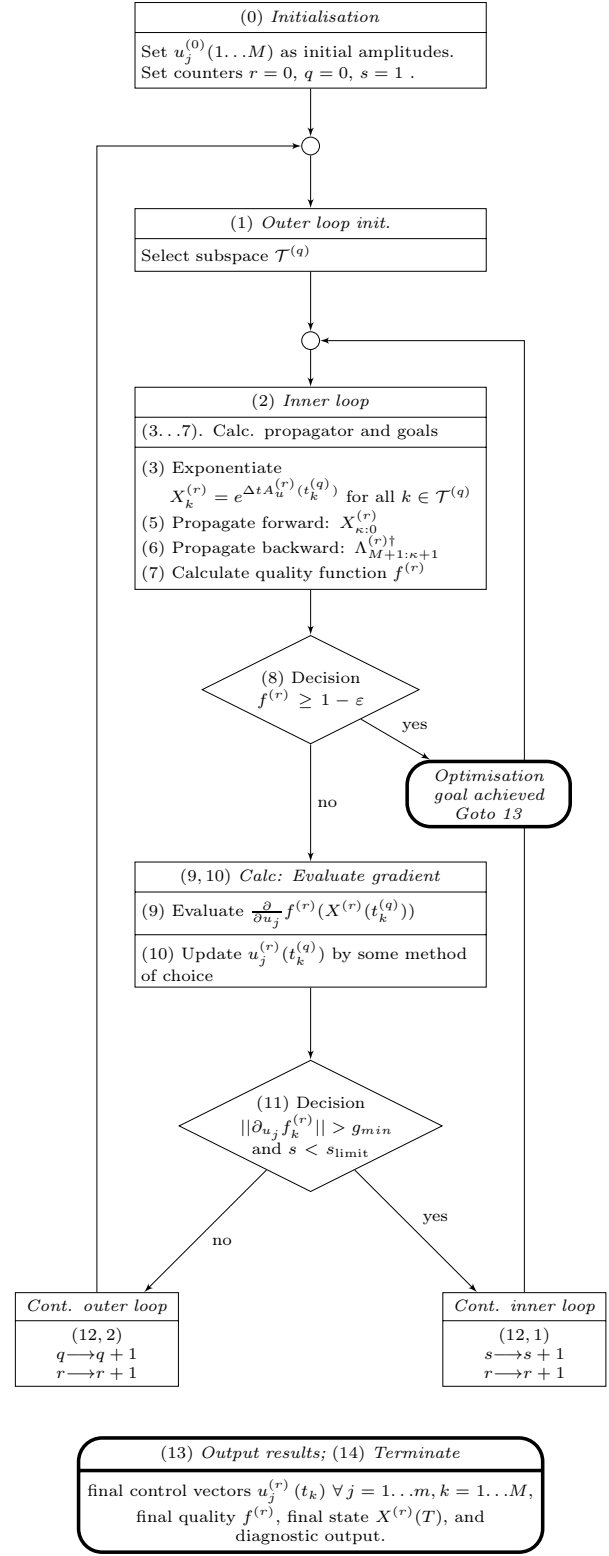}
\caption{Flow diagram for the generalised \dynamo optimal control search
	embracing standard \grape and \krotov methods as limiting special cases.
    \label{fig:flowchart}}
\end{center}
\end{figure}

With the above stipulations, one may readily characterise the core algorithm by the following
steps, also illustrated in Fig.~\ref{fig:GKH-algs} and the flowchart in Fig. ~\ref{fig:flowchart}.
%%%%%%%%%%%%%%%%%%%%%%%%%%%%%%%%%%%%%%
%%%%%%%%%%%%%%%%%%%%%%%%%%%%%%%%%%%%%%

\bigskip
\squishlist
    \item[0.] Set initial control amplitudes $u_j^{(0)}(t_k)\in\mathcal U \subseteq \mathbb R$
        for all times $t_k$ with $k\in\mathcal T^{(0)}:=\{1,2,\dots,M\}$ then set counters 
	$r=0$, $q=0$, $s=1$; fix $s_{\rm limit}$ and $f'$.
    \item[1.] \emph{Outer loop start}, enumerated by $q$: \\
        Unless $r=q=0$, choose a selection of \timeslices, i.e.\ a \subspace, $\mathcal T^{(q)}$, 
	on which to perform the next stage of the search
        will update only $u_{j}^{(r)}(t_k^{(q)})$ for $t_{k\in\{1{\ldots}M^{(q)}\}}^{(q)}\in\mathcal{T}^{(q)}$.
        \squishlist
            \item[2.] \emph{Inner loop,} enumerated by $s$: \\
		Take one or more gradient-based steps within the \subspace.
                Depending on \subspace choice, number of matrix operations may be reduced as compared to the naive implementation of the algorithm.
            \squishlist
                 \item[3.] {\em Exponentiate:} $X_k^{(r)} = e^{\Delta t {A_u^{(r)}}(t_k^{(q)})}$ for all $k\in\mathcal T^{(q)}$
                        with ${A}^{(r)}_u(t_k^{(q)}) := {A} + \sum_j u^{(r)}_j(t_k^{(q)}) {B}_j$
                 \item[4.] Compute goal function at some $k=\kappa$:
                 \squishlist
                        \setlength{\leftmargin}{0.0em}
                        \item[5.] {\em Forward propagation:} \\
				  $X^{(r)}_{\kappa:0} := X^{(r)}_k  X^{(r)}_{k-1}\cdots X^{(r)}_1 X_0$
                        \item[6.] {\em Backward propagation:} \\
				  $\Lambda^{(r)\dagger}_{M+1:\kappa+1} := {X}^{\dagger}_{\rm target}
						 X^{(r)}_M  X_{M-1}^{(r)}\cdots X^{(r)}_{k+1}$
                        \item[7.] {\em Evaluate current fidelity:} \\
                           %\mbox{$f^{(r)}=\tfrac{1}{N^2} \Re \tr\big\{ \Lambda^{(r)\dagger}_{M+1:{\kappa}+1} 
			%		X^{(r)\phantom{ \dagger}}_{{\kappa}:0} \big\} $}\\[0.5mm]
			%		\phantom{XXJ}{$=\tfrac{1}{N^2} \Re \tr\big\{\widehat{X}^\dagger_{\rm tar}%
			%			X^{(r)\phantom{ \dagger}}_{M:0}\big\}$} for some $k$\\
						{\mbox{$f^{(r)}=\tfrac{1}{N} | \tr\big\{ \Lambda^{(r)\dagger}_{M+1:{\kappa}+1} 
					X^{(r)\phantom{ \dagger}}_{{\kappa}:0} \big\} $}|\\[0.5mm]
					\phantom{XXJ}{$=\tfrac{1}{N} | \tr\big\{{X}^\dagger_{\rm tar}%
						X^{(r)\phantom{ \dagger}}_{M:0}\big\}$}| for some $k$}
                 \squishend % Compute goal function
                 \item[8.] If $f^{(r)}\geq 1-\varepsilon_{\rm threshold}$, done: goto step 13.
                 \item[9.] Else, {\em calculate gradients} $\frac{\partial f^{(r)}(X^{(r)}(t_k^{(q)}))}%
									{\partial u_j(t_k^{(q)})}$ 
				for all $k\in\mathcal T^{(q)}$
                 \item[10.] {\em Gradient-based update} step: $u^{(r)}_j(t_k^{(q)})\mapsto u^{(r+1)}_j(t_k^{(q)})$ 
				for all $k\in\mathcal T^{(q)}$ by a method of choice
                                (e.g., Newton, quasi-Newton, \bfgs or \lbfgs, conjugate gradient etc.)
                  \item[11.] If $s < s_{\rm limit}$ and $||\frac{\partial f^{(r)}_k}{\partial u_j}|| < f'_{\rm limit}\ 
				\forall k\in\mathcal{T}_k^{(r)}$, then set and $s{\longrightarrow}s+1,\ r{\longrightarrow}r+1$  
				and return to step 3
            \squishend % inner loop
            \item[12.] $q{\longrightarrow}q+1$. Choose a new \subspace $\mathcal T^{(q)}$ and return to step 2
        \squishend % outer loop
\item[13.] {\em Output:}\\ final control vectors $\{u^{(r)}_j(t_k)|k=1,2,\dots,M\}$ for all controls $j$, 
			final quality $f^{(r)}$, final state $X^{(r)}(T)$, and diagnostic output.
\item[14.] {\em Terminate.}
\squishend % algorithm

\bigskip
Having set the frame, one may now readily compare the \krotov and \grape approaches:
In \krotov-type algorithms, we make use of a sequential update scheme,
where $\mathcal{T}^{(q)}=\{q \mod M\}$ and $s_{\rm limit}=1$, implying the inner loop is degenerate,
as only a single step is performed per \subspace selection, giving $s{\equiv}1, r=s$. With \grape,
a concurrent update scheme, $\mathcal{T}^{(q)}=\{1{\ldots}M\}$, i.e.\ the entire parameter set is
updated in each step of the inner loop, implying $q{\equiv}1,\ r=s$ and the outer loop is degenerate.

The above construction naturally invites hybrids: algorithms where the \subspace size
is arbitrary in the $1{\ldots}M$ range and where the size of the \subspace to be updated in each step $q$
as well as the number of steps within each \subspace, $s$, can vary dynamically with iteration, depending,
e.g., on the magnitude of the gradient and the distance from the goal fidelity.
This is a subject of on-going research.

\subsection{Overview of the \dynamo Package and Its Programming Modules}
\dynamo provides a flexible framework for optimal-control
algorithms with the purpose of allowing
(i) quick and easy optimisation for a given problem using the existing set of optimal-control
     search methods as well as
(ii) flexible environment for development of and research into new algorithms.

For the first-use case, the design goal is to make optimal-control techniques available to a
broad audience, which is eased as \dynamo is implemented in \matlab. 
Thus to generate an optimised control sequence to a specific problem, 
one only needs modify one of the provided examples, specifying the drift and control Hamiltonians 
of interest, choose \grape, \krotov, or one of the other hybrid algorithms provided, and wait for 
the calculation to complete. Wall time, \cpu time, gradient-size and iteration-number constraints 
may also be imposed.

For the second use case---developing optimal-control algorithms---\dynamo provides a flexible framework 
allowing researchers to focus on aspects of immediate interest, allowing \dynamo to handle 
all other issues, as well as providing facilities for benchmarking and comparing performance of 
the new algorithms to the current cadre of methods.

\subsubsection*{Why a Modular Programming Framework ?}
\noindent
The explorative findings underlying this work make a strong case
for setting up a programming framework in a modular way.
They can be summarised as follows:
\medskip

(a) There is no universal single optimal-control algorithm that serves
all types of tasks at a time. For quantum computation, unitary gate synthesis,
or state-to-state transfer of (non)-pure states require
accuracies beyond the error-correction threshold, while for spectroscopy
improving robustness of controls for state-to-state transfer may well come
at the expense of lower maximal fidelities.

(b) Consequently, for a {\em programming framework} to be universal, it
has to have a {\em modular structure} allowing to switch between different
update schemes (sequential, concurrent and hybrids) with task-adapted
parameter settings.

(c) In particular, the different update schemes have to be matched with
the higher-order gradient module (conjugate gradients, Newton, quasi-Newton).
For instance, with increasing dimension the inverse Hessian for a Newton-type
algorithm becomes computationally too costly to be still calculated exactly as one
may easily afford to do in low dimensions. Rather, it is highly advantageous to approximate
the inverse Hessian and the gradient {\em iteratively} by making use of
previous runs within the same inner loop (see flow diagram to Fig.~\ref{fig:GKH-algs}, Fig. ~\ref{fig:flowchart}).
This captures the
spirit of the well-established limited-memory Broyden-Fletcher-Goldfarb-Shanno (\lbfgs)
approach \cite{NocWri06,Noc80,BLS94}. The pros of \lbfgs, however, are rather incompatible
with restricting the number of inner loops to $s_{\rm max}=1$ as is
often done in sequential approaches. Therefore in turn,
gradient modules scaling favourably with problem dimension
may ask for matched update schemes.

(d) It is a common misconception to extrapolate from very few iterations
needed for convergence in low dimensions that the same algorithmic setting will
also perform best in high dimensional problems. Actually, {\em effective \cpu time}
and {\em number of iterations} needed for convergence are far from
being one-to-one. --- The same feature may be illustrated by recent results
in the entirely different field of tensor approximation, where again in low dimensions,
exact Newton methods outperform any other by number of iteration as well as by
\cpu time, while in higher dimensions, exact Newton steps cannot be calculated at all
(see Figs.~11.2 through 11.4 in Ref.~\cite{SL10}).

\bigskip
It is for these reasons we discuss the key steps of the algorithmic framework in
terms of their constituent modules.

%%%%%%%%%%%%%%%
\subsubsection{Gradient-Based Update Modules}\label{gradient-update-modules}
%%%%%%%%%%%%%%%
Here we describe the second-order and first-order control-update modules
used by the respective algorithms.
\bigskip

%%%%%
{\em Second-Order (Quasi)Newton Methods:}\/
%%%%%
The array of piecewise constant control amplitudes
(in the $r^{\rm th}$ iteration),
$\{u_j^{(r)}(t_k^{(q)})\,|\, j=1,2,\dots, m\;\text{and}\; k=1,2,\dots,M^{(q)}\}$
are concatenated to a control vector written \ket{u^{(r)}}
for convenience (in slight abuse of notation).
Thus the standard Newton update takes the form
\begin{equation}
\ket{u^{(r+1)}} = \ket{u^{(r)}} + \alpha_r \Hess_r^{-1} \ket{\grad f^{(r)}}.
\end{equation}
Here $\alpha_r$ is again a step size and $\Hess_r^{-1}$ denotes the
inverse Hessian, where \ket{\grad f^{(r)}} is the gradient vector.
For brevity we also introduce shorthands
for the respective differences of control vectors
and gradient vectors
\begin{equation*}
\begin{split}
\ket{x_r} &:= \ket{u^{(r+1)}} - \ket{u^{(r)}} \quad\text{and}\quad\\
\ket{y_r} &:= \ket{\grad f^{(r+1)}}-\ket{\grad f^{(r)}}\,.
\end{split}
\end{equation*}
Now in the Broyden-Fletcher-Goldfarb-Shanno standard algorithmic
scheme referred to as \bfgs \cite{NocWri06}, the inverse Hessian is conveniently
approximated by making use of previous iterations via
\begin{equation}
\Hess_{r+1}^{-1} = V^t_r \Hess_r^{-1} V_r + \pi_r \ketbra{x_r}{x_r}
\end{equation}
with the definitions
\begin{equation*}
\pi_r := \braket{y_r}{x_r}^{-1} \quad\text{and}\quad V_r:=\unity - \pi_r \ketbra{y_r}{x_r}\quad.
\end{equation*}
By its recursive construction, (i) \bfgs introduces time non-local information into
the optimisation procedure as soon as the inverse Hessian has off-diagonal components 
and (ii) \bfgs perfectly matches {\em concurrent updates}
within the inner loop: using
second-order information makes up for its high initialisation costs
by iterating over the same \subspace of controls throughout the optimisation.
Note that the \matlab routine {\sf fminunc} uses the standard \bfgs scheme,
while the routine {\sf fmincon} uses the standard limited-memory variant
\lbfgs \cite{Noc80,BLS94,BLN95,NocWri06}.  
%% SGS
Another advantage of the \bfgs scheme is that the approximate Hessian is by construction
positive-definite, allowing for straightforward Newton updates.

%%% SGS
In contrast, for {\em sequential updates}, \bfgs is obviously far from
being the method of choice, because sequential updates iterate over a
changing subset of controls.  In principle, direct calculation of the
Hessian is possible.  However, this is relatively expensive and the local
Hessian is not guaranteed to be positive definite, necessitating the need
for more complex trust-region Newton updates.  A detailed analysis of
optimal strategies for sequential update methods is necessary and 
is presented in \cite{SF11}.  Preliminary numerical data 
(see Sec.~\ref{sec:mimick2order})
suggest that the gain from such higher-order methods for sequential update
schemes is limited and not sufficient to offset the increased computational costs
per iteration in general. Thus we shall restrict ourselves here to 
sequential updates based on first-order gradient information.

\bigskip
%%%%%
{\em First-Order Gradient Ascent:}\/
%%%%%
The simplest case of a gradient-based sequential-update algorithm amounts to
steepest-ascent in the control vector, whose elements follow
\begin{equation}\label{eqn:recursion}
u_j^{(r+1)}(t_k^{(q)}) = u_j^{(r)}(t_k^{(q)}) + \alpha_r \frac{\partial f^{(r)}}{\partial u_j(t_k^{(q)})}\quad,
\end{equation}
where $\alpha_r$ is an appropriate step size or search length.  For gate
optimization problems of the type considered here it can be shown that
sequential gradient update with suitable step-size control can match the
performance of higher order methods such as sequential Newton updates
while avoiding the computational overhead of the latter~\cite{SF11}.
Although choosing a small constant $\alpha_r$ ensures convergence (to a
critical point of the target function) this is usually a bad choice.  We
can achieve much better performance with a simple heuristic based on a
quadratic model $\alpha_r(2-\alpha_r)$ of $f$ along the gradient
direction in the step-size parameter $\alpha_r$.  Our step-size control
is based on trying to ensure that the actual gain in the fidelity
$\Delta f=f(\alpha_r)-f(0)$ is at least $2/3$ of the maximum gain
achievable based on the current quadratic model.  Thus, we start with an
initial guess for $\alpha_r$, evaluate $\Delta f(\alpha_r)$ and use the
quadratic model to estimate the optimal step size 
$\alpha_*(r)$.  
If the
current $\alpha_r$ is less than $2/3$ of the optimum step size
then we increase $\alpha_r$ by a small factor, e.g., $1.01$; if
$\alpha_r$ is greater than $4/3$ of the estimated optimal $\alpha_*(r)$
then we decrease $\alpha_r$ by a small factor, e.g., $0.99$.  Instead of
applying the change in $\alpha_r$ immediately, i.e., for the current time
step, which would require re-evaluating the fidelity, we apply it only in
the next time step to give 
\begin{equation}
\alpha_{r+1} = \begin{cases} %
	1.01\;\, \alpha_r\quad \text{for $\alpha_r < \tfrac{2}{3} \alpha_*(r)$}\\[1mm]
        0.99\;\, \alpha_r \quad \text{for $\alpha_r > \tfrac{4}{3} \alpha_*(r)$}\\[1mm]
        \phantom{1.011}\alpha_r \quad \text{else}\\[1mm]\end{cases}.
\end{equation}
For sequential update with many time steps,
avoiding the computational overhead of multiple fidelity evaluations is
usually preferable compared to the small gain achieved by continually
adjusting the step size $\alpha_{r}$ at the current time step.  This
deferred application of the step size change is justified in our case as
for unitary gate optimization problems of the type considered here, as
$\alpha_r$ usually quickly converges to an optimal (problem-specific)
value and only varies very little after this initial adjustment period,
regardless of the initial $\alpha_r$~\cite{SF11}.  

As has been mentioned above, 
this step-size control scheme for sequential update comes close to 
a direct implementation of trust-region Newton 
(see Fig.~\ref{fig:mimick2s} in Sec.~\ref{sec:mimick2order}), a detailed
analysis of which is given in \cite{SF11}.

%%%%%%%%%%%%%%%
\subsubsection{Gradient Modules}
%%%%%%%%%%%%%%%

{\em Exact Gradients:}\/
In the module used for most of the subsequent comparisons,
exact gradients to the exponential maps of total Hamiltonians with piecewise
constant control amplitudes over the time interval $\Delta t$ are to be evaluated.
Here we use exact gradients as known from various applications \cite{NS09,TILO96}.
Their foundations were elaborated in \cite{Aizu63,Wilcox67}, so here we give
a brief sketch along the lines of \cite{TILO96,Aizu63}
(leaving more involved scenarios beyond piecewise constant
controls to be dwelled upon elsewhere).
For
\begin{equation}\label{eqn:prop-closed}
X:= \exp\{-i\Delta t H_u\} = \exp\{-i \Delta t (H_d + \sum_j u_j H_j)\}
\end{equation}
the derivative invokes the spectral theorem to take
the form
\begin{equation}
\label{eqn:exact-derivative}
\begin{split}
&\braket{\lambda_l}{\Partial{X}{u_j}\lambda_m} =\\
	 &\qquad 
    \begin{cases} 
	-i\Delta t\,\braket{\lambda_l}{H_j|\lambda_m}\,  e^{-i\Delta t\lambda_l} &\text{if}\; \lambda_l=\lambda_m\\[2mm]
	-i\Delta t\,\braket{\lambda_l}{H_j|\lambda_m}\, %
	\frac{e^{-i\Delta t\lambda_l}-e^{-i\Delta t\lambda_m}}{-i\Delta t\,(\lambda_l-\lambda_m)}&\text{if}\;\lambda_l\neq\lambda_m \;,
    \end{cases}
\end{split}
\end{equation}
where in the second identity we have deliberately kept the factor $-i\Delta t$ for clarity.
Thus the derivative is given elementwise in the orthonormal eigenbasis $\{\ket{\lambda_i}\}$ to the real eigenvalues
$\{\lambda_i\}$ of the Hamiltonian $H_u$. Details are straightforward, yet lengthy, and are thus
relegated to Appendix~A. %\ref{sec:exactGradientDerivation}. 

\medskip

{\em Approximate Gradients}:
In Ref.~\cite{GRAPE} we took
an approximation valid as long as the respective digitisation \timeslices
are small enough in the sense $\Delta t \ll 1/||H_u||_2$ with
$H_u$ as in Eqn.~\eqref{eqn:prop-closed}
\begin{equation}
    \frac{\partial X}{\partial u_j} \approx -i\,\Delta t\; H_j\;e^{-i\,\Delta t\; H_u}\quad.
\end{equation}
This approximation can be envisaged as replacing the average value
brought about by the time integral over the duration $\Delta t = t_k - t_{k-1}$,
which in the above eigenbasis takes the form
\begin{equation}
\begin{split}
&\braket{\lambda_l} {\Partial{X}{u_j}\lambda_m} =\\
\quad &= -i\int\limits_{t_{k-1}}^{t_k} d\tau\; e^{-i\lambda_l(t_k-\tau)} \braket{\lambda_l}{H_j|\lambda_m}\,
				 e^{-i\lambda_m (\tau-t_{k-1})} \\[1mm]
\quad & \approx -i\,\Delta t\; \braket{\lambda_l} {H_j|\lambda_m} \;e^{-i\lambda_m \Delta t} %(t_k-t_{k-1})
\end{split}
\end{equation}
by the value of the integrand at the right-hand side of the time interval 
$\tau\in[t_{k-1},t_k]$. 
Clearly, this approximation ceases to be exact as soon as the time evolution
$U(t_k, t_{k-1})=e^{-i\Delta t H_u}$ fails to commute with $H_j$.
Generically this is the case and the error scales with $|\lambda_l-\lambda_m|\Delta t$.

\bigskip
{\em Finite Differences} provide another standard alternative, which may be
favourable particularly in the case of pure-state transfer, see \cite{Diss-Sander}.
%%%%%%%%%%%%%%%

%%%%%%%%%%%%%%%
\subsubsection{Exponentiation Module}
%%%%%%%%%%%%%%%
Matrix exponentials are a notorious problem in computer science \cite{dubious1,dubious2}.
Generically, the standard \matlab module takes the matrix exponential
via the {\sc Pad\protect{\'e}}-approximation, while in special cases
(like the Hermitian one pursued throughout this paper)
the eigendecomposition is used \footnote{In view of future optimisation,
however, note that our parallelised C++ version of \grape already
uses faster methods based on Chebychev polynomials as described in \cite{Wald07,HLRB07}.}.

From evaluating exact gradients (see above) 
the eigendecomposition of the Hamiltonian is already available.
Though in itself the eigendecomposition typically comes at
slightly higher computational overhead than the
{\sc Pad\protect{\'e}} matrix exponential, this additional computational cost is
outweighed by the advantage that evaluating the matrix exponential now becomes trivial
by exponentiation of the eigenvalues and a matrix multiplication.

Thus as long as the eigendecompositions are available, the matrix exponentials
essentially come for free. Since in the sequential-update algorithm, the gradient needed
for the exponential in {\timeslice} $k$ requires an update in {\timeslice} $k-1$,
the exponential occurs in the {\em inner loop} of the 
algorithm, while obviously the concurrent-update algorithm takes its
exponentials only in the {\em outer loops}. 
The total number of exponentials required by the two algorithms are basically the same.

\subsubsection{Reducing the Number of Matrix Operations}
As described above, the search for an optimal control sequence
proceeds on two levels: an outer loop choosing the \timeslices
to be updated (a decision which may imply choice of gradient-based step method, as well
as other control parameters), and an inner loop which computes
gradients and advances the search point. With \dynamo, significant effort has been made to optimise
the overall number of matrix operations.

For a general hybrid scheme, where $\mathcal{T}^{(q)}$ is a subset
of \timeslices $\{t_1^{(q)}{\ldots}t_{M^{(q)}}^{(q)}\}$  approach is as follows:
Given \timeslices $X_1,\dots, X_M$, of which in hybrid update schemes we select
for updating any general set
$X_{t_1},\dots,X_{t_p}$, we can collapse multiple consecutive non-updating $X$
into a single effective $Y$. For example, consider $X_1,\dots, X_{10}$ of which we
update $X_2$,$X_5$ and $X_6$. Before proceeding with the inner loop,
we generate concatenated products $Y_1,\dots,Y_4$ such that $Y_1=X_1$, $Y_2=X_4X_3$, and $Y_3=X_{10}X_9X_8X_7$.
Now the heart of the expression to optimise for is $Y_3 X_6 X_5 Y_2 X_2 Y_1$.

As a result, computation of forward and backward propagators can be done with the minimal number
of matrix multiplications. Matrix exponentiation is also minimised by way of caching and making
use of the fact that for some gradient computation schemes eigendecomposition is required,
thus allowing for light-weight exponentiation.

Moreover, the \dynamo platforms isolates the problem of minimising matrix operations to a
specific module, which is aware of which $H_u$-s, $X$-s and $\Lambda$-s are needed for
the next step, compares these with the \timeslices which have been updated, and attempts
to provide the needed data with the minimal number of operations. And while for some
hybrid update schemes the current number of operations performed in the outer loop is not strictly optimal in all cases, optimality is reached
for \krotov, \grape and schemes which update consecutive blocks of \timeslices.

%%%%
\subsubsection{Modularisation Approach in \dynamo}
%%%%
To allow for flexibility in design and implementation of new optimal control techniques, 
the framework is modularised by way of function pointers,
allowing, e.g., the second-order search method to receive a pointer to a function which
calculates the gradient, which in-turn may receive a pointer to a function which
calculates the exponential. The cross-over algorithm described Fig.~\ref{fig:handover}, e.g.,
is implemented by a search method receiving as input two search-method modules and a
cross-over condition, which is used as a termination condition for the first search method.
The first-order hybrids described in Fig.~\ref{fig:hybrid} are
similarly implemented by a block-wise \subspace selection function (generalisation
of the sequential versus concurrent selection schemes) receiving a pointer to the search
function to be used within each block. \dynamo is provided with many such examples.

If one is exploring, e.g., second-order search methods appropriate
for serial update schemes, one only needs to write the update-rule function.
\dynamo will provide both the high-level \subspace-selection logic and the
low-level book-keeping that is entrusted with tracking which controls have been updated. 
When given a demand for gradients, propagators or the value function, it performs the needed
calculations while minimising the number of matrix operations.
Moreover, once a new algorithm is found, \dynamo makes it easy both to compare its performance to that
of the many schemes already provided as examples and to do so for a wide set of problems described
in this paper. Thus \dynamo serves as a valuable benchmarking tool for current and future algorithms.
%%%%
%%%%

%%%%
\section{Standard Scenarios for Quantum Applications}\label{sec:appl}
%%%%
We have discussed the versatile features of the framework embracing all
standard scenarios of bilinear quantum control problems listed in Tab.~\ref{tab:glossary}. 
Here we give the (few!) necessary adaptations for applying our algorithms to such a broad variety
of paradigmatic applications, while our test 
suite is confined to unitary gate synthesis and cluster-state preparation in closed quantum systems.

\subsection{Closed Quantum Systems}
The most frequent standard tasks for optimal control of closed systems comprise different ways of gate synthesis 
as well as state transfer of pure or non-pure quantum states. 
More precisely, sorted for convenient development from the general case,
they amout to\\[2mm]
\phantom{X}{\bf Task 1:} unitary gate synthesis up to a global phase,\\
\phantom{X}{\bf Task 2:} unitary gate synthesis with fixed global phase,\\
\phantom{X}{\bf Task 3:} state transfer among pure-state vectors,\\
\phantom{X}{\bf Task 4:} state transfer among density operators.\\[2mm]
As will be shown, all of them can be treated by common propagators that are of the form
\begin{equation}
\begin{split}
X_k &= \exp\{-i\Delta t\,{H}_u(t_k)\}\\
    &=\exp\{-i \Delta t (H_d + \sum_j u_j(t_k) H_j)\}\quad.
\end{split}
\end{equation}
Algorithmically, this is very convenient, because then the specifics of the problem just enter 
via the boundary conditions as given in Tab.~\ref{tab:glossary-ini}: clearly, the data type %and dimension 
of the state evolving in time via the propagators $X_k$ is induced %or determined 
by the initial state being a vector or a matrix represented in Hilbert space or (formally) in Liouville space.

Indeed for seeing interrelations, it is helpful to formally consider some problems in Liouville space, 
before breaking them down to a Hilbert-space representation for all practical purposes, which is obviously 
feasible in any closed system.

\medskip
\noindent
{\bf Task 1} {\em projective phase-independent gate synthesis:}\\
In Tab.~\ref{tab:glossary-ini} the target projective gate $\widehat U_{\rm target}$ can be taken
in the phase-independent superoperator representation $\widehat X := \bar X\otimes X$
to transform the quality function
\begin{equation}
\begin{split}
f^2_{PSU} &= \tfrac{1}{N^2} \Re\, \tr\big\{\widehat{U}^\dagger_{\rm target} \widehat X(T)\big\}\\
        %&= \tfrac{1}{N^2} \Re \,\tr\big\{(\bar U_{\rm tar}\otimes U_{\rm target})^\dagger (\bar X_T\otimes X_T)\big\}\\
        &= \tfrac{1}{N^2} \Re \,\tr\{(U^t_{\rm target}\bar X_T)\otimes(U_{\rm target}^\dagger X_T)\big\}\\
        &= \tfrac{1}{N^2}  |\tr\{U_{\rm tar}^\dagger X_T\big\}|^2\qquad\text{so}\\[1mm]
f_{PSU} &= \tfrac{1}{N}  |\tr\{U_{\rm tar}^\dagger X_T\big\}|
	= \tfrac{1}{N}  |\tr\big\{\Lambda^\dagger_{M+1:k+1} X_{k:0} \big\}|\quad,
\end{split}
\end{equation}
where the last identity recalls the forward and backward propagations
$X(t_k) := X_k X_{k-1}\cdots X_2 X_1 X_0$
and
$\Lambda^\dagger(t_k) := {U}^\dagger_{\rm target} X_M X_{M-1}\cdots
		X_{k+2}  X_{k+1}$.

\medskip
So with the overlap 
$g := \tfrac{1}{N} \tr \{\Lambda^{\dagger}_{M+1:k+1}  U_{k:0}\}$
of Eqn.~\eqref{eq:overlap},
the derivative of the squared fidelity with respect to the control amplitude $u_j(t_k)$ becomes
\begin{equation}\label{eqn:grad-psu2}
\tPartial {f^2_{PSU}(X(t_k))}{u_j} = \tfrac{2}{N} \Re \,\tr \{g^* \Lambda^{\dagger}_{M+1:k+1}
		\big(\tPartial{X_k}{u_j}\big) X_{k-1:0}\}\;,\qquad
\end{equation}
where $\tPartial{X_k}{u_j}$ is given by Eqn.~\eqref{eqn:exact-derivative}.
The term $g^*$ arises via $f^2(u)=|g(u)|^2$, %cp.~Ref.~\cite{PRA05}
so that by $\Partial{f^2}{u}=2\;|g(u)| ( \Partial{}{u}|g(u)|)$ one gets (for $|g(u)|\neq0$)
$\Partial{f}{u}=\Partial{}{u}|g(u)|=\tfrac{1}{2|g(u)|} \Partial{f^2}{u}$ 
to arrive at
\begin{equation}\label{eqn:grad-psu}
\tPartial {f_{PSU}(X(t_k))}{u_j} = \tfrac{1}{N} \,\Re\,\tr \{e^{-i\phi_g} \Lambda^{\dagger}_{M+1:k+1}
		\big(\tPartial{X_k}{u_j}\big) X_{k-1:0}\}\;,\\
\end{equation}\\[-1mm]
where $e^{-i\phi_g}:={g^*}/{|g|}$ uses the polar form $g=|g|\, e^{+i\phi_g}$ for a
numerically favourable formulation.

\medskip
Thus, in closed systems, the superoperator representation is never used in the algorithm explicitly, 
yet it is instructive to apply upon derivation, because Task 2 now follows immediately.

%%%%%%%%%%%%%%%%%%%
%%%%%%%%%%%%%%%%
%%%%%%%%%%
\begin{table}[Ht!]
\begin{center}
\caption{Boundary Conditions for Standard Scenarios}
\label{tab:glossary-ini}
\begin{tabular}{ll|cccc}
\hline\hline\\[-1mm]
\multicolumn{2}{l}{Conditions }&\phantom{.}& Initial &\phantom{.}& Final \\[0mm]
\multicolumn{2}{l} {\phantom{XXXXXXXXXXXXXXX}} &\phantom{.}& $X_0$ &\phantom{.}&  $X_{M+1}$ \\[1mm]
\hline\\[-3.7mm]
{} &          &&      &&  \\[-2mm]
{\em closed systems:} &          &&      &&  \\[0mm]
\; pure-state transfer &  \hspace{0mm} && $\ket{\psi_0}$  && $\ket{\psi}_{\rm target}$  \\[0mm]
\; gate synthesis I &  && $\unity_N$  && $U_{\rm target}$  \\[0mm]
\; state transfer &  && $\rho_0$  && $\rho_{\rm target}$  \\[0mm]
\; gate synthesis II &  \hspace{0mm} && $\unity_{N^2}$  && $ \widehat{U}_{\rm target}$  \\[2mm]
\hline%\\[-1mm]
{} &          &&      &&  \\[-2mm]
{\em open systems:}   &         &&              &&          \\[0mm]
\; state transfer & \hspace{0mm} &&  $\rho_0$  && $\rho_{\rm target}$   \\[0mm]
\; map synthesis &  && $\unity_{N^2}$ &&  $F_{\rm target}$  \\[2mm]
\hline\\[-2mm]
\multicolumn{6}{l}{{\em state of the system:} evolution of initial state as} \\[1mm]
\multicolumn{6}{l}{\; $X(t_k) = X_{k:0} := X_k X_{k-1}\cdots X_1 {\mathbf X_0}$ }\\[0.5mm]
\multicolumn{6}{l}{\; with propagators  $X_\nu = e^{\Delta t (A + \sum_j u_j(t_\nu) B_j)}$ for $\nu=1{\dots}k$ }\\[0.5mm]
\multicolumn{6}{l}{\; and with $A, B_j$ as defined in Tab.~\ref{tab:glossary} }\\[1mm]
\hline\hline
\end{tabular}\hspace{15mm}
\end{center}
\end{table}
%%%%%%%%%%
%%%%%%%%%%%%%%%%%

\medskip
\noindent
{\bf Task 2} {\em phase-dependent gate synthesis:}\\
In Tab.~\ref{tab:glossary-ini} the target gate $U_{\rm target}$ now directly enters
the quality function
\begin{equation}
f_{SU}  = \tfrac{1}{N}  \Re \,\tr\big\{U_{\rm tar}^\dagger X_T\big\} 
	= \tfrac{1}{N}  \Re \,\tr\big\{\Lambda^\dagger_{M+1:k+1} X_{k:0} \big\}\quad.
\end{equation}
So the derivative of the fidelity with respect to the control amplitude $u_j(t_k)$ 
with reference to $\tPartial{X_k}{u_j}$ of Eqn.~\eqref{eqn:exact-derivative} reads
\begin{equation}\label{grad-su}
\tPartial {f_{SU}(X(t_k))}{u_j} = \tfrac{1}{N} \Re \,\tr \{\Lambda^{\dagger}_{M+1:k+1}
		\big(\tPartial{X_k}{u_j}\big) X_{k-1:0}\}\;.\qquad
\end{equation}
It is in entire analogy to Eqn.~\eqref{eqn:grad-psu}.

\medskip
Actually, this problem can be envisaged as the lifted operator version
of the pure-state transfer in the subsequent Task 3, which again thus follows immedialtely
as a special case.

\medskip
\medskip
\noindent
{\bf Task 3} {\em transfer between pure-state vectors:}\\
Target state and propagated initial state from Tab.~\ref{tab:glossary-ini},
$\ket{\psi}_{\rm target}$, $X(T)\ket{\psi_0}$ form the scalar product in
the quality function
\begin{equation}
f  = \tfrac{1}{N}  \Re \,\braket{\psi_{\rm target}}{X_T}
   = \tfrac{1}{N}  \Re \,[\tr] \big\{\Lambda^\dagger_{M+1:k+1} X_{k:0} \big\}\quad,
\end{equation}
where the latter identity treats the propagated column vector $X_{k:1} \ket{X_0}$ as $N\times 1$ 
matrix $X_{k:0}$ and likewise the back-propagated final state $\bra{\psi_{\rm tar}}(X_{M:k+1})^\dagger$ 
as $1\times N$ matrix $\Lambda^\dagger_{M+1:k+1}$ so the trace can be ommited.
Hence the derivative of the fidelity with respect to the control amplitude $u_j(t_k)$ 
remains 
\begin{equation}
\tPartial {f_{SU}(X(t_k))}{u_j} = \tfrac{1}{N} \Re \,[\tr] \{\Lambda^{\dagger}_{M+1:k+1}
                \big(\tPartial{X_k}{u_j}\big) X_{k-1:0}\}\;\qquad
\end{equation}
with $\tPartial{X_k}{u_j}$ of Eqn.~\eqref{eqn:exact-derivative}.

\medskip
\noindent
{\bf Task 4} {\em state transfer between density operators:}\\
The quality function normalised with respect to the (squared) norm of the
target state $c:=||\rho_{\rm tar}||_2^2$ reads
\begin{equation}
\begin{split}
f &= \tfrac{1}{c} \Re\,\tr\{X^\dagger_{M+1} \Adr_{X_T}(X_0)\}\\
  &\equiv \tfrac{1}{c} \Re\,\tr\{X^\dagger_{M+1} X_T X_0 X^\dagger_T\}\\
  &= \tfrac{1}{c} \Re\,\tr\{X^\dagger_{M+1} X_M X_{M-1}\cdots X_k\cdots X_2 X_1 X_0 \times\\
		&\qquad\qquad \times X^\dagger_1 X^\dagger_2\cdots X^\dagger_k\cdots X^\dagger_{M-1} X^\dagger_M\}\quad.
\end{split}
\end{equation}
Hence the derivative of the quality function with respect to the control amplitude
$u_j(t_k)$ takes the somewhat lengthy form
\begin{equation}
\begin{split}
\tPartial {f(X(t_k))}{u_j} &= \tfrac{1}{c} 
     \Re\,\Big(\tr\{X^\dagger_{M+1} X_M \cdots \big(\tPartial{X_k}{u_j}\big) \cdots X_2 X_1 X_0 \times\\
		&\qquad\qquad \times X^\dagger_1 X^\dagger_2\cdots X^\dagger_k\cdots  X^\dagger_M\}\\
	&\qquad  + \tr\{X^\dagger_{M+1} X_M \cdots X_k \cdots X_2 X_1 X_0 \times\\
                &\qquad\qquad \times X^\dagger_1 X^\dagger_2\cdots \big(\tPartial{X^\dagger_k}{u_j}\big) \cdots  X^\dagger_M\}\Big)\quad, 
\end{split}
\end{equation}
where the exact gradient $\tPartial{X_k}{u_j}$ again follows Eqn.~\eqref{eqn:exact-derivative}.

Notice that Task 1 can be envisaged as the lifted operator analogue to Task 4
if {\em phase independent} projective representations $\ketbra{\psi_\nu}{\psi_\nu}$ 
of pure states $\ket{\psi_\nu}$ are to be transferred.

\subsection{Open Quantum Systems}
\noindent
{\bf Task 5} {\em quantum map synthesis in Markovian systems:}\\
The superoperator $\widehat H_u(t_k)$ to the Hamiltonian above
can readily be augmented by the relaxation operator $\Gamma$.
Thus one obtains the generator to the quantum map
\begin{equation}
X_k = \exp\{-\Delta t\,(i\widehat{H}_u(t_k) + \Gamma(t_k))\}
\end{equation}
following the Markovian equation of motion
\begin{equation}\label{eq:dgl_map}
\dot X(t) = -(i\widehat{H}_u + \Gamma)\; X(t)\quad.
\end{equation}

By the (super)operators
$X(t_k) := X_k X_{k-1}\cdots %X_2\cdot 
X_1 X_0$
and
$\Lambda^\dagger(t_k) := {F}^\dagger_{\rm target}  X_M\cdot X_{M-1}\cdots
		X_{k+2}  X_{k+1}$
the derivative of the trace fidelity at fixed final time $T$
\begin{equation*}
f=   \tfrac{1}{N^2} \Re\,\tr\{{F}_{\rm target}^\dagger X(T)\}=\tfrac{1}{N^2} \Re\,\tr\{\Lambda^\dagger(t_k) X(t_k)\}
\end{equation*}
with respect to the control amplitude $u_j(t_k)$ formally reads
\begin{equation}
\tfrac{\partial f}{\partial u_j(t_k)} = \tfrac{1}{N^2}\Re\ \tr
	\big\{\Lambda^\dagger(t_k) \big(\tfrac{\partial X_k}{\partial u_j(t_k)}\big) X(t_{k-1})\big\}
\end{equation}
Since in general $\Gamma$ and $i\widehat{H}_u$ do not commute, the
semigroup generator $(i\widehat{H}_u + \Gamma)$ is not normal,
so taking the exact gradient as in Eqn.~\eqref{eqn:exact-derivative} via the spectral decomposition
has to be replaced by other methods. There are two convenient alternatives,
(i) approximating the gradient for sufficiently small 
$\Delta t \ll 1/||i\widehat{H}_u + \Gamma||_2$ by
\begin{equation}\label{eq:grad}
\tfrac{\partial X_k}{\partial u_j(t_k)} \approx 
 	- \Delta t \big(i {\widehat{H}}_{u_j}+\;
	\tfrac{\partial\Gamma(u_j(t_k))}{\partial u_j(t_k)}\big) X_k
\end{equation}
or (ii) via finite differences.

% In this form, one readily checks that in closed systems ($\Gamma=0$)
% the derivative vanishes as soon as the forward propagation coincides with the inverse of the
% backward propagation, i.e. $X(t_k)\Lambda^\dagger(t_k)=\unity$, because all
% Hamiltonians ${\widehat{H}}_{u_j}$ can be constructed as to be traceless.
% Moreover, in the context of optimal control theory, Eqn.~\eqref{eq:grad} naturally links
% to the Lagrange term $\braket {\Lambda(t)} {\dot X(t)}$ ensuring the quantum map
% follows the differential equation of motion \eqref{eq:dgl_map}, as has been outlined in Ref.~\cite{GRAPE}.
% 

\bigskip
This standard task devised for Markovian systems \cite{PRL_decoh}
can readily be adapted to address also {\em non-Markovian systems}, provided the latter 
can be embedded into a (numerically manageable) larger system that in turn interacts with 
its environment in a Markovian way \cite{PRL_decoh2}.

\bigskip
\noindent
{\bf Task 6} {\em state transfer in open Markovian systems:}\\
This problem can readily be solved as a special case of Task 5 when
envisaged as the vector version of it.

To this end it is convenient to resort to the so-called $vec$-notation \cite{HJ1} of
a matrix $M$ as the column vector $\vec(M)$ collecting all columns of $M$.
Now, identifying $X_0:=\vec(\rho_0)$ and  
$X^\dagger_{\rm target}:= \vec^t(\rho_{\rm target}^\dagger)$ 
%from Tab.~\ref{tab:glossary-ini}
one obtains the propagated initial state
$X(t_k) := X_k X_{k-1}\cdots X_1 X_0$
and
$\Lambda^\dagger(t_k) := {X}^\dagger_{\rm target}  X_M X_{M-1}\cdots
		X_{k+2}  X_{k+1}$
as back propagated target state. In analogy to Task~3, they take 
the form of $N^2\times 1$ and $1\times N^2$ vectors, respectively.
Thus the derivative of the trace fidelity at fixed final time $T$
\begin{equation*}
f=   \tfrac{1}{N} \Re\,[\tr]\{{X}_{\rm target}^\dagger X(T)\}=\tfrac{1}{N} \Re\,[\tr]\{\Lambda^\dagger(t_k) X(t_k)\}
\end{equation*}
with respect to the control amplitude $u_j(t_k)$ reads
\begin{equation}
\tfrac{\partial f}{\partial u_j(t_k)} = \tfrac{1}{N}\Re\ [\tr]
	\big\{\Lambda^\dagger(t_k) \big(\tfrac{\partial X_k}{\partial u_j(t_k)}\big) X(t_{k-1})\big\}\quad,
\end{equation}
where for $\tfrac{\partial X_k}{\partial u_j(t_k)}$ the same gradient approximations apply as in Task~5.

\medskip
For the sake of completeness, Appendix~C gives all the key steps of the
standard {\bf Tasks 1} through {\bf 6} in a nutshell.

%%%%%%%%%%%%%%%%%
\section{Results on Update Schemes: Concurrent and Sequential}
%%%%%%%%%%%%%%%%%
\subsection{Specification of Test Cases}

We studied the 23 systems listed in Tab.~\ref{tab:test-def} as test cases
for our optimisation algorithms. This test suite includes
spin chains, a cluster state system whose effective Hamiltonian represents a $C_4$
graph, an NV-centre system and two driven spin-$j$ systems with $j=3,6$. Attempting to cover 
many systems of practical importance (spin chains, cluster-state preparation,
NV-centres)
with a range of coupling topologies and control schemes, the study includes large
sets of parameters like system size, final time, number of \timeslices,
and target gates. We therefore anticipate our suite of test cases will
provide good guidelines for choosing an appropriate algorithm in many practical cases.

\subsubsection{Spin Chains with Individual Local Controls\label{sec:ExampleSpinChains}}
Explorative {\bf problems 1-12} are Ising-$ZZ$ spin chains of various length in which
the spins are addressable by individual $x$- and $y$-controls. The Hamiltonians
for these systems take the following form:
\begin{eqnarray}
	H_d &=& \tfrac{J}{2} \sum_{k=1}^{n-1}\sigma_k^z\sigma_{k+1}^z
	\label{eqn:sc-ham-ising-drift}\\
	H_j^{x,y} &=& \tfrac{1}{2}\; \sigma_j^{x,y}
	\label{eqn:sc-ham-ising-control}
\end{eqnarray}
where $J=1$, $n = 1, \dots, 5$ and $j = 1, \dots, n$.\\

In example 1 we also consider linear crosstalk (e.g., via off-resonant excitation), leading to the
control Hamiltonians
\begin{align}
H_{1,2} & =\alpha_{1,2}\sigma_{1}^{x}+\alpha_{2,1}\sigma_{2}^{x}\\
H_{3,4} & =\beta_{2,1}\sigma_{1}^{y}+\beta_{1,2}\sigma_{2}^{y}
\end{align}
where $u_k$ are independent control fields and $\alpha_k$ and $\beta_k$ are
crosstalk coefficients. We chose $\alpha_1=\beta_2=1$ and $\alpha_2 = \beta_1 = 0.1$.\\

%%%%%%%%%%%
\subsubsection{Cluster State Preparation in Completely Coupled Spin Networks}\label{sec:c4graph}
The effective Hamiltonian of test {\bf problems 13} and {\bf 14},
\begin{equation}
H_{CS} = \tfrac{J}{2}(\sigma_1^z\sigma_{2}^z+\sigma_2^z\sigma_{3}^z+\sigma_3^z\sigma_{4}^z+\sigma_4^z\sigma_{1}^z)\;,
\end{equation}
represents a $C_4$ graph of Ising-$ZZ$ coupled qubits which can be used
for cluster state preparation according to~\cite{Wunderlich09}. The
underlying physical system is a completely Ising-coupled set of 4 ions
that each represents a locally addressable qubit:
\begin{align}
	H_d &= \tfrac{J}{2} \sum_{k=1}^{3}\sum_{l=k+1}^{4}\sigma_k^z\sigma_{l}^z
	\label{eqn:c4-ham-drift}\\
	H_j^{x,y} &= \tfrac{1}{2}\; \sigma_j^{x,y} \hspace{1.5cm} (j = 1,\dots,4).
	\label{eqn:c4-ham-control}
\end{align}
Again, the coupling constant $J$ was set to $1$. The following unitary was chosen as a target gate,
which applied to the state $\ket{\psi_1}=((\ket{0}+\ket{1})/\sqrt{2})^{\otimes 4}$ generates a cluster state
\begin{equation}
U_G = \exp(-i\frac{\pi}{2}H_{CS})\;.
\end{equation}

\subsubsection{NV-Centre in Isotopically Engineered Diamond}
In test {\bf problems 15} and {\bf 16} we optimised for a CNOT gate on two strongly coupled nuclear spins at an
nitrogen-vacancy (NV) centre in diamond as described in~\cite{NeumannScience08}. 

In the eigenbasis of the coupled system, after a transformation into the rotating frame, the Hamiltonians are of the form
\begin{eqnarray}
	H_d &=& \diag(E_1, E_2, E_3, E_4) + \omega_c \diag(1, 0, 0, -1)\quad
	\label{eqn:nvc-ham-drift}\\[1mm]
	H_1 &=& \tfrac 1 2 \big( \mu_{12}\sigma_{12}^x + \mu_{13}\sigma_{13}^x
	+ \mu_{24}\sigma_{24}^x +\mu_{34}\sigma_{3,4}^x \big)
	\label{eqn:nvc-ham-control1}\\[1mm]
	H_2 &=& \tfrac 1 2 \big( \mu_{12}\sigma_{12}^y + \mu_{13}\sigma_{13}^y
	+ \mu_{24}\sigma_{24}^y +\mu_{34}\sigma_{34}^y \big)\;.
	\label{eqn:nvc-ham-control2}
\end{eqnarray}
Here $E_1\dots E_4$ are the energy levels, $\omega_c$ is the carrier frequency of the driving field and
$\mu_{\alpha,\beta}$ is the relative dipole moment of the transition between levels $\alpha$ and $\beta$.
We chose the following values for our optimisations: $\{E_1, E_2, E_3, E_4\} =  2\pi\{-134.825,-4.725, 4.275,135.275\}$ MHz, 
$\omega_c = 2\pi\times135$ MHz, $\{\mu_{12}, \mu_{13}, \mu_{24}, \mu_{34}\} = \{1,1/3.5, 1/1.4,1/1.8\}$
in accordance with \cite{NeumannScience08}.

\subsubsection{Special Applications of Spin Chains\label{sec:ExampleSpinChains2}}
Test {\bf problems 17} and {\bf 18} are modified 
five-qubit Ising chains extended by a local Stark-shift term being
added in the drift Hamiltonian $H_d$ resembling a gradient.
The control consists of simultaneous $x$- and $y$-rotations on all spins
\begin{align}
H_{d} & =\frac{J}{2}\sum_{i=1}^{4}\sigma_{i}^{z}\sigma_{i+1}^{z}-(i+2)\sigma_{i}^{z}\\
H_{1} & =\frac{1}{2}\sum_{i=1}^{5}\sigma_{i}^{x}\quad\text{and}\quad
H_{2}  =\frac{1}{2}\sum_{i=1}^{5}\sigma_{i}^{y}\;.
\end{align}

%%%%%%%%%
{\bf Problem 19} is a Heisenberg-XXX coupled chain of five spins extended by global
permanent fields inducing simultaneous $x$-rotations on all spins:
\begin{equation}
H_{d}=\tfrac{J}{2}\sum_{i=1}^{4}\sigma_{i}^{x}\sigma_{i+1}^{x}+\sigma_{i}^{y}\sigma_{i+1}^{y}+\sigma_{i}^{z}\sigma_{i+1}^{z}-10\sigma_{i}^{x}\;.
\end{equation}
Control is exerted by switchable local Stark shift terms,
\begin{equation}
H_{i}=\sigma_{i}^{z}\qquad(i=1,\dots,5)\;.
\end{equation}

%%%%%%%
Spin chains may be put to good use as quantum wires \cite{Bose03,Burg05,Bose07,Kay09,SP09}.
The idea is to control just the input end of the chain using the remainder to passively transfer
this input to the other end of the chain.
To embrace such applications, in {\bf problems 20} and {\bf 21}, the spins are coupled by an 
isotropic Heisenberg-$XXX$ interaction and the
chains are subject to $x$- and $y$-controls only at one end (at one or two spins, respectively):
\begin{align}
	H_{d} & =\tfrac{J}{2}\sum_{i=1}^{n-1}\sigma_{i}^{x}\sigma_{i+1}^{x}+\sigma_{i}^{y}\sigma_{i+1}^{y}+\sigma_{i}^{z}\sigma_{i+1}^{z}\\
H_{1,2} & =\tfrac{1}{2}\;\sigma_{1}^{x,y}\quad\text{and}\quad
(H_{3,4}  =\frac{1}{2}\;\sigma_{2}^{x,y})
\end{align}
Here $J=1$ and $n = 3,4$. Restricting the controls in this way makes the
systems harder to steer and thus raises the bar for numerical optimisation.

%%%%%%%
\subsubsection{Spin-$3$ and Spin-$6$ Systems}
As an example beyond spin-$1/2$ systems, 
in test {\bf problems 22} and {\bf 23} we consider a Hamiltonian of the following form \cite{HigherSpin}
\begin{equation}
H_u = J_z^2 + u_1J_z + u_2J_x,
\end{equation}
where the $J_i$ are angular momentum operators in spin-$j$ representation. The $J_z^2$ term represents
the drift Hamiltonian and the other two terms function as controls. We chose $j=6$
for {\bf problem 22} and $j=3$ for {\bf problem 23}.

%%%%%
\subsection{Test Details}
%%%%%
As shown in Tab.~\ref{tab:test-def}, we optimised each test system for one of
four quantum gates: a CNOT, a quantum Fourier transformation, a random unitary,
or a unitary for cluster state preparation according to section~\ref{sec:c4graph}.
Random unitary gates generated according to the Haar measure \cite{Mez07}
are meant to be numerically more demanding than the other gates.
The final times $T$ were always chosen sufficiently long to ensure the respective problem
is solvable with full fidelity (hence the times should not be mistaken as underlying
time-optimal solutions). All results were averaged over 20 runs with different
initial pulse sequences (control vectors), i.e.~randomly generated vectors with a
mean value of $mean(u_{ini})=0$ and a standard deviation of $std(u_{ini})=1$ in units of $1/J$
unless specified otherwise (as in Tab.~\ref{tab:results-overview-highamps}, where $std(u_{ini})=10$
to study the influence of the initial conditions). 
The maximum number of loops was set to
3000 for the concurrent update scheme and to 300000 for the sequential update.
All systems were optimised with a target fidelity of $f_{\rm target}=1-10^{-4}$.
%\RED{WATCH-OUT: (PROBLEM-22: MODIFIED thresholds apply as well to the data listed in Tab.~\ref{tab:results-overview-smallamps}).}
As an additional stopping criterion the change of the function value from one
iteration to the next (concurrent update) or between the last iteration and the
average of the previous $M$ iterations (sequential update) was introduced. The threshold
value in this case was set to $10^{-8}$. For the concurrent update algorithm, the
optimisation stopped when the smallest change in the control vector was below $10^{-8}$.
We measured the wall times of our optimisations to give a measure for the actual
running time form start to completion (including, e.g., memory loads and communication
processes) instead of only measuring the time spent on the \cpu.
The optimisations were carried out under \matlab R2009b (64bit, single-thread mode) on an
{\sc amd} Opteron dual-core \cpu at $2.6$ {\sc gh{\small z}} with $8$ {\sc gb} of {\sc ram}.
(The \dynamo hybrids ran later with an extension to $32$ {\sc gb} of {\sc ram} under 
features of \matlab R2010b).
The wall time was measured using the {\sf tic} and {\sf toc} commands in \matlab.
Pure \krotov vs \grape comparisons (Tabs.~\ref{tab:results-overview-smallamps} through 
\ref{tab:results-overview-smallamps-con}) were carried
out on separate optimised \matlab implementations thus avoiding any overhead 
(e.g., loops and checks) required for more flexibility in \dynamo, where the hybrids 
(Figs.~\ref{fig:hybrid}, \ref{fig:hybrid2}) were run.
%%%%%%%%%%

%%%%%%%%%%
\begin{figure*}[ht]
\begin{center}
$\begin{array}{c@{\hspace{.3cm}}c}
\includegraphics[width=0.4\textwidth]{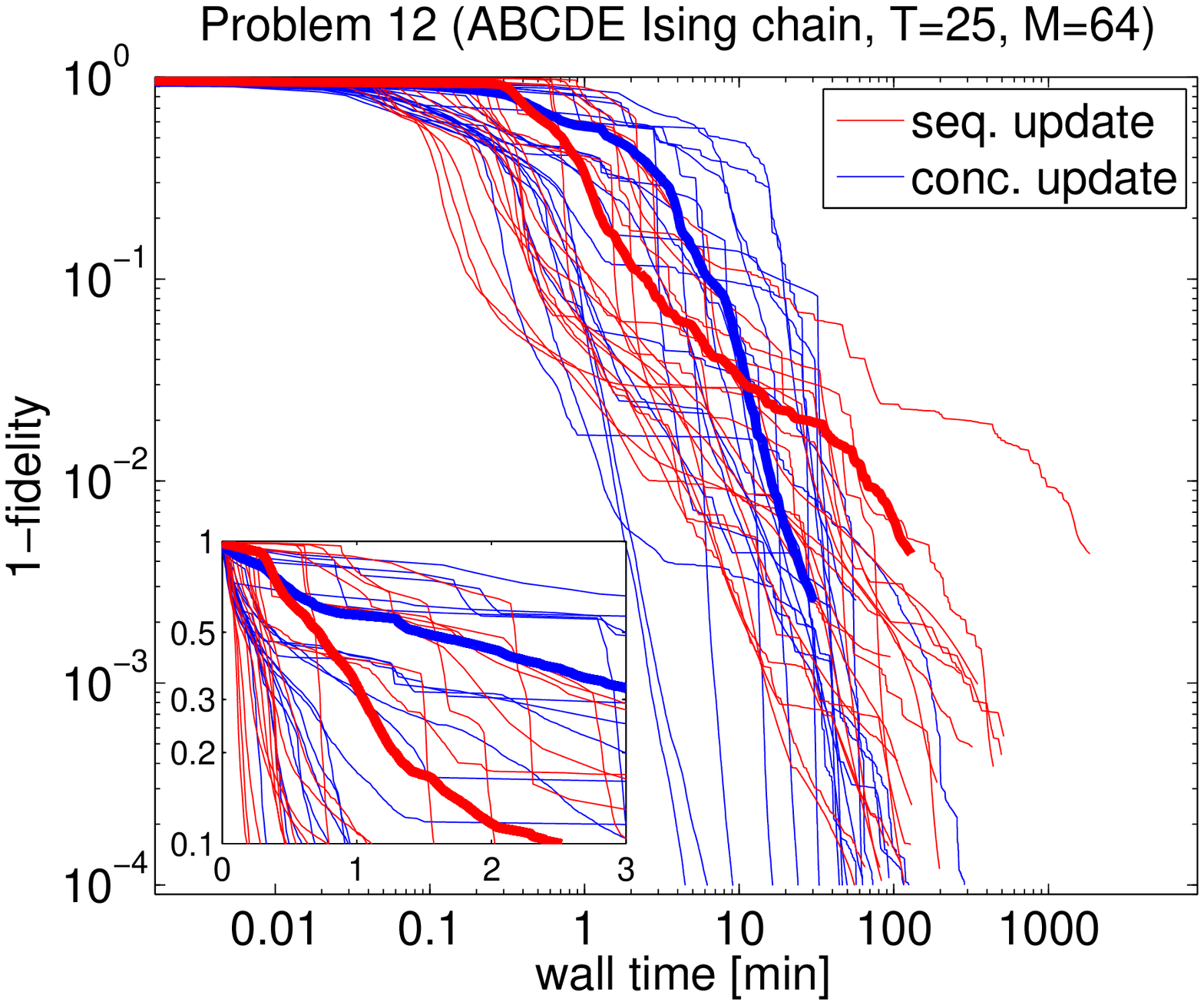}  &
\includegraphics[width=0.4\textwidth]{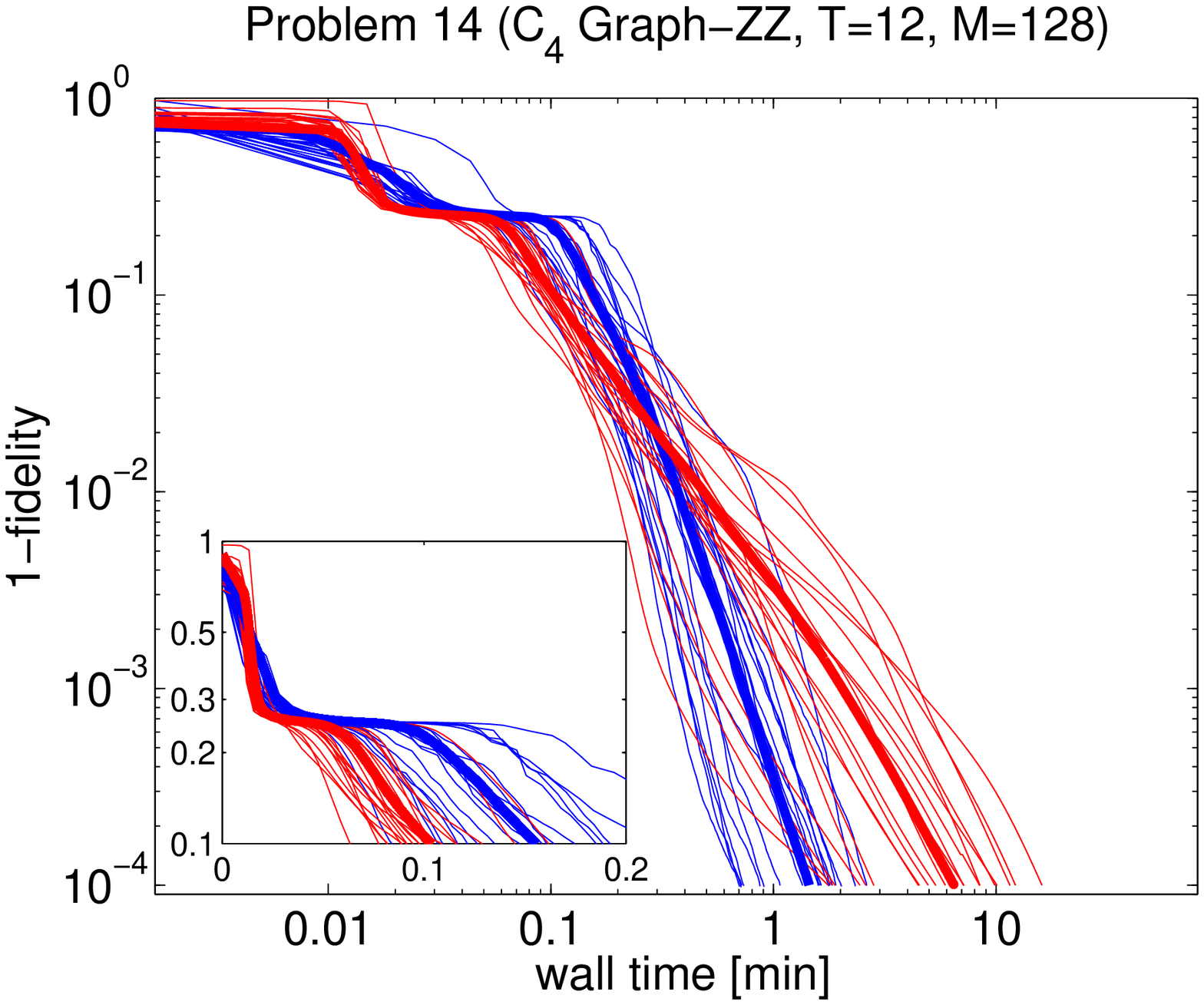} \\[0.2cm]
\includegraphics[width=0.4\textwidth]{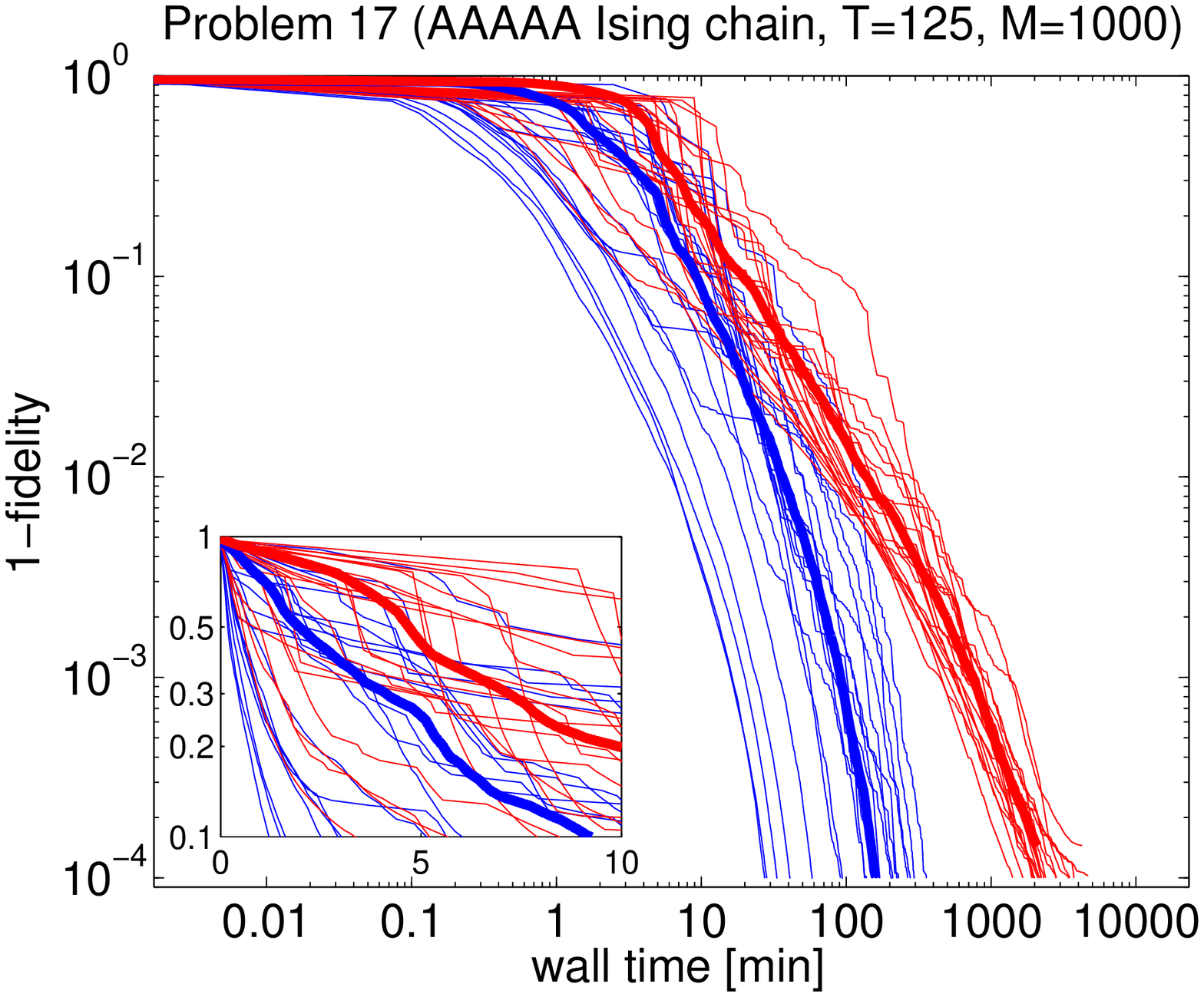}  &
\includegraphics[width=0.4\textwidth]{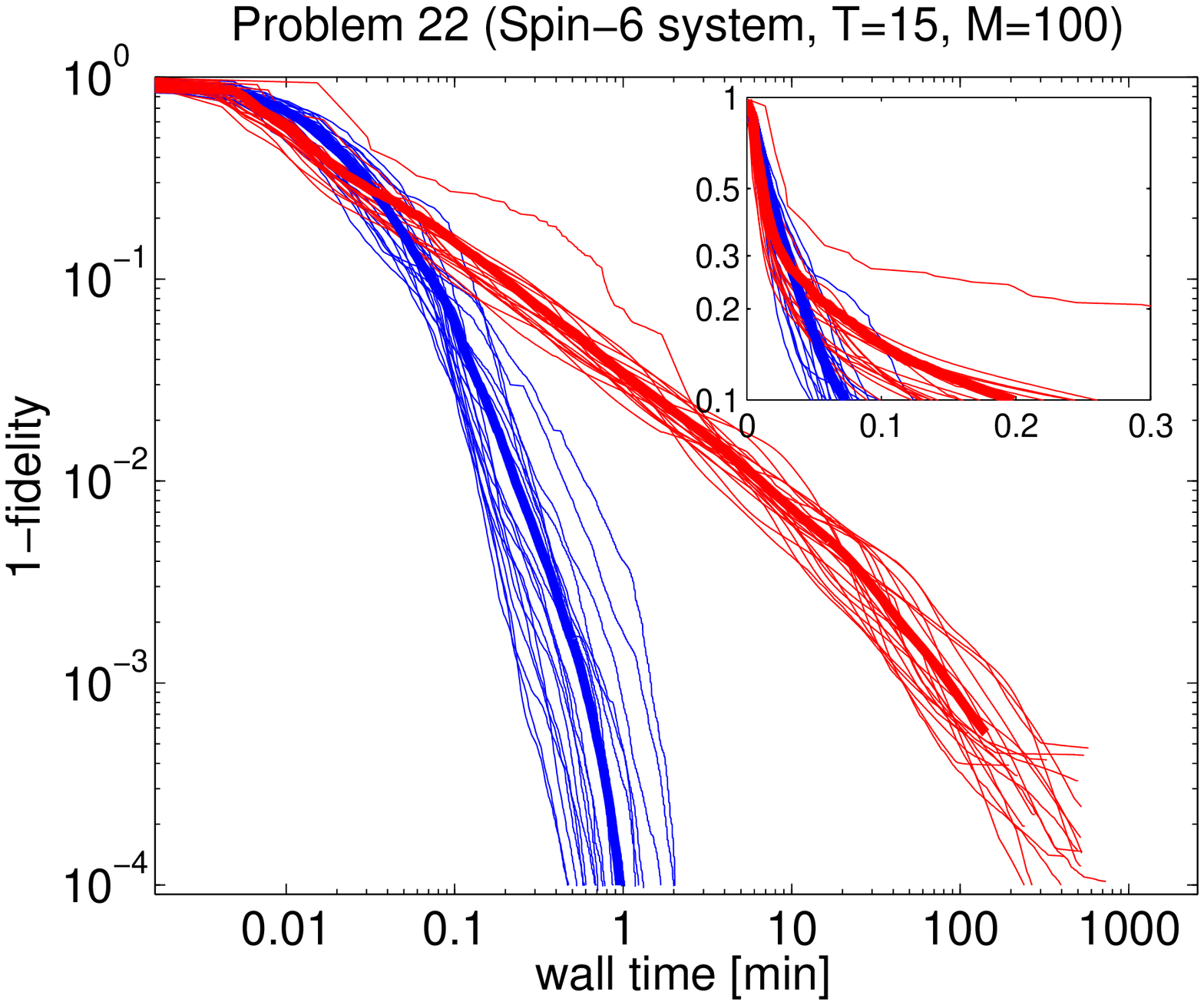} \\
\end{array}$
\end{center}
\caption{(Colour) Optimisation results for problems 12, 14, 17, and 22 shown in
doubly logarithmic plots; each optimisation is run with twenty random initial conditions;
the trace of mean values is given in boldface.
The blue (concurrent) and red (sequential) lines depict the deviation of the
quality from the maximum of 1 as a function of the wall time. Each line
represents one optimisation.  The insets show the initial behaviours and crossing
points in a log-linear scale. For the sequential-update algorithm in problem 22 (last panel), 
the thresholds for the change in the control and function values have to be lowered
(to $10^{-10}$ instead of the standard $10^{-8}$ [see test conditions]) 
for reaching qualities comparable to the ones the concurrent scheme arrives at under standard conditions.
{(Note the altered thresholds apply as well to the data listed in Tab.~\ref{tab:results-overview-smallamps}).}
\label{fig:spaghetti-overview}}
\end{figure*}
%%%%%%%%%%

%%%%%%%%%%
\begin{figure}[ht]
\begin{center}
\includegraphics[width=0.4\textwidth]{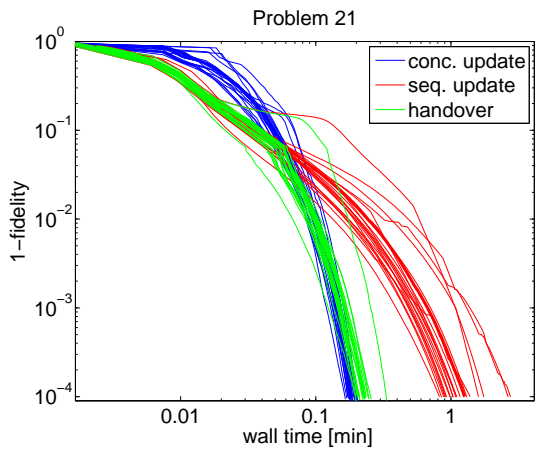}
\caption{(Colour) Example of a handover (green) from a sequential- (red) to a concurrent-update (blue) scheme.
The sequential algorithm is run up to a handover quality of $0.93$, 
where the resulting pulse sequence is then used as input to
the concurrent algorithm for optimisation up to the target quality. 
This type of handover is supported by the modular structure of \dynamo.
\label{fig:handover}}
\end{center}
\end{figure}
%%%%%%%%%%

%%%%%%%%%%
\begin{figure}[ht]
\begin{center}
\includegraphics[width=0.4\textwidth]{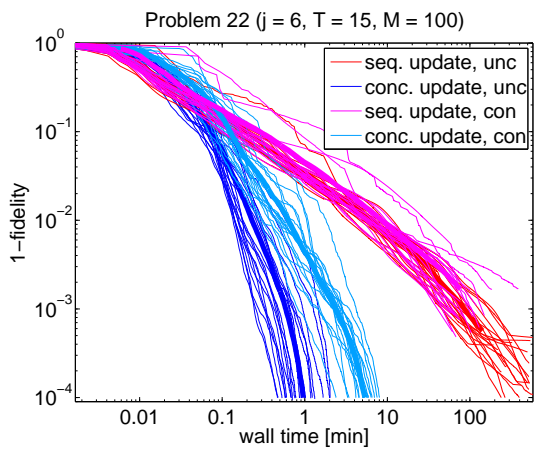}
\caption{(Colour) Comparison of unconstrained and (loosly) constrained optimisations. 
The concurrent-update algorithm uses the standard \matlab-toolbox functions
\texttt{fminunc} and \texttt{fmincon} with the latter being slower
than the former, as it may switch between different internal routines.
The sequential-update algorithm uses a very basic cut-off method for respecting the constraints,
which shows little effect on the performance. 
\label{fig:conunccomp}}
\end{center}
\end{figure}
%%%%%%%%%%

%%%%%%%%%%
\subsection{Test Results and Discussion}
%%%%%%%%%%

From the full set of data presented in Tab.~\ref{tab:results-overview-smallamps},
Fig.~\ref{fig:spaghetti-overview} selects a number of representatives for further illustration.
Note the following results:
First, in most of the problems, sequential and concurrent-update algorithms reach similar
final fidelities, the target set to $1-10^{-4}$ being in the order of a conservative
estimate for the error-correction threshold \cite{Knill05}.
Out of the total of 23 test problems, this target is met
within the limits of iterations specified above 
except in problems 5, 7, 10, 12 and 13. Only in problem 23
the sequential-update algorithm yields average residual errors (1-fidelity) up to two orders of
magnitude higher than in the concurrent optimisation. Remarkably enough, the average running times
differ substantially in most of the test problems, with the concurrent-update algorithm being faster.
Only in problems 3, 4, 15 and 16 the final wall times are similar. 
Note that in all but the very easy problems 3, 4,
and 16, the sequential algorithm needs a larger total number of matrix multiplications and eigendecompositions.
In particular, due to the slower convergence near the critical points,
the sequential-update scheme requires more iterations in order to reach the target fidelity
of $1-10^{-4}$ thus resulting in a greater number of matrix multiplications and eigendecompositions.

In many problems (3, 5, 6, 8, 9, 11, 12, 14, 18, 19, 21, and 22), we observe a crossing point in
time course of the fidelity of the two algorithms. The sequential-update algorithm is overtaken by the
concurrent-update scheme between a quality of $0.9$ and $0.99$ (see, e.g.,
Problem 21 in Fig.~\ref{fig:spaghetti-overview}).
Therefore, exploiting the modular framework of the programming package to 
dynamically change from a sequential to a concurrent-update scheme at a medium fidelity
can be advantageous. This is exemplified in the (constrained) optimisation 
shown in Fig.~\ref{fig:handover}: here
the sequential method is typically faster at the beginning of the optimisation,
whereas the concurrent method overtakes at higher fidelities near the end of the optimisation. --- 
Moreover with regard to dispersion of the final wall times required to achieve the
target fidelity, in problems 5, 7, 10, 11, 12, 13 and 23 the
sequential-update algorithm shows a larger standard deviation thus indicating
higher sensitivity to the initial controls.

Also on a more general scale,
we emphasise that the run-times may strongly depend on the {\em choice of initial conditions}. 
Results for larger initial pulse amplitudes with a higher standard deviation
can be found in Tab.~\ref{tab:results-overview-highamps}.
Increasing mean value and standard
deviation of the initial random control-amplitude vectors typically translates into 
longer run times. This effect is more pronounced for sequential than for
concurrent-update algorithms. Consequently, the performance differences
between the two algorithms may increase and crossing or handover points may
change as well. 

Finally, as shown in Fig.~\ref{fig:conunccomp}, 
the performance of the concurrent-update scheme also differs between constrained
and unconstrained optimisation, i.e.~between the standard \matlab subroutines
\texttt{fmincon} and \texttt{fminunc} (see \matlab documentation). 
In contrast, the sequential-update algorithm uses the same set of routines for
both types of optimisations, where a basic cut-off method for respecting the
constraints has almost no effect, as also illustrated by Fig.~\ref{fig:conunccomp}.

%%%%%%%%%%%%
\begin{figure}[Ht!]
\begin{center}
\includegraphics[width=0.45\textwidth]{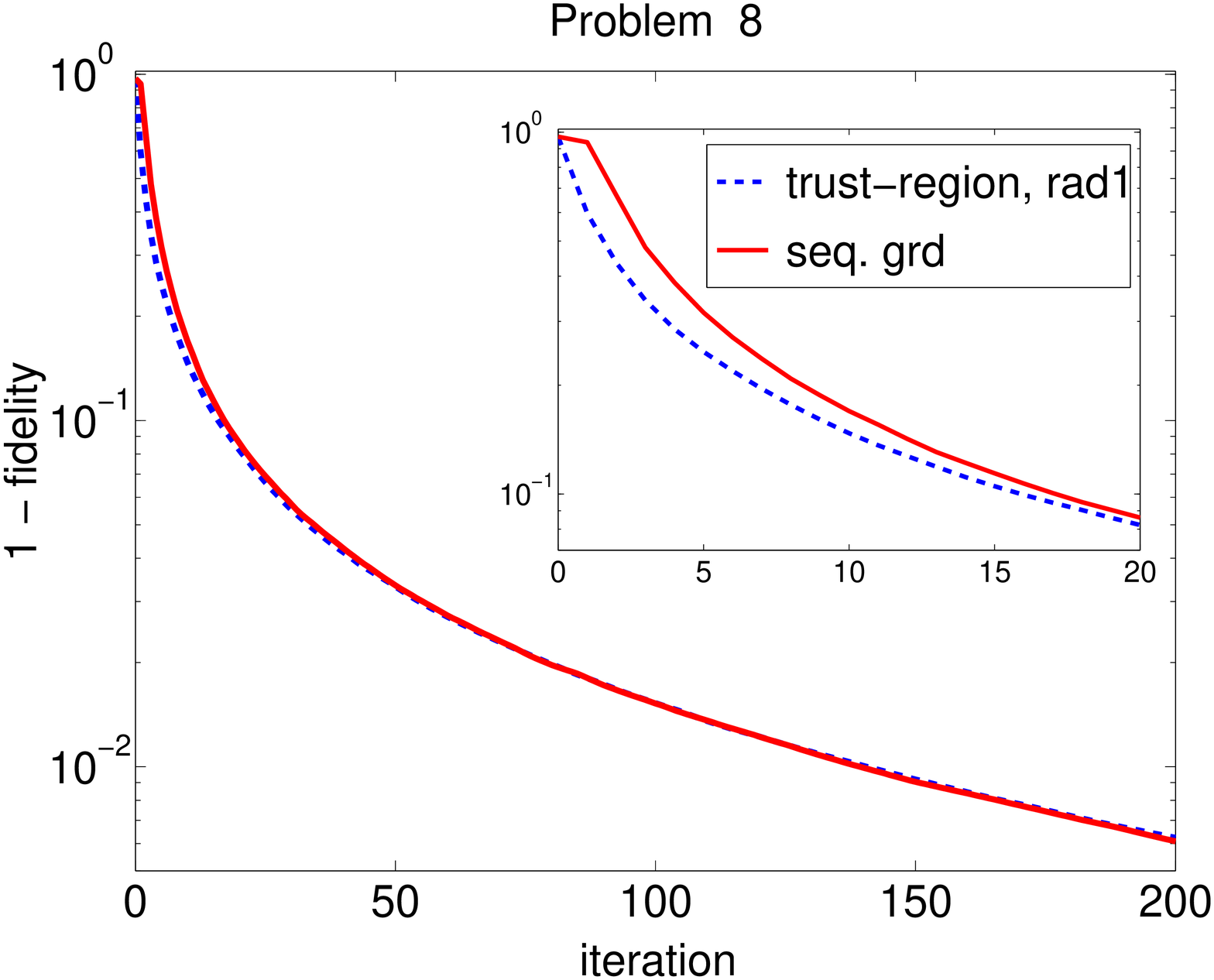}
\caption{\label{fig:mimick2s}
(Colour) 
Comparison of sequential-update methods with first-order gradient
information (red track) and with a direct implementation of a trust-region Newton 
method (blue dotted track) showing that per iteration the gains are similar, in particular
in the long run. The curves represent averages over $100$ trajectories with random initial
conditions. 
}
\end{center}
\end{figure}
%%%%%%%%%%%%
\subsection{Preliminaries on Trust-Region Newton Methods for Sequential-Update Algorithms}\label{sec:mimick2order}
Fig.~\ref{fig:mimick2s} shows that the sequential-update method with first-order gradient
information used in this work already achieves a quality gain per iteration that comes closest to the one 
obtained by a direct implementation of a trust-region Newton method. However, 
as is analysed in detail on a larger scale in \cite{SF11},
the small initial advantage {\em per iteration} of latter against the former is outweighed 
\cpu-timewise by more costly calculations, which is why we have used the first-order gradients
for comparison.

\subsection{Comparing Gradient Methods}
We compare the performance of four different methods to compute gradients for the
concurrent algorithm: in addition to the standard approximation and the exact
procedure described in section \ref{gradient-update-modules}, we
follow Ref.~\cite{Kuprov09} and study a Taylor series to compute the
exponential and a Hausdorff series to compute the gradient, while the fourth
method is standard finite-differences. Note that Hausdorff series and finite
differences can be taken to a numerical precision exceeding that of the standard
approximation.
%%%%%%%%%%%%
\begin{figure}[Ht!]
\begin{center}
\includegraphics[width=0.4\textwidth]{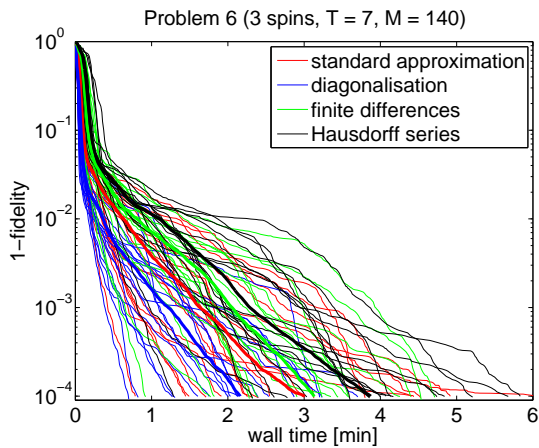}
\caption{\label{fig:gradcomp}
(Colour) Comparison of four different methods for computing gradients in $20$ unitary optimisations
of problem $6$. Apart from the standard approximation, all methods compute exact gradients.
By making use of the spectral decomposition,
diagonalising the total Hamiltonian to give exact parameter derivatives \cite{TILO96,Aizu63,Wilcox67}
is the fastest among these methods, because by the eigen-decomposition the matrix exponential
can be settled as well (i.e.~in the same go).
In case of optimising controls for (pure) state-to-state transfer, the standard
approximation can be shown to be competitive.} 
\end{center}
\end{figure}
%%%%%%%%%%%%

An example of the performance results found for these four methods is given in
Fig.~\ref{fig:gradcomp}, 
where we optimise controls for a QFT on the 4-spin system of problem 6.
Unitary optimisations on other systems yield similar results, with the diagonalisation
being the fastest methods in all cases. For state-to-state transfer (pure states), however,
the standard approximation performs well enough as to be competitive with exact gradients by
diagonalisation.  Note that for unitary gate synthesis of generic gates, 
one cannot use sparse-matrix techniques, for which the Hausdorff
series is expected to work much faster as demonstrated in the software package
{\sc spinach}~\cite{Kuprov07}.

\subsection{Hybrid Schemes}

%%%%
\begin{figure}[Ht!]
\begin{center}
\hspace{10mm}\sf{(a)}$\hfill$\\
\includegraphics[width=0.39\textwidth]{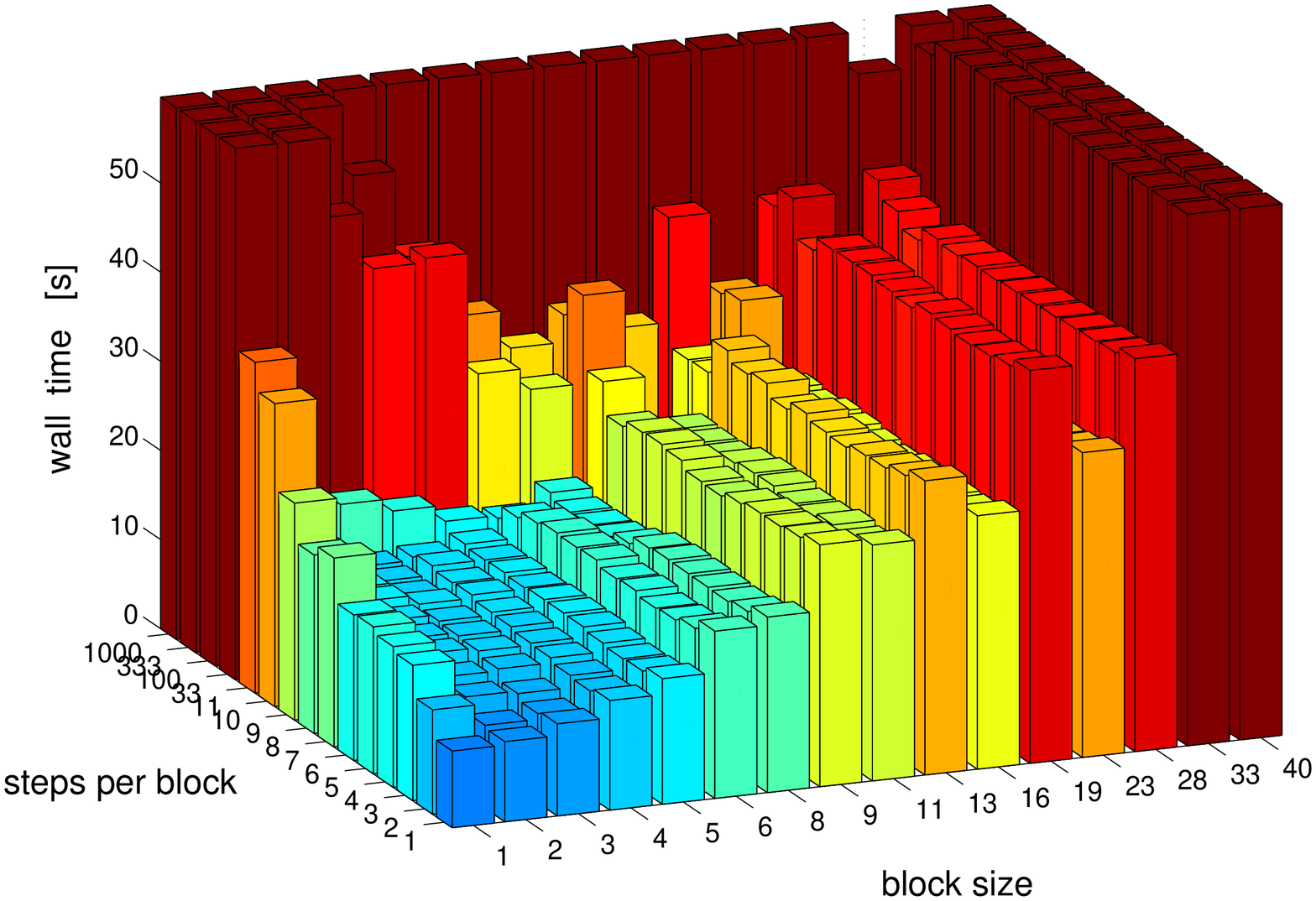}\\
\hspace{10mm}\sf{(b)}$\hfill$\\
\includegraphics[width=0.39\textwidth]{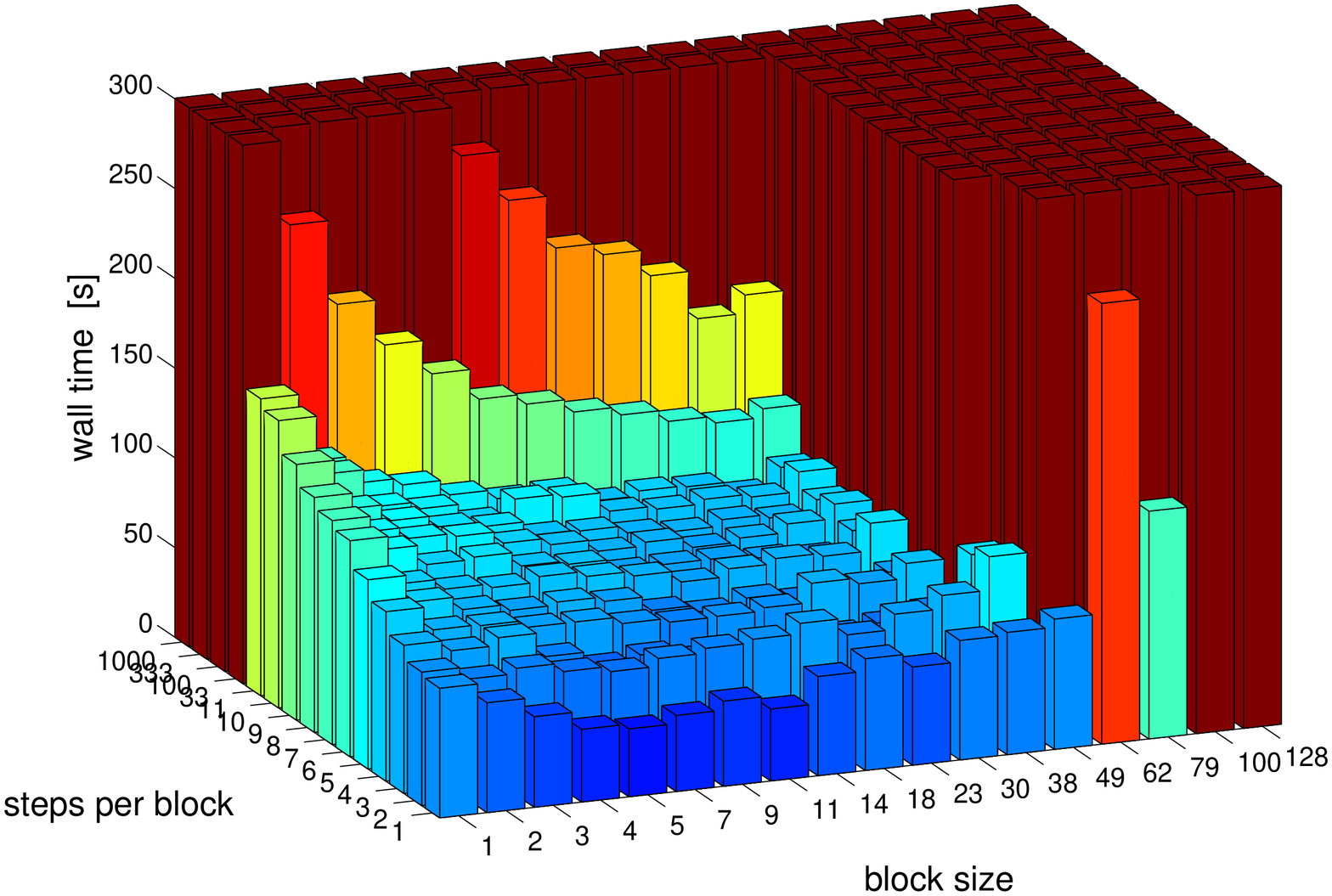}
\caption{
(Colour) Performance of generalising the {\em first-order-gradient 
sequential scheme} to updating blocks of joint \timeslices
and allowing for multiple iteration steps within each block  ($s_{limit} > 1$), 
as applied to (a) test problem 2 and (b) 21 (see Tab.~\ref{tab:test-def} and Sec.~\ref{sec:ExampleSpinChains}). 
Original \krotov modifies one \timeslice in a single iteration ($s=1$) before moving to the
subsequent \timeslice to be updated: this special case
is shown in the lower corner of the plot, while the upper right is the first-order
variant of \grape (in a suboptimal setting, since the step-size handling is
taken over from the one optimised for \krotov). Wall times represent the average over
$42$ runs with random initial control vectors 
(again with $mean(u_{ini})=0$ and $std(u_{ini})=1$ in units of $1/J$); times are cut off at 
$60$ (resp.~$300$) sec.
Note that in problem 2 the hybrid first-order versions are not faster than the original Krotov, while
in problem 21 it pays to concurrently update four or five time slots by a single step 
before moving on to the next set of time slots. --- Note that in other cases also the
first-order concurrent update can be fastest, see Fig.~\ref{fig:moreSpaghettis}.
\label{fig:hybrid}}
\end{center}
\end{figure}
%%%%

\begin{figure}[Ht!]
\begin{center}
\hspace{10mm}\sf{(a)}$\hfill$\\
\includegraphics[width=0.39\textwidth]{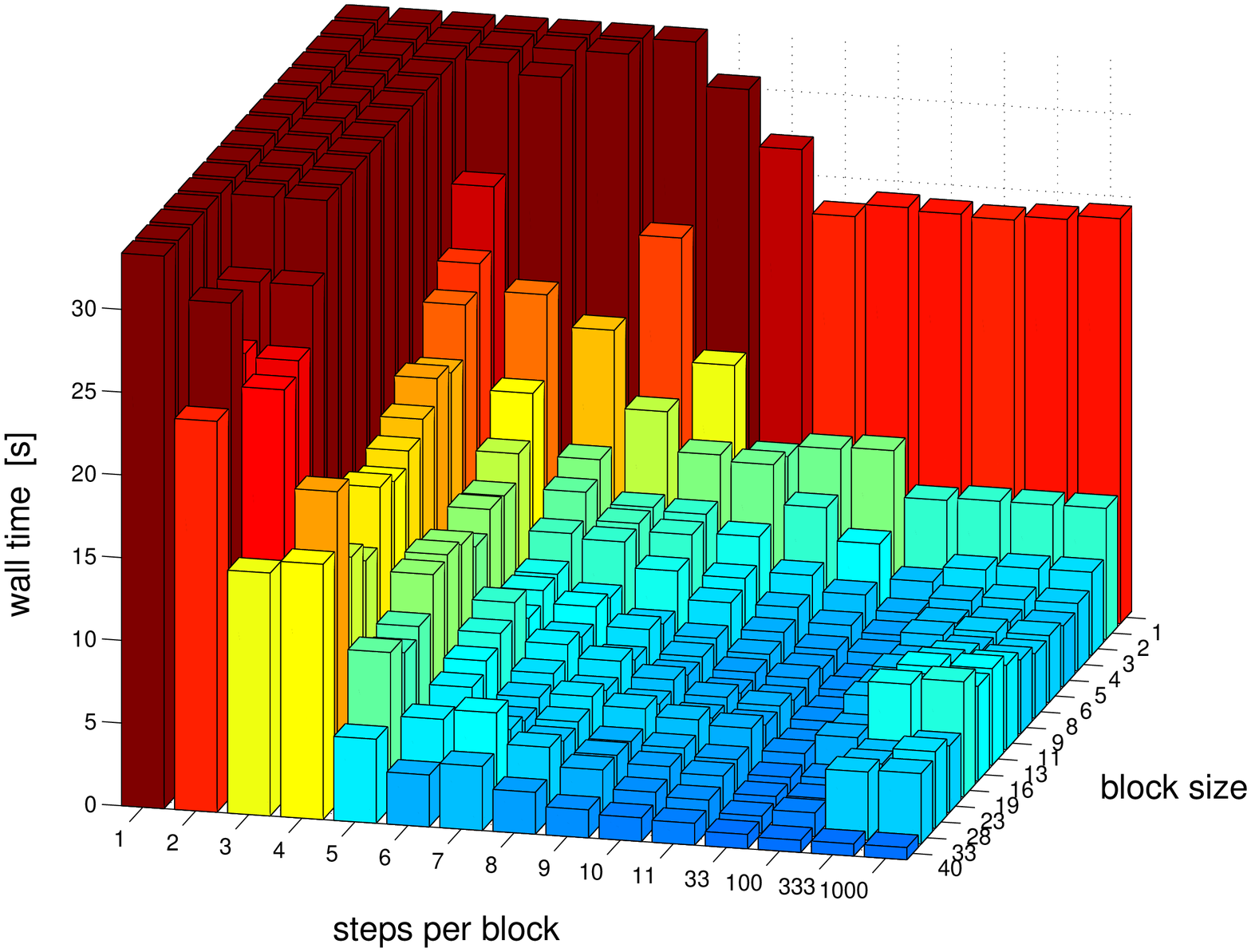}\\
\hspace{10mm}\sf{(b)}$\hfill$\\
\includegraphics[width=0.39\textwidth]{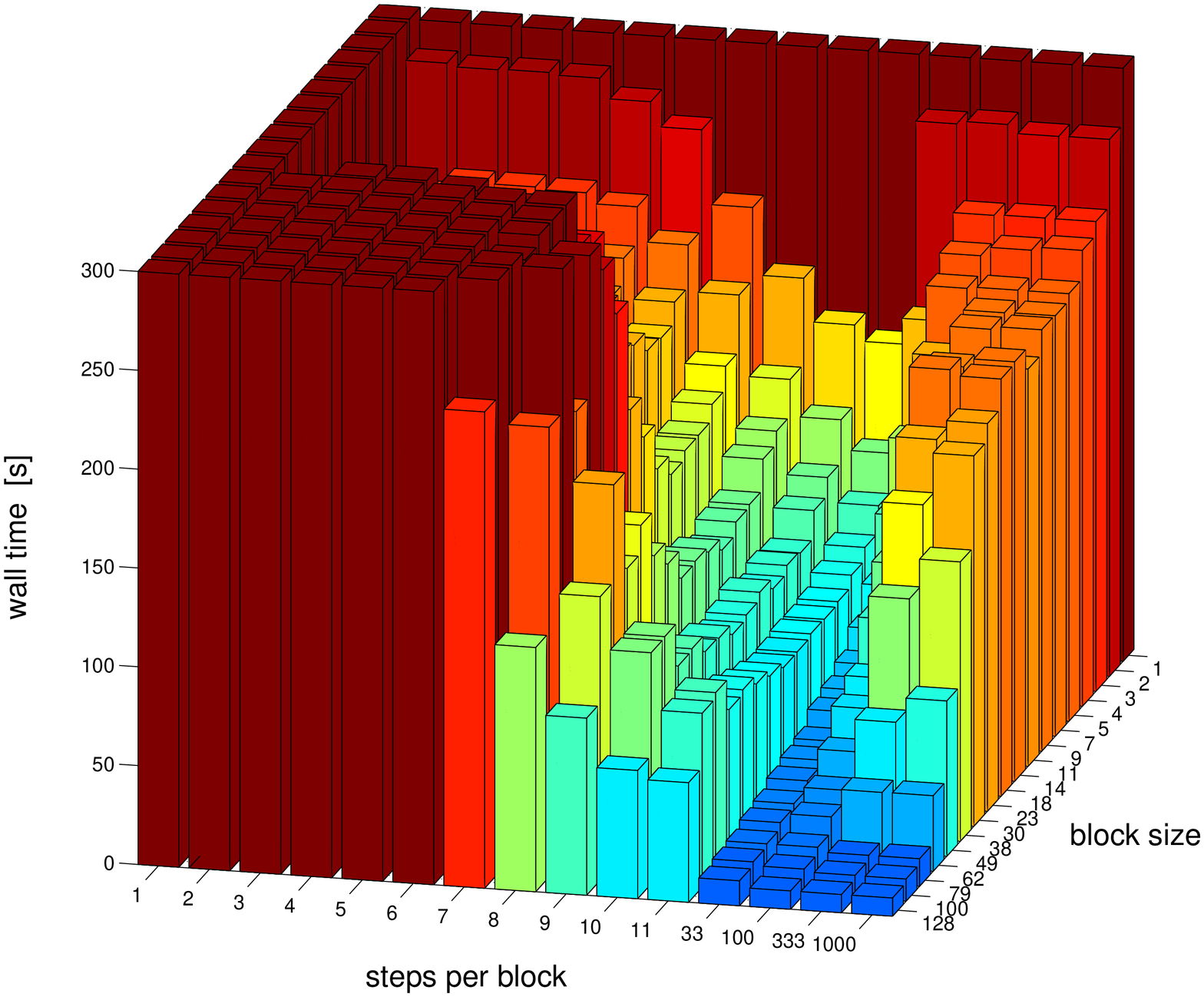}
\caption{
(Colour) Performance of generalising the {\em second-order} (\bfgs) {\em concurrent scheme} 
to updating blocks of joint \timeslices
and allowing for fewer iteration steps within each block  ($s_{limit}$),
as applied to (a) problem~2 and (b) problem~21  
(see Tab.~\ref{tab:test-def} and Sec.~\ref{sec:ExampleSpinChains}).
Original \grape modifies all time slices in each iteration: this special case is shown in 
the lower right corner of the plot, while the upper left corner is the crude second-order variant of \krotov 
(for the sake of comparison here in the unrecommendable setting of \bfgs). 
It is part of the obvious {\em no-go area} of single iterations ($s=1$) on a single time slice ($n=1$),
or just few, shown for completeness.
Wall times represent the average over $42$ runs with random initial control vectors
again with $mean(u_{ini})=0$ and $std(u_{ini})=1$ in units of $1/J$; 
times are cut off at $35$ (resp.~$300$) sec.
\label{fig:hybrid2}}
\end{center}
\end{figure}

Using \dynamo, we have just begun to explore the multitude of possible hybrid schemes.
Here we present first-order (Fig.~\ref{fig:hybrid}) as well as second-order (Fig.~\ref{fig:hybrid2}) schemes, 
where the hybrids are taken with respect to sequential {\em versus} concurrent \subspace-selection.
More precisely, this amounts to an outer-loop subspace-selection scheme which picks consecutive 
blocks of $n$ \timeslices to be updated in the inner-loop using either a first-order or a second-order-method 
update scheme each allowing to take at most $s_{limit}$ steps within each block. 
The results of these explorations, as applied to the two-spin case of problem~2 
and problem~21 (see Tab. \ref{tab:test-def} and Sec. \ref{sec:ExampleSpinChains} with the same
initial conditions as in Tab.~\ref{tab:results-overview-smallamps}), are depicted in 
Fig.~\ref{fig:hybrid} for first-order gradient update and in Fig.~\ref{fig:hybrid2} for
second-order \bfgs update. 
They provide illuminating guidelines for further investigation, as the \krotov method 
taking a single timeslice ($n=1$) sequentially after the other for
a single update step ($s_{\rm limit}=1$)
may not always be the best-performing use of the first-order update scheme.
%They are surprising on several fronts: in first-order updates\\
%(i) the \krotov method taking a single timeslice ($n=1$) sequentially after the other for 
%	a single update step ($s_{\rm limit}=1$)
%       are clearly not the best-performing use of the first-order update scheme;\\
%(ii) for pure sequential updates (i.e.\ selecting a single \timeslice each time), a
%	slight advantage arises by taking two steps in each \timeslice instead of one (as in \krotov);\\
%(iii)  the best performance (significantly outpacing plain-vanilla \krotov and the first-order \grape variant) 
%	is achieved by setting the block size to $n=10$ (for a total of $M=50$ time slices), 
%	while performing about $s=9$ iterations within each block before continuing 
%	on to the next $n$ \timeslices. --- 

On the other hand, in second-order \bfgs methods %\\
%(iv) 
the \grape scheme with totally concurrent update cannot be accelerated by allowing
for smaller blocks of concurrent update in the sense of a \/`compromise towards \krotov\/';
rather it is an optimum within a broader array of similarly performing schemes. This is
remarkable, while the incompatibility of \bfgs with sequential update rules is to
be expected on the grounds of the discussion above.

Further explorative numerical results on first-order hybrids between sequential and concurrent
update as compared to the second-order concurrent update can be found as Fig.~\ref{fig:moreSpaghettis}
in the appendix. They show that in simpler problems first-order sequential update (as in \krotov)
is faster than the (highly suboptimal) {\em first-order variants} of hybrid or concurrent update, while
in more complicated problems already the first-order variant of concurrent update is slightly faster. 
In any case, all first-order methods are finally outperformed by second-order concurrent update (as in \grape-\bfgs).

Clearly, these explorative results are by no means the last word on the subject. Rather
they are meant to invite further
studies over a wider selection of problems. But even at this early stage we
can state that there are hints that hybrid methods hold a yet untapped potential,
and follow-up work is warranted.

%%%%
%%%%

%%%%%%%%%%%%%%%%%%%%%%%%%%%%%%%%%
\section{Conclusions and Outlook}
%%%%%%%%%%%%%%%%%%%%%%%%%%%%%%%%%
We have provided a unifiying modular programming framework, \dynamo, for numerically 
addressing bilinear quantum control systems. It allows for benchmarking, 
comparing, and optimising numerical algorithms that constructively 
approximate optimal quantum controls. Drawing from the modular 
structure, we have compared the performance of gradient-based algorithms with 
sequential update of the \timeslices in the control vector (\krotov-type) versus
algorithms with concurrent update (\grape-type) with focus on synthesising unitary
quantum gates with high fidelity. ---
For computing gradients,
exact methods using the eigendecomposition have on average proven superior
to gradient approximations by finite differences, series expansions, or time averages.

When it comes to implementing second-order schemes, the different construction 
of sequential update and concurrent update translates into different performance: 
in contrast to the former, recursive concurrent updates
match particularly well with quasi-Newton methods and their iterative approximation of 
the (inverse) Hessian as in standard \bfgs implementations. Currently, however,
there seems to be no standard Newton-type second-order routine that would match with
sequential update in a computationally fast and efficient way such as to significantly
outperform our implementation of first-order methods. 
Finding such a routine is rather an open research problem. At this stage, we have employed
efficient implementations, i.e.\ first-order gradient ascent for sequential
update and a second-order concurrent update (\grape-\bfgs). As expected from
second-order versus first-order methods, at higher fidelities (here typically $90-99\%$), 
\grape-\bfgs overtakes \krotov. For reaching
a fidelity in unitary gate synthesis of $1-10^{-4}$, \grape-\bfgs is faster, 
in a number of instances even by more than one order of magnitude on average. 
Yet at lower qualities the computational speeds 
are not that different and sequential update typically has a (small) advantage.

By its flexibility, the \dynamo framework answers a range of needs,
reaching from quantum information processing to coherent spectroscopy.
For the primary focus of this study, namely gate synthesis with
high fidelities beyond the error-correction threshold of some $10^{-4}$ \cite{Knill05},
fidelity requirements
significantly differ from pulse-engineering for state-transfer, where often for the sake of robustness
over a broad range of experimental parameters, some fidelity (say $5\%$) may readily
be sacrificed. Thus for optimising robustness, sequential update schemes are
potentially advantageous, while for gate synthesis sequential methods can
be a good start, but for reaching high fidelities, we recommend to change
to concurrent update. More precisely, since \dynamo allows for efficient handover from
one scheme to the other, this is our state-of-the-art recommendation.

On-going and future comparisons are expected to profit from this framework,
e.g., when trying update modules with non-linear conjugate gradients \cite{HZ05,HZ06}.

\medskip
\subsubsection*{Research Perspectives}
We have presented a first step towards establishing a ``best of breed'' toolset
for quantum optimal control. It is meant to provide the platform for future improvements
and follow-up studies, e.g., along the following lines:\\[-2.3mm]

{\em Further Types of Applications:} We have focussed on the synthesis of high-fidelity unitary
quantum gates in closed systems. --- Yet, follow-up comparisons should extend to open systems or to
spectroscopic state transfer, where it is to be anticipated that different demand of fidelity
may lead to different algorithmic recommendations.\\[-2.3mm]

{\em Initial Conditions:} Currently there is no systematic way how to choose 
good initial control vectors in a problem-adapted way. 
Scaling of initial conditions has been shown to 
translate into computational speeds differing significantly (i.e.\ up to an order of magnitude). 
--- Yet, good guidelines for selecting initial controls are still sought for.\\[-2.3mm]

{\em Second-Order methods for sequential update:} As has been mentioned, we have indications
that sequential update methods are most efficient when matched to first-order gradient
procedures. --- Yet, this issue is subject of follow-up work.\\[-2.3mm]

{\em Hybrid Algorithms:} We have focused on the two extremes of the
update scheme spectrum: the sequential and the fully concurrent. --- Yet, hybrid schemes which
intelligently select the subset of \timeslices to update at each iteration, and dynamically
decide on the number of steps and appropriate gradient-based stepping methodology for the
inner loop may even achieve better results than the established two extremes. The success,
however, depends on developing alternatives to \bfgs matching with
sequential update schemes (s.a.).\\[-2.3mm]

{\em Control Parametrisation Methods:} We
have looked exclusively at piece-wise-constant discretisation of the control
function in the time domain. --- Yet, although also frequency-domain methods exist (e.g. \cite{CRAB}),
there is both ample space to develop further methods and need for comparative benchmarking.\\[-2.3mm]

{\em Algorithms for Super-Expensive Goal Functions:} For
many-body quantum systems, ascertaining the time-evolved state of the system
requires extremely costly computational resources. --- Yet, algorithms described in this 
manuscript all require some method of ascertaining the gradient, by finite differences if no other
approach is available. Such requirements, however, are mal-adapted to super-expensive goal functions.
Further research to discover new search algorithms excelling in such use cases is required.

\bigskip
\begin{acknowledgments}
%%%%%%%
We wish to acknowlwdge useful discussions at the {\sc Kavli}-Institute; 
in particular, we are indebted to exchange within 
the informal \/`optimal-control comparison group\/' hosted and supported by 
Tommaso ~Calarco through the {\sc eu} project {\sc acute}.
We thank Seth Merkel and Frank Wilhelm for suggesting to test higher spin-$j$ systems
and Ilya Kuprov for helpful discussions.
%Also helpful discussions with Ilya Kuprov at the Kavli Institute Santa Barbara are gratefully
%acknowledged. ---
This work was supported %in part
by the Bavarian PhD programme of excellence {\sc qccc},
by the {\sc eu} projects {\sc qap, q-essence}, exchange with {\sc coquit},
by {\em Deutsche Forschungsgemeinschaft}, {\sc dfg}, in {\sc sfb}~631.
S.S. gratefully acknowledges the {\sc epsrc} {\sc arf} grant {\sc ep/do7192x/1}. 
P.d.F. is supported by {\sc epsrc} and Hitachi ({\sc case/cna/07/47}).
S.M. wishes to thank the {\sc eu} project {\sc corner} and the Humboldt foundation.
The calculations were carried out mostly on the Linux cluster of {\em Leibniz Rechenzentrum}
({\sc lrz}) of the Bavarian Academy of Sciences.
\end{acknowledgments}
%%%
%%%%%%%%%%%%%%
%\bibliography{control21}
%%%
%%%%%%%%%%%%%%
%%%%%%%%%%%%%%

%%%%%%%%%%%%%%
%%%%%%%%%%%%%%
%%%%%%%%%%%%%%%%%%%%%%%%%%%%%%%%%%%%%%%%%%%%%%%%%%%%%%%%%%%%%%%%%%%%%%%%%%%%%%%%%%%%%%%%%%%%%%%%
%\subsection{Tables}
%%%%%%%%%%%%%%%%%%%%%%%%%%%%%%%%%%%%%%%%%%%%%%%%%%%%%%%%%%%%%%%%%%%%%%%%%%%%%%%%%%%%%%%%%%%%%%%%
%%%%%%%%%
\begin{table*}[Ht!]
\begin{center}
\caption{Specification of Test Problems}
\label{tab:test-def}
\begin{tabular}{c|cccccccc}
\hline\hline\\
Problem & Quantum System & Matrix     & No. of    & Final Time & Target &\\
        &                & Dimensions & Time Slices & [1/J]      & Gate  &\\[0.3cm]
\hline\\[0.1cm]
1 & $AB$ Ising-ZZ chain & 4 & 30 & 2 & CNOT &\\[0.3cm]
2 & $AB$ Ising-ZZ chain & 4 & 40 & 2 & CNOT &\\[0.3cm]
3 & $AB$ Ising-ZZ chain & 4 & 128 & 3 & CNOT &\\[0.3cm]
4 & $AB$ Ising-ZZ chain & 4 & 64 & 4 & CNOT &\\[0.1cm]\hline\\[-0.1cm]
5 & $ABC$ Ising-ZZ chain & 8 & 120 & 6 & QFT &\\[0.3cm]
6 & $ABC$ Ising-ZZ chain & 8 & 140 & 7 & QFT &\\[0.1cm]\hline\\[-0.1cm]
7 & $ABCD$ Ising-ZZ chain & 16 & 128 & 10 & QFT &\\[0.3cm]
8 & $ABCD$ Ising-ZZ chain & 16 & 128 & 12 & QFT &\\[0.3cm]
9 & $ABCD$ Ising-ZZ chain & 16 & 64 & 20 & QFT  &\\[0.1cm]\hline\\[-0.1cm]
10 & $ABCDE$ Ising-ZZ chain & 32 & 300 & 15 & QFT &\\[0.3cm]
11 & $ABCDE$ Ising-ZZ chain & 32 & 300 & 20 & QFT &\\[0.3cm]
12 & $ABCDE$ Ising-ZZ chain & 32 & 64 & 25 & QFT  &\\[0.1cm]\hline\\[-0.1cm]
13 & $C_4$ Graph-ZZ & 16 & 128 & 7 & $U_{CS}$ &\\[0.3cm]
14 & $C_4$ Graph-ZZ & 16 & 128 & 12 & $U_{CS}$  &\\[0.1cm]\hline\\[-0.1cm]
15 & NV-centre & 4 & 40 & 2 & CNOT  &\\[0.3cm]
16 & NV-centre & 4 & 64 & 5 & CNOT &\\[0.1cm]\hline\\[-0.1cm]
17 & $AAAAA$ Ising-ZZ chain & 32 & 1000 & 125 & QFT &\\[0.3cm]
18 & $AAAAA$ Ising-ZZ chain & 32 & 1000 & 150 & QFT &\\[0.1cm]\hline\\[-0.1cm]
19 & $AAAAA$ Heisenberg-XXX chain & 32 & 300 & 30 & QFT &\\[0.1cm]\hline\\[-0.1cm]
20 & $A00$ Heisenberg-XXX chain & 8 & 64 & 15 & rand U  &\\[0.1cm]\hline\\[-0.1cm]
21 & $AB00$ Heisenberg-XXX chain & 16 & 128 & 40 & rand U &\\[0.1cm]\hline\\[-0.1cm]
22 & driven spin-$6$ system & 13 & 100 & 15 & rand U &\\[0.1cm]\hline\\[-0.1cm]
23 & driven spin-$3$ system & 7 & 50 & 5 & rand U &\\[0.3cm]
\hline\hline
\multicolumn{5}{l}{}\\
\multicolumn{5}{l}{A notation \textit{ABC} means the spin chain consists of three spins that are addressable %
			each by an }\\
\multicolumn{5}{l}{individual set of $x$- and $y$-controls. We write $A0$ for a locally controllable spin $A$ which is coupled}\\
\multicolumn{5}{l}{to a neighbour $0$ not accessible by any control field.}
\end{tabular}
\hspace{15mm}
\end{center}
\end{table*}
%%%%%%%%%

%%%%%%%%%
\begin{table*}[!Ht]
\begin{center}
\caption{Test results obtained from 20 {\em unconstrained optimisations}
(\texttt{fminunc} in \matlab)
for each problem of Tab.~\ref{tab:test-def} using the sequential
or the concurrent-update algorithm. {\em Small initial pulse amplitudes} were used
($\mathrm{mean}(u_{ini}) = 0$, $\mathrm{std}(u_{ini}) = 1$).
}\vspace{2mm}
\label{tab:results-overview-smallamps}
\footnotesize
\begin{tabular}{c|ccccccc}
\hline\hline\\
Problem & Algorithm & Final Fidelity & Wall Time [min] & \#Eigendecs/1000 & \#Matrix Mults/1000  \\
& & (\textbf{mean}/min/max) & (\textbf{mean}/min/max) & (\textbf{mean}/min/max) & (\textbf{mean}/min/max) \\[0.3cm]
\hline\\[-1mm]
1 & conc. & \textbf{0.9999}/0.9999/1.0000 & \textbf{0.02}/0.01/0.03 & \textbf{2.02}/1.35/2.94 & \textbf{38}/25/56 \\ 
  & seq.  & \textbf{0.9999}/0.9999/0.9999 & \textbf{0.19}/0.13/0.34 & \textbf{6.29}/4.36/11.68 & \textbf{88}/61/163 \\[0.3cm]
2 & conc. & \textbf{0.9999}/0.9999/1.0000 & \textbf{0.05}/0.03/0.08 & \textbf{2.68}/1.76/4.44 & \textbf{50}/33/84  \\ 
  & seq. & \textbf{0.9999}/0.9999/0.9999 & \textbf{0.16}/0.11/0.27 & \textbf{5.43}/3.80/9.04 & \textbf{76}/53/126  \\[0.3cm]
3 & conc. & \textbf{0.9999}/0.9999/1.0000 & \textbf{0.05}/0.04/0.08 & \textbf{4.61}/3.46/7.04 & \textbf{85}/63/132 \\ 
  & seq. & \textbf{0.9999}/0.9999/0.9999 & \textbf{0.07}/0.05/0.12 & \textbf{2.29}/1.56/4.12 & \textbf{32}/22/57 \\[0.3cm]
4 & conc. & \textbf{0.9999}/0.9999/1.0000 & \textbf{0.02}/0.01/0.02 & \textbf{1.70}/1.28/2.43 & \textbf{31}/23/45  \\ 
  & seq. & \textbf{0.9999}/0.9999/0.9999 & \textbf{0.05}/0.03/0.11 & \textbf{1.72}/1.08/3.84 & \textbf{24}/15/54 \\[0.1cm]\hline\\[-0.1cm]
5 & conc. & \textbf{0.9978}/0.9973/0.9990 & \textbf{7.19}/5.84/7.86 & \textbf{362}/310/367 & \textbf{9774}/8364/9917 \\ 
  & seq. & \textbf{0.9973}/0.9918/0.9986 & \textbf{34}/22/58 & \textbf{1976}/1320/3292 & \textbf{35542}/23729/59209  \\[0.3cm]
6 & conc. & \textbf{0.9999}/0.9999/0.9999 & \textbf{0.85}/0.34/2.21 & \textbf{35}/17/76 & \textbf{954}/450/2050  \\ 
  & seq. & \textbf{0.9999}/0.9999/0.9999 & \textbf{5.14}/1.14/18.72 & \textbf{310}/68/1143 & \textbf{5574}/1216/20554 \\[0.1cm]\hline\\[-0.1cm]
7 & conc. & \textbf{0.9970}/0.9886/0.9999 & \textbf{9.42}/2.32/18.48 & \textbf{229}/63/391 & \textbf{8028}/2210/13679 \\ 
  & seq. & \textbf{0.9945}/0.9825/0.9999 & \textbf{242}/50/491 & \textbf{3975}/1242/7753 & \textbf{87385}/27313/170455\\[0.3cm]
8 & conc. & \textbf{0.9999}/0.9999/0.9999 & \textbf{2.90}/0.83/10.39 & \textbf{72}/21/275 & \textbf{2530}/735/9627 \\ 
  & seq. & \textbf{0.9999}/0.9999/0.9999 & \textbf{16.11}/2.36/65.72 & \textbf{500}/89/2223 & \textbf{11002}/1953/48876 \\[0.3cm]
9 & conc. & \textbf{0.9999}/0.9999/0.9999 & \textbf{0.61}/0.33/0.92 & \textbf{20}/11/30 & \textbf{685}/381/1052 \\ 
  & seq. & \textbf{0.9999}/0.9999/0.9999 & \textbf{4.83}/1.91/7.81 & \textbf{161}/63/259 & \textbf{3536}/1375/5696 \\[0.1cm]\hline\\[-0.1cm]
10 & conc. & \textbf{0.9982}/0.9740/0.9999 & \textbf{376}/12/918 & \textbf{435}/82/917 & \textbf{18694}/3510/39442 \\ 
   & seq. & \textbf{0.9959}/0.9661/0.9999 & \textbf{2591}/244/8458 & \textbf{13123}/1312/40136 & \textbf{341107}/34116/1043279 \\[0.3cm]
11 & conc. & \textbf{0.9999}/0.9991/0.9999 & \textbf{148}/11/1236 & \textbf{189}/71/919 & \textbf{8114}/3045/39519 \\ 
   & seq. & \textbf{0.9998}/0.9988/0.9999 & \textbf{786}/72/4817 & \textbf{4041}/427/17767 & \textbf{105031}/11097/461821 \\[0.3cm]
12 & conc. & \textbf{0.9996}/0.9974/0.9999 & \textbf{62.22}/4.48/286.60 & \textbf{89}/22/192 & \textbf{3818}/942/8276 \\ 
   & seq. & \textbf{0.9994}/0.9956/0.9999 & \textbf{284}/56/1842 & \textbf{987}/245/4563 & \textbf{25637}/6360/118491 \\[0.1cm]\hline\\[-0.1cm]
13 & conc. & \textbf{0.9989}/0.9936/0.9999 & \textbf{5.20}/1.52/14.89 & \textbf{138}/41/390 & \textbf{4833}/1434/13634 \\ 
   & seq. & \textbf{0.9759}/0.9373/0.9999 & \textbf{129.00}/6.39/439.92 & \textbf{4174}/215/15103 & \textbf{91773}/4719/332029 \\[0.3cm]
14 & conc. & \textbf{0.9999}/0.9999/0.9999 & \textbf{1.45}/0.70/2.62 & \textbf{35}/19/60 & \textbf{1235}/677/2089 \\ 
   & seq. & \textbf{0.9999}/0.9999/0.9999 & \textbf{6.47}/1.76/16.06 & \textbf{219}/59/547 & \textbf{4813}/1292/12033 \\[0.1cm]\hline\\[-0.1cm]
15 & conc. & \textbf{0.9999}/0.9999/1.0000 & \textbf{0.01}/0.00/0.01 & \textbf{0.90}/0.64/1.24 & \textbf{9.57}/6.67/13.30 \\ 
   & seq. & \textbf{0.9999}/0.9999/1.0000 & \textbf{0.02}/0.01/0.04 & \textbf{1.76}/0.72/3.28 & \textbf{17.53}/7.16/32.64 \\[0.3cm]
16 & conc. & \textbf{0.9999}/0.9999/1.0000 & \textbf{0.01}/0.00/0.01 & \textbf{0.80}/0.70/1.28 & \textbf{8.19}/7.13/13.48 \\ 
   & seq. & \textbf{0.9999}/0.9999/1.0000 & \textbf{0.01}/0.01/0.02 & \textbf{0.67}/0.51/1.54 & \textbf{6.64}/5.10/15.31 \\[0.1cm]\hline\\[-0.1cm]
17 & conc. & \textbf{0.9999}/0.9999/0.9999 & \textbf{160}/27/357 & \textbf{684}/616/773 & \textbf{7516}/6767/8495 \\ 
   & seq. & \textbf{0.9999}/0.9999/0.9999 & \textbf{2582}/1411/4638 & \textbf{16577}/11490/27082 & \textbf{165733}/114877/270766 \\[0.3cm]
18 & conc. & \textbf{0.9999}/0.9999/0.9999 & \textbf{88}/13/220 & \textbf{394}/286/620 & \textbf{4330}/3137/6811 \\ 
   & seq. & \textbf{0.9999}/0.9999/0.9999 & \textbf{492}/247/1520 & \textbf{2954}/2434/3985 & \textbf{29535}/24335/39842 \\[0.1cm]\hline\\[-0.1cm]
19 & conc. & \textbf{0.9999}/0.9999/0.9999 & \textbf{45.85}/8.49/213.95 & \textbf{170}/103/264 & \textbf{3896}/2354/6060 \\ 
   & seq. & \textbf{0.9999}/0.9999/0.9999 & \textbf{128}/76/217 & \textbf{1124}/809/1490 & \textbf{17978}/12945/23822 \\[0.1cm]\hline\\[-0.1cm]
20 & conc. & \textbf{0.9999}/0.9999/0.9999 & \textbf{0.06}/0.04/0.08 & \textbf{6.92}/4.80/9.47 & \textbf{76}/52/104 \\ 
   & seq. & \textbf{0.9999}/0.9999/0.9999 & \textbf{0.45}/0.21/1.02 & \textbf{26}/15/43 & \textbf{258}/148/431 \\[0.1cm]\hline\\[-0.1cm]
21 & conc. & \textbf{0.9999}/0.9999/0.9999 & \textbf{0.18}/0.16/0.20 & \textbf{8.56}/7.81/9.60 & \textbf{161}/146/180 \\ 
   & seq. & \textbf{0.9999}/0.9999/0.9999 & \textbf{1.26}/0.82/2.76 & \textbf{39}/29/57 & \textbf{551}/410/804 \\[0.1cm]\hline\\[-0.1cm]
22 & conc. & \textbf{0.9999}/0.9999/0.9999 & \textbf{0.96}/0.47/2.02 & \textbf{68}/42/105 & \textbf{750}/459/1154 \\ 
   & seq.$^{***}$ & \textbf{0.9998}/0.9994/0.9999 & \textbf{407}/112/732 & \textbf{21692}/6473/30000 & \textbf{216483}/64599/299399 \\[0.1cm]\hline\\[-0.1cm]
23 & conc. & \textbf{0.9999}/0.9999/0.9999 & \textbf{0.60}/0.24/1.64 & \textbf{53}/25/141 & \textbf{588}/279/1559 \\ 
   & seq. & \textbf{0.9951}/0.9797/0.9995 & \textbf{39.03}/9.39/111.74 & \textbf{2992}/744/7163 & \textbf{29796}/7408/71343 \\[1mm]
\hline\hline
\multicolumn{7}{l}{$^{***}$ Here the stopping conditions were changed for Fig.~\ref{fig:spaghetti-overview}(d), %
	so the data are no longer comparable to Tabs.~\ref{tab:results-overview-highamps} and \ref{tab:results-overview-smallamps-con}.}\\
\end{tabular}
\end{center}
\end{table*}
%%%%%%%%%

\begin{table*}[!Ht]
\begin{center}
\caption{Test results obtained from 20 {\em unconstrained optimisations}
(\texttt{fminunc} in \matlab)
for each problem of Tab.~\ref{tab:test-def} using sequential
or concurrent-update. {\em Higher initial pulse amplitudes}
($\mathrm{mean}(u_{ini}) = 0$, $\mathrm{std}(u_{ini}) = 10$) 
than in Tab.~\ref{tab:results-overview-smallamps} were used.
} \vspace{2mm}
\label{tab:results-overview-highamps}
\footnotesize
\begin{tabular}{c|ccccccc}
\hline\hline\\
Problem & Algorithm & Final Fidelity & Wall Time [min] & \#Eigendecs/1000 & \#Matrix Mults/1000  \\
& & (\textbf{mean}/min/max) & (\textbf{mean}/min/max) & (\textbf{mean}/min/max) & (\textbf{mean}/min/max) \\[0.3cm]
\hline\\[-1mm]
1 & conc. & \textbf{0.9999}/0.9999/0.9999 & \textbf{0.04}/0.02/0.06 & \textbf{4.25}/2.61/6.60 & \textbf{80}/49/125  \\ 
  & seq. & \textbf{0.9999}/0.9998/0.9999 & \textbf{1.54}/0.43/3.78 & \textbf{118}/33/289 & \textbf{1645}/459/4023 \\[0.3cm]
2 & conc. & \textbf{0.9999}/0.9999/0.9999 & \textbf{0.05}/0.02/0.08 & \textbf{5.39}/2.76/8.56 & \textbf{102}/52/162 \\ 
  & seq. & \textbf{0.9999}/0.9999/0.9999 & \textbf{1.38}/0.42/3.28 & \textbf{109}/33/261 & \textbf{1520}/464/3635 \\[0.3cm]
3 & conc. & \textbf{0.9999}/0.9999/1.0000 & \textbf{0.07}/0.05/0.10 & \textbf{6.06}/4.35/8.45 & \textbf{113}/80/158  \\ 
  & seq. & \textbf{0.9999}/0.9999/0.9999 & \textbf{0.22}/0.12/0.49 & \textbf{17.29}/9.73/38.53 & \textbf{242}/136/539 \\[0.3cm]
4 & conc. & \textbf{0.9999}/0.9999/1.0000 & \textbf{0.02}/0.01/0.03 & \textbf{2.50}/1.60/3.39 & \textbf{46}/29/63  \\
  & seq. & \textbf{0.9999}/0.9999/0.9999 & \textbf{0.18}/0.05/0.54 & \textbf{13.79}/4.22/42.18 & \textbf{193}/59/589 \\[0.1cm]\hline\\[-0.1cm]
5 & conc. & \textbf{0.9976}/0.9959/0.9986 & \textbf{6.97}/6.24/7.32 & \textbf{364}/362/370 & \textbf{9839}/9784/9995  \\ 
  & seq. & \textbf{0.9969}/0.9952/0.9983 & \textbf{73}/51/105 & \textbf{4246}/2954/6021 & \textbf{76349}/53121/108271 \\[0.3cm]
6 & conc. & \textbf{0.9999}/0.9999/0.9999 & \textbf{2.23}/1.36/3.82 & \textbf{105}/60/180 & \textbf{2842}/1623/4860  \\ 
  & seq. & \textbf{0.9999}/0.9999/0.9999 & \textbf{16}/11/32 & \textbf{935}/614/1866 & \textbf{16823}/11049/33567 \\[0.1cm]\hline\\[-0.1cm]
7 & conc. & \textbf{0.9893}/0.9366/0.9999 & \textbf{15}/14/16 & \textbf{386}/385/387 & \textbf{13499}/13469/13563  \\ 
  & seq. & \textbf{0.9928}/0.9444/0.9993 & \textbf{325}/129/730 & \textbf{8053}/4190/16630 & \textbf{177047}/92125/365595 \\[0.3cm]
8 & conc. & \textbf{0.9998}/0.9984/0.9999 & \textbf{9.86}/4.27/15.02 & \textbf{257}/110/386 & \textbf{9000}/3832/13495  \\ 
  & seq. & \textbf{0.9990}/0.9851/0.9999 & \textbf{158}/43/645 & \textbf{3511}/1400/13269 & \textbf{77189}/30788/291716 \\[0.3cm]
9 & conc. & \textbf{0.9999}/0.9999/0.9999 & \textbf{2.81}/1.80/4.62 & \textbf{87}/57/142 & \textbf{3056}/2007/4982  \\ 
  & seq. & \textbf{0.9996}/0.9995/0.9998 & \textbf{41}/22/104 & \textbf{1057}/693/2033 & \textbf{23223}/15223/44653 \\[0.1cm]\hline\\[-0.1cm]
10 & conc. & \textbf{0.9978}/0.9834/0.9999 & \textbf{638}/91/1566 & \textbf{798}/402/944 & \textbf{34315}/17276/40590  \\ 
   & seq. & \textbf{0.9995}/0.9974/0.9999 & \textbf{3213}/529/8282 & \textbf{16045}/6914/33844 & \textbf{417076}/179721/879708 \\[0.3cm]
11 & conc. & \textbf{0.9999}/0.9998/0.9999 & \textbf{449}/37/1165 & \textbf{613}/335/906 & \textbf{26342}/14386/38939  \\ 
   & seq. & \textbf{0.9999}/0.9998/0.9999 & \textbf{1557}/863/2935 & \textbf{8408}/4895/14865 & \textbf{218564}/127232/386383 \\[0.3cm]
12 & conc. & \textbf{0.9984}/0.9948/0.9999 & \textbf{197}/16/416 & \textbf{196}/192/202 & \textbf{8426}/8273/8678  \\ 
   & seq. & \textbf{0.9974}/0.9911/0.9990 & \textbf{883}/255/2060 & \textbf{4320}/1994/9522 & \textbf{112192}/51780/247266 \\[0.1cm]\hline\\[-0.1cm]
13 & conc. & \textbf{0.9999}/0.9999/0.9999 & \textbf{2.65}/1.74/4.43 & \textbf{64}/47/107 & \textbf{2247}/1636/3729  \\ 
   & seq. & \textbf{0.9999}/0.9999/0.9999 & \textbf{16.31}/9.63/31.65 & \textbf{520}/310/1040 & \textbf{11427}/6804/22872 \\[0.3cm]
14 & conc. & \textbf{0.9999}/0.9999/0.9999 & \textbf{1.48}/1.09/1.94 & \textbf{40}/33/48 & \textbf{1405}/1152/1676  \\ 
   & seq. & \textbf{0.9999}/0.9999/0.9999 & \textbf{5.25}/3.90/6.50 & \textbf{166}/126/207 & \textbf{3654}/2772/4550 \\[0.1cm]\hline\\[-0.1cm]
15 & conc. & \textbf{0.9999}/0.9999/1.0000 & \textbf{0.00}/0.00/0.01 & \textbf{0.55}/0.40/0.76 & \textbf{5.70}/4.02/8.00  \\ 
   & seq. & \textbf{0.9999}/0.9999/1.0000 & \textbf{0.01}/0.01/0.02 & \textbf{0.75}/0.52/1.60 & \textbf{7.44}/5.17/15.92 \\[0.3cm]
16 & conc. & \textbf{0.9999}/0.9999/1.0000 & \textbf{0.00}/0.00/0.01 & \textbf{0.76}/0.58/1.22 & \textbf{7.73}/5.71/12.77  \\ 
   & seq. & \textbf{0.9999}/0.9999/1.0000 & \textbf{0.01}/0.01/0.02 & \textbf{0.58}/0.45/1.15 & \textbf{5.77}/4.47/11.48 \\[0.1cm]\hline\\[-0.1cm]
17 & conc. & \textbf{0.9999}/0.9999/0.9999 & \textbf{162}/28/320 & \textbf{616}/536/750 & \textbf{6768}/5887/8242  \\ 
   & seq. & \textbf{0.9999}/0.9999/0.9999 & \textbf{1346}/763/2603 & \textbf{9238}/7502/13230 & \textbf{92361}/75005/132274 \\[0.3cm]
18 & conc. & \textbf{0.9999}/0.9999/0.9999 & \textbf{118}/24/309 & \textbf{522}/400/652 & \textbf{5736}/4391/7163  \\ 
   & seq. & \textbf{0.9999}/0.9999/0.9999 & \textbf{897}/428/1152 & \textbf{6788}/5799/8207 & \textbf{67863}/57978/82054 \\[0.1cm]\hline\\[-0.1cm]
19 & conc. & \textbf{0.9999}/0.9999/0.9999 & \textbf{41.77}/5.48/120.52 & \textbf{65}/58/71 & \textbf{1481}/1332/1636  \\ 
   & seq. & \textbf{0.9999}/0.9999/0.9999 & \textbf{59}/32/89 & \textbf{460}/398/547 & \textbf{7354}/6362/8747 \\[0.1cm]\hline\\[-0.1cm]
20 & conc. & \textbf{0.9998}/0.9991/0.9999 & \textbf{1.91}/0.51/6.98 & \textbf{139}/47/202 & \textbf{1529}/517/2225  \\ 
   & seq. & \textbf{0.9980}/0.9879/0.9996 & \textbf{64}/29/103 & \textbf{3508}/2098/6647 & \textbf{34972}/20910/66265 \\[0.1cm]\hline\\[-0.1cm]
21 & conc. & \textbf{0.9999}/0.9999/0.9999 & \textbf{1.50}/1.18/2.07 & \textbf{70}/53/96 & \textbf{1328}/1005/1818  \\ 
   & seq. & \textbf{0.9998}/0.9998/0.9999 & \textbf{118}/84/164 & \textbf{4269}/2860/5791 & \textbf{59697}/39992/80980 \\[0.1cm]\hline\\[-0.1cm]
22 & conc. & \textbf{0.9999}/0.9999/0.9999 & \textbf{0.58}/0.32/0.90 & \textbf{51}/29/74 & \textbf{563}/317/820  \\ 
   & seq. & \textbf{0.9995}/0.9982/0.9998 & \textbf{81}/35/137 & \textbf{4128}/1981/6114 & \textbf{41194}/19771/61017 \\[0.1cm]\hline\\[-0.1cm]
23 & conc. & \textbf{0.9999}/0.9999/0.9999 & \textbf{0.06}/0.05/0.07 & \textbf{7.58}/6.20/9.85 & \textbf{83}/68/108  \\ 
   & seq. & \textbf{0.9999}/0.9999/0.9999 & \textbf{1.20}/0.59/2.24 & \textbf{93}/46/171 & \textbf{925}/458/1702 \\[1mm]
\hline\hline
\end{tabular}
\end{center}
\end{table*}

%%%%%%%%%%%%%%%%%%%%%%%

\begin{table*}[!Ht]
\begin{center}
\caption{Test results obtained from 20 {\em constrained optimisations}
(\texttt{fmincon} in \matlab)
for each problem of Tab.~\ref{tab:test-def} using the sequential
or the concurrent-update algorithm. {\em Small initial pulse amplitudes} were used
($\mathrm{mean}(u_{ini}) = 0$, $\mathrm{std}(u_{ini}) = 1$).
}\vspace{2mm}
\label{tab:results-overview-smallamps-con}
\footnotesize
\begin{tabular}{c|cccccccc}
\hline\hline\\
Problem & Algorithm & Final Fidelity & Wall Time [min] & \#Eigendecs/1000 & \#Matrix Mults/1000  \\
& & (\textbf{mean}/min/max) & (\textbf{mean}/min/max) & (\textbf{mean}/min/max) & (\textbf{mean}/min/max) \\[0.3cm]
\hline\\[-1mm]
1 & conc. & \textbf{0.9999}/0.9999/1.0000 & \textbf{0.11}/0.02/0.26 & \textbf{1.43}/1.23/1.68 & \textbf{27}/23/31  \\ 
  & seq. & \textbf{0.9999}/0.9999/0.9999 & \textbf{0.10}/0.04/0.22 & \textbf{6.17}/2.70/10.92 & \textbf{86}/38/152 \\[0.3cm]
2 & conc. & \textbf{0.9999}/0.9999/0.9999 & \textbf{0.05}/0.03/0.10 & \textbf{2.10}/1.64/2.52 & \textbf{39}/31/47 \\ 
  & seq. & \textbf{0.9999}/0.9999/0.9999 & \textbf{0.08}/0.05/0.09 & \textbf{5.74}/4.04/6.96 & \textbf{80}/56/97 \\[0.3cm]
3 & conc. & \textbf{0.9999}/0.9999/1.0000 & \textbf{0.09}/0.08/0.13 & \textbf{4.49}/3.71/5.50 & \textbf{83}/68/102 \\ 
  & seq. & \textbf{0.9999}/0.9999/0.9999 & \textbf{0.08}/0.03/0.15 & \textbf{5.13}/2.05/8.06 & \textbf{72}/29/113 \\[0.3cm]
4 & conc. & \textbf{0.9999}/0.9999/1.0000 & \textbf{0.03}/0.02/0.04 & \textbf{1.60}/1.34/1.98 & \textbf{29}/24/37 \\ 
  & seq. & \textbf{0.9999}/0.9999/0.9999 & \textbf{0.03}/0.01/0.04 & \textbf{1.89}/1.02/2.75 & \textbf{26}/14/38 \\[0.1cm]\hline\\[-0.1cm]
5 & conc. & \textbf{0.9877}/0.9322/0.9990 & \textbf{44}/24/67 & \textbf{364}/361/368 & \textbf{9828}/9759/9947  \\ 
  & seq. & \textbf{0.9973}/0.9918/0.9986 & \textbf{34}/22/61 & \textbf{1976}/1320/3292 & \textbf{35542}/23729/59209 \\[0.3cm]
6 & conc. & \textbf{0.9999}/0.9999/0.9999 & \textbf{1.87}/0.73/3.61 & \textbf{24}/18/40 & \textbf{650}/473/1074  \\ 
  & seq. & \textbf{0.9999}/0.9999/0.9999 & \textbf{5.34}/1.15/19.60 & \textbf{310}/68/1143 & \textbf{5574}/1216/20554 \\[0.1cm]\hline\\[-0.1cm]
7 & conc. & \textbf{0.9958}/0.9808/0.9999 & \textbf{29.82}/5.74/69.99 & \textbf{244}/69/390 & \textbf{8552}/2411/13634 \\ 
  & seq. & \textbf{0.9945}/0.9825/0.9999 & \textbf{123}/39/244 & \textbf{3978}/1242/7749 & \textbf{87443}/27313/170360 \\[0.3cm]
8 & conc. & \textbf{0.9999}/0.9999/0.9999 & \textbf{5.39}/2.08/20.83 & \textbf{49}/29/198 & \textbf{1697}/995/6925  \\ 
  & seq. & \textbf{0.9999}/0.9999/0.9999 & \textbf{15.19}/2.66/67.58 & \textbf{500}/89/2223 & \textbf{11002}/1953/48876 \\[0.3cm]
9 & conc. & \textbf{0.9999}/0.9999/0.9999 & \textbf{1.21}/0.69/1.68 & \textbf{14.94}/8.19/21.06 & \textbf{521}/285/735  \\ 
  & seq. & \textbf{0.9999}/0.9999/0.9999 & \textbf{4.90}/1.91/7.99 & \textbf{161}/63/259 & \textbf{3536}/1375/5696 \\[0.1cm]\hline\\[-0.1cm]
10 & conc. & \textbf{0.9998}/0.9985/0.9999 & \textbf{281}/21/1753 & \textbf{355}/94/904 & \textbf{15251}/4052/38874 \\ 
   & seq. & \textbf{0.9991}/0.9864/0.9999 & \textbf{1942}/122/12208 & \textbf{11549}/988/84232 & \textbf{300198}/25679/2189468 \\[0.3cm]
11 & conc. & \textbf{0.9999}/0.9999/0.9999 & \textbf{153.76}/7.80/1104.29 & \textbf{144}/68/566 & \textbf{6182}/2890/24346 \\ 
   & seq. & \textbf{0.9998}/0.9988/0.9999 & \textbf{1141}/86/9848 & \textbf{6990}/427/74922 & \textbf{181706}/11097/1947465 \\[0.3cm]
12 & conc. & \textbf{0.9999}/0.9992/0.9999 & \textbf{76.27}/5.76/948.99 & \textbf{70}/22/194 & \textbf{3017}/936/8331  \\ 
   & seq. & \textbf{0.9997}/0.9970/0.9999 & \textbf{1108}/39/5566 & \textbf{5054}/245/19200 & \textbf{131246}/6360/498598 \\[0.1cm]\hline\\[-0.1cm]
13 & conc. & \textbf{0.9844}/0.9102/0.9999 & \textbf{13.91}/2.49/75.62 & \textbf{120}/27/398 & \textbf{4189}/950/13926 \\ 
   & seq. & \textbf{0.9759}/0.9373/0.9999 & \textbf{128.25}/6.64/454.86 & \textbf{4174}/215/15103 & \textbf{91773}/4719/332029 \\[0.3cm]
14 & conc. & \textbf{0.9973}/0.9867/0.9999 & \textbf{7.47}/2.06/14.20 & \textbf{49}/29/80 & \textbf{1720}/1000/2801 \\ 
   & seq. & \textbf{0.9999}/0.9999/0.9999 & \textbf{6.58}/1.77/16.61 & \textbf{219}/59/547 & \textbf{4813}/1292/12033 \\[0.1cm]\hline\\[-0.1cm]
15 & conc. & \textbf{0.9999}/0.9999/1.0000 & \textbf{0.05}/0.02/0.10 & \textbf{0.71}/0.52/0.92 & \textbf{7.49}/5.35/9.77 \\ 
   & seq. & \textbf{0.9999}/0.9999/1.0000 & \textbf{0.02}/0.01/0.04 & \textbf{1.76}/0.72/3.28 & \textbf{17.53}/7.16/32.64 \\[0.3cm]
16 & conc. & \textbf{0.9999}/0.9999/1.0000 & \textbf{0.04}/0.01/0.07 & \textbf{0.96}/0.77/1.34 & \textbf{9.99}/7.83/14.19 \\ 
   & seq. & \textbf{0.9999}/0.9999/1.0000 & \textbf{0.01}/0.01/0.02 & \textbf{0.67}/0.51/1.54 & \textbf{6.64}/5.10/15.31 \\[0.1cm]\hline\\[-0.1cm]
17 & conc. & \textbf{0.9999}/0.9999/0.9999 & \textbf{531}/58/1443 & \textbf{1224}/1032/1551 & \textbf{13454}/11344/17054 \\ 
   & seq. & \textbf{0.9999}/0.9999/0.9999 & \textbf{2284}/1054/3898 & \textbf{16774}/11490/27294 & \textbf{167710}/114877/272885 \\[0.3cm]
18 & conc. & \textbf{0.9999}/0.9999/0.9999 & \textbf{157}/26/754 & \textbf{574}/530/655 & \textbf{6300}/5821/7196  \\ 
   & seq. & \textbf{0.9999}/0.9999/0.9999 & \textbf{386}/175/690 & \textbf{2953}/2434/3985 & \textbf{29524}/24335/39842 \\[0.1cm]\hline\\[-0.1cm]
19 & conc. & \textbf{0.9999}/0.9999/0.9999 & \textbf{105}/16/335 & \textbf{166}/141/186 & \textbf{3807}/3244/4273  \\ 
   & seq. & \textbf{0.9999}/0.9999/0.9999 & \textbf{143}/64/328 & \textbf{1130}/996/1465 & \textbf{18064}/15925/23433 \\[0.1cm]\hline\\[-0.1cm]
20 & conc. & \textbf{0.9999}/0.9999/0.9999 & \textbf{0.53}/0.12/1.14 & \textbf{5.15}/4.16/6.91 & \textbf{56}/45/76  \\ 
   & seq. & \textbf{0.9999}/0.9999/0.9999 & \textbf{0.45}/0.23/0.77 & \textbf{30}/16/51 & \textbf{302}/160/511 \\[0.1cm]\hline\\[-0.1cm]
21 & conc. & \textbf{0.9999}/0.9999/0.9999 & \textbf{0.49}/0.36/0.89 & \textbf{9.53}/9.09/10.37 & \textbf{179}/171/195  \\ 
   & seq. & \textbf{0.9999}/0.9999/0.9999 & \textbf{1.39}/0.79/3.80 & \textbf{39}/29/57 & \textbf{551}/410/804 \\[0.1cm]\hline\\[-0.1cm]
22 & conc. & \textbf{0.9999}/0.9999/0.9999 & \textbf{5.34}/2.39/8.03 & \textbf{131}/93/193 & \textbf{1444}/1026/2128  \\ 
   & seq. & \textbf{0.9991}/0.9983/0.9995 & \textbf{108}/59/386 & \textbf{4702}/3317/6780 & \textbf{46924}/33106/67669 \\[0.1cm]\hline\\[-0.1cm]
23 & conc. & \textbf{0.9999}/0.9999/0.9999 & \textbf{2.26}/0.42/9.57 & \textbf{38}/13/83 & \textbf{420}/148/913  \\ 
   & seq. & \textbf{0.9951}/0.9797/0.9995 & \textbf{37.38}/9.40/97.52 & \textbf{2991}/744/7163 & \textbf{29786}/7408/71343 \\[1mm]

\hline\hline
\end{tabular}
\end{center}
\end{table*}
%%%%%%%%%
\clearpage

\section{Appendix}

\subsection{Exact Gradients (Eqn.~\eqref{eqn:exact-derivative})}
\label{sec:exactGradientDerivation}
\noindent
For deriving the gradient expression in Eqn.~\eqref{eqn:exact-derivative},
we follow \cite{Aizu63,TILO96}.
Note that by
\begin{equation}
\begin{split}
\Partial{X}{u_j} &= \Partial{}{u}\exp\{-i\Delta t(H_d + (u_j+ u) H_j + \sum_{\nu\neq j} u_\nu H_\nu)\}\Big|_{u=0} \\
                 &= \Partial{}{u}\exp\{-i\Delta t(H_u + u H_j)\}\Big|_{u=0}
\end{split}
\end{equation}
one may invoke the spectral theorem in a standard way and calculate matrix functions
via the eigendecoposition.  For an arbitrary pair of
{\em Hermitian} (non-commuting) matrices $A,B$ and $x\in\R{}$,
take $\{\ket{\lambda_\nu}\}$ as the orthonormal eigenvectors to
the eigen\-values $\{\lambda_\nu\}$ of $A$ to
obtain the following straightforward yet lengthy series of identities
\begin{equation}
\begin{split}
D &= \braket{\lambda_l}{\Partial{}{x}\;e^{A+xB} |\lambda_m}\Big|_{x=0}\\
  &= \braket{\lambda_l}{\Partial{}{x}\;\sum_{n=0}^\infty \frac{1}{n!} \big(A+xB\big)^n|\lambda_m}\Big|_{x=0}\\
  &= \braket{\lambda_l}{\sum_{n=0}^\infty \frac{1}{n!}\sum_{q=1}^n \big(A+xB\big)^{q-1} B \big(A+xB\big)^{n-q}|\lambda_m}\Big|_{x=0}\\
  &= \braket{\lambda_l}{\sum_{n=0}^\infty \frac{1}{n!}\sum_{q=1}^n A^{q-1} B A^{n-q}|\lambda_m}\\
  &= {\sum_{n=0}^\infty \frac{1}{n!}\sum_{q=1}^n \lambda_l^{q-1} \braket{\lambda_l}{B|\lambda_m} \lambda_m^{n-q}}\\
  &= \braket{\lambda_l}{B|\lambda_m} \sum_{n=0}^\infty \frac{1}{n!}\sum_{q=1}^n \lambda_l^{q-1} \lambda_m^{n-q}
\end{split}
\end{equation}
already explaining the case $\lambda_l=\lambda_m$, while for $\lambda_l\neq\lambda_m$ we have
\begin{equation}
\begin{split}
D  &= \braket{\lambda_l}{B|\lambda_m} \sum_{n=0}^\infty \frac{1}{n!} \lambda_m^{n-1} \sum_{q=1}^n \left(\frac{\lambda_l}{\lambda_m}\right)^{q-1} \\
  &= \braket{\lambda_l}{B|\lambda_m} \sum_{n=0}^\infty \frac{1}{n!} \lambda_m^{n-1} \frac{(\lambda_l/\lambda_m)^n-1}{(\lambda_l/\lambda_m)-1} \\
  &= \braket{\lambda_l}{B|\lambda_m} \sum_{n=0}^\infty \frac{1}{n!} \frac{\lambda_l^n-\lambda_m^n}{\lambda_l-\lambda_m} \\
  &= \braket{\lambda_l}{B|\lambda_m} \frac{e^{\lambda_l}-e^{\lambda_m}}{\lambda_l-\lambda_m} \quad.
\end{split}
\end{equation}
An analogous result holds for skew-Hermitian $iA, iB$. So
substituting $A\mapsto -i\Delta t H_u$ and $xB\mapsto -i\Delta t\, u H_j$ 
as well as $\lambda_\nu \mapsto -i\Delta t\lambda_\nu$
for $\nu=l,m$ while keeping the eigenvectors $\ket{\lambda_\nu}$
readily recovers Eqn.~\eqref{eqn:exact-derivative}.
Note that we have explicitly made use of the orthogonality of eigenvectors
to different eigenvalues in Hermitian (or more generally {\em normal}\/) matrices $A,B$. Hence
in generic open quantum systems with a non-normal Lindbladian, there is no such simple
extension for calculating exact gradients.
%%%%%%%%%%%%%%%%%%%%%%%%%%%%%%%%%%%%%%%%%%%%
\subsection{Standard Settings in a Nutshell}\label{sec:nutshell}
%%%%%%%%%%%%%%%%%%%%%%%%%%%%%%%%%%%%%%%%%%%%
%%%%%%%%%%
For convenience, here we give the details for six standard tasks
of optimising state transfer or gate synthesis. The individual steps
give the key elements of the core algorithm in Sec.~\ref{sec:Alg-core} and
its representation as a flow chart.
%%%%%%%%%%

\medskip
\noindent
 {\bf Task 1:} {\em Approximate Unitary Target Gate up to Global Phase in Closed Systems}\\[2mm]
  Define boundary conditions $X_0:= \unity$, $X_{M+1}:= U_{\rm target}$;
	fix final time $T$ and digitisation $M$ so that $T=M \Delta t$.
 \begin{itemize}
  \item[(a)] set initial control amplitudes $u_j^{(0)}(t_k)\in\mathcal U \subseteq \mathbb R$
 		for all times $t_k$ with $k\in\mathcal T^{(0)}:=\{1,2,\dots,M\}$;
  \item[(b)] exponentiate
       $U_k = e^{-i \Delta t H(t_k)}$ for all $k\in\mathcal T^{(r)}$
         with $H_k := H_d + \sum_j u_j(t_k) H_j$ ;
  \item[(c)] calculate forward-propagation \\
         \mbox{$ U_{k:0} := U_k  U_{k-1}\cdots U_1 U_0$}
  \item[(d)] calculate  back-propagation \\
         \mbox{$\Lambda^\dagger_{M+1:k+1}
		:= {U}^\dagger_{\rm tar}  U_M  U_{M-1}\cdots U_{k+1}$}
  \item[(e)] evaluate fidelity $f=|g|$, where\\[1mm]
    \mbox{$g := \tfrac{1}{N} \tr\big\{ \Lambda^{\dagger}_{M+1:k+1} U^{\phantom{ \dagger}}_{k:0} \big\}%
	=\tfrac{1}{N} \tr\big\{{U}^\dagger_{\rm tar} U^{\phantom{ \dagger}}_{M:0}\big\}$}\\[1mm]
	and stop if ${f}\geq 1-\varepsilon_{\rm threshold}$ or iteration $r>r_{\rm limit}$.
  \item[(f)] evaluate gradients for all $k\in\mathcal T^{(r)}$\\[1mm]
    \mbox{$\tfrac{\partial f(U(t_k))}{u_j} = \tfrac{1}{N} \Re \tr \big\{e^{-i\phi_g} \Lambda^{\dagger}_{M+1:k+1}
		\big(\Partial{U_k}{u_j}\big) U_{k-1:0}\big\}$}
		with $\Partial{U_k}{u_j}$ of Eqn.~\eqref{eqn:exact-derivative} and $e^{-i\phi_g}:=g^*/|g|$;
  \item[(g)] update amplitudes for all $k\in\mathcal T^{(r)}$ by quasi-Newton\\
     $u_j^{(r+1)}(t_k) = u_j^{(r)}(t_k) + \alpha_k\, \Hess_k^{-1}\, \frac{\partial f(X(t_k))}{\partial u_j}$ 
		or other methods (as in the text);
  \item[(h)] while $\tfrac{\partial f_k}{\partial u_j} > f'_{\rm limit}$ for some $k\in\mathcal T^{(r)}$ re-iterate;\\
	else re-iterate with new set $\mathcal T^{(r+1)}$.
 \end{itemize}
{\em Comments:} Algorithmic scheme for synthesising a unitary gate $U(T)$ 
such as to optimise the gate fidelity
$f:= | \tfrac{1}{N}  \tr \{U^\dagger_{\rm target} U(T)\}|$.
This setting automatically absorbs global phase factors as immaterial:
tracking for minimal times $T_*$ to realise $U_{\rm target}$ up to an
undetermined phase automatically gives a $e^{i\phi_*}U_{\rm target}\in {\rm SU(N)}$
of fastest realisation.
%%%%%%%%%%%%%%%%%%%

\bigskip 
\newpage
%%%%%%%%%%%%%%%%%%%%
\noindent
 {\bf Task 2:} {\em Approximate Unitary Target Gate Sensitive to Global Phase in Closed Systems }\\[2mm]
   Fix boundary conditions $X_0:= \unity$, $X_{M+1}:= e^{i\phi}U_{\rm target}$ by
   choosing global phase to ensure $\det (e^{i\phi} {U}_{\rm target}) = +1$
 	so that $e^{i\phi} {U}_{\rm target}\in {\rm SU(N)}$; there are $N$ such choices \cite{PRA05};
 	fix final time $T$ and digitisation $M$ so $T=M \Delta t$.
 \begin{itemize}
   \item[(a)] through (d) as in Task 1.
   \item[(e)] evaluate fidelity \\[1mm]
    \mbox{\hspace{-5mm}$f= \tfrac{1}{N} \Re \tr\big\{ \Lambda^{\dagger}_{M+1:k+1} U^{\phantom{\dagger}}_{k:0} \big\}%
 	=\tfrac{1}{N} \Re \tr\big\{e^{-i\phi}U^{\dagger}_{\rm tar} U^{\phantom{\dagger}}_{M:0} \big\}$}\\[1mm]
 	and stop if ${f}\geq 1-\varepsilon_{\rm threshold}$ or iteration $r>r_{\rm limit}$.
  \item[(f)] evaluate gradients for all $k\in\mathcal T^{(r)}$\\[1mm]
    \mbox{$\tfrac{\partial f(U(t_k))}{u_j} = \tfrac{1}{N} \Re \tr \big\{\Lambda^{\dagger}_{M+1:k+1}
		\big(\Partial{U_k}{u_j}\big) U_{k-1:0}\big\}$}
		with $\Partial{U_k}{u_j}$ of Eqn.~\eqref{eqn:exact-derivative};

   \item[(g)] and (h) as in Task 1.
 \end{itemize}
{\em Comments:} Algorithmic scheme for
 synthesising a unitary gate $U(T)$ in closed quantum systems
 such as to optimise the gate fidelity
 $f:=\tfrac{1}{N} \Re \tr \{e^{-i\phi}U^\dagger_{\rm target} U(T)\}$.
 This setting is sensitive to global phases $\phi$ that have to be specified in advance.
 {\em Warning:} whenever drift and control Hamiltonians operate on different time scales, the minimal
 time $T_*$ required to realise $e^{i\phi}U_{\rm target}\in {\rm SU(N)}$ will typically and significantly
 depend on $\phi$ as demonstrated in \cite{PRA05}, a problem eliminated by Task~1.
%%%%%%%%%%%%%%%%%%%

\bigskip
%%%%%%%%%%%%%%%%%%%
\noindent
 {\bf Task 3:} {\em Optimise State Transfer between Pure-State Vectors}\\[2mm]
  Define boundary conditions $X_0:= \ket{\psi_0}$, $X_{M+1}:= \ket{\psi}_{\rm target}$;
	fix final time $T$ and digitisation $M$ so that $T=M \Delta t$.
 \begin{itemize}
  \item[(a)] set initial control amplitudes $u_j^{(0)}(t_k)\in\mathcal U \subseteq \mathbb R$
 %		for all times $t_k$ with $k\in\mathcal T^{(0)}:=\{1,2,\dots,M\}$;
  \item[(b)] exponentiate
       $U_k = e^{-i \Delta t H(t_k)}$ for all $k\in\mathcal T^{(r)}$
         with $H_k := H_d + \sum_j u_j(t_k) H_j$ ;
  \item[(c)] calculate forward-propagation \\
         \mbox{$ \ket{\psi_0(t_k)} := U_k  U_{k-1}\cdots U_1 \ket{\psi_0}$}
  \item[(d)] calculate  back-propagation \\
         \mbox{$\bra{\psi_{\rm tar}(t_k)}
		:= \bra{\psi_{\rm tar}}  U_M  U_{M-1}\cdots U_{k+1}$}
  \item[(e)] evaluate fidelity $f$, where\\[1mm]
    {$f :=  \Re\;\braket{\psi_{\rm tar}(t_k)} {\psi_0(t_k)}\\[1mm]
	\phantom{X} =\Re\; \braket{\psi_{\rm tar}} {( U_M\cdots U_k\cdots U_1\,|\psi_0}) $ }\\[1mm]
	and stop if $f\geq 1-\varepsilon_{\rm threshold}$ or iteration $r>r_{\rm limit}$.
  \item[(f)] evaluate gradients for all $k\in\mathcal T^{(r)}$\\[1mm]
    $\tfrac{\partial f(U(t_k))}{u_j} =  \\[-1mm]
		\phantom{X}= \Re \; \braket{\psi_{\rm tar}}
		{(U_M\cdots U_{k+1} \big(\Partial{U_k}{u_j}\big) U_{k-1}\cdots U_1\, |\psi_0})$\\
		again with $\Partial{U_k}{u_j}$ of Eqn.~\eqref{eqn:exact-derivative};
  \item[(g)] update amplitudes for all $k\in\mathcal T^{(r)}$ by quasi-Newton\\
     $u_j^{(r+1)}(t_k) = u_j^{(r)}(t_k) + \alpha_k \,\Hess_k^{-1}\, \frac{\partial f(X(t_k))}{\partial u_j}$ 
		or other methods (as in the text);
  \item[(h)] while $\tfrac{\partial f_k}{\partial u_j} > f'_{\rm limit}$ for some $k\in\mathcal T^{(r)}$ re-iterate;\\
	else re-iterate with new set $\mathcal T^{(r+1)}$.
 \end{itemize}
{\em Comments:} Algorithmic scheme for
optimising (pure) state-to-state transfer in closed quantum systems.
This setting is sensitive to global phases $\phi$ in $e^{i\phi}\ket\psi$
that have to be specified in advance.
%%%%%%%%%%%%%%%%%%%

\bigskip
%%%%%%%%%%%%%%%%%%%
\noindent
 {\bf Task 4:} {\em Optimise State Transfer between Density Operators in Closed Systems}\\[2mm]
  Define boundary conditions $X_0:= \rho_0$, $X_{M+1}:= \rho_{\rm target}$;
	fix final time $T$ and digitisation $M$ so that $T=M \Delta t$.
 \begin{itemize}
  \item[(a)] and (b) as in Tasks 1 through 3.
  \item[(c)] calculate forward-propagation \\
         \mbox{$ \rho_0(t_k) := U_k  U_{k-1}\cdots U_1 \rho_0 U_1^\dagger \cdots U_{k-1}^\dagger U_k^\dagger$}
  \item[(d)] calculate  back-propagation \\
         \mbox{$\rho^\dagger_{\rm tar}(t_k)
		:= U_{k+1}^\dagger\cdots U_{M-1}^\dagger U_M^\dagger {\rho}^\dagger_{\rm tar} U_M  U_{M-1}\cdots U_{k+1}$}
  \item[(e)] evaluate fidelity $f$ with normalisation $c:=||\rho_{\rm tar}||_2^2$\\[1mm]
    $f := \tfrac{1}{c} \Re \tr\big\{\rho^\dagger_{\rm tar}(t_k) \rho^{\phantom{ \dagger}}_0(t_k)\big\}$\\[1mm]
	and stop if $f\geq 1-\varepsilon_{\rm threshold}$ or iteration $r>r_{\rm limit}$.
  \item[(f)] evaluate gradients for all $k\in\mathcal T^{(r)}$\\[1mm]
    $\tfrac{\partial f(U(t_k))}{u_j} = \tfrac{1}{c} \Re \big(\tr \big\{\rho^{\dagger}_{\rm tar}(t_k)
		\big(\Partial{U_k}{u_j}\big) \rho_0(t_{k-1})U^\dagger_k\big\}
		+ \tr \big\{\rho^{\dagger}_{\rm tar}(t_k)
                {U_k} \rho_0(t_{k-1}) \big(\Partial{U^\dagger_k}{u_j}\big) \big\}\big)$
		with $\Partial{U_k}{u_j}$ of Eqn.~\eqref{eqn:exact-derivative};
  \item[(g)] and (h) as in Tasks 1 through 3;
 \end{itemize}

{\em Comments:} Algorithmic scheme for
optimising state-to-state transfer of density operators in closed quantum systems.
%%%%%%%%%%%%%%%%%%%

\bigskip
%%%%%%%%%%%%%%%%%%%%
\noindent
{\bf Task 5:} {\em Approximate Unitary Target Gate by Quantum Map in Open Markovian Systems}\\[2mm]
   Define boundary conditions $X_0:= \unity$, $X_{M+1}:= \widehat{U}_{\rm target}$;
 	fix final time $T$ and digitisation $M$ so that $T=M \Delta t$.
 \begin{itemize}
  \item[(a)] set initial control amplitudes $u_j^{(0)}(t_k)\in\mathcal U \subseteq \mathbb R$
 		for all times $t_k$ with $k\in\mathcal T^{(0)}:=\{1,2,\dots,M\}$;
  \item[(b)] exponentiate
        $X_k = e^{-i \Delta t \widehat{H}(t_k) + \Gamma}$ for all $k\in\mathcal T^{(r)}$
        with $\widehat{H}_k := \widehat{H}_0 + \sum_j u_j(t_k) \widehat{H}_j$ ;
  \item[(c)] calculate forward-propagation \\
         \mbox{$ X_{k:0} := X_k X_{k-1}\cdots X_1 (X_0=\unity)$};
  \item[(d)] calculate back-propagation \\
         \mbox{$\Lambda^\dagger_{M+1:k+1}
		:= \widehat{U}^\dagger_{\rm tar} X_M  X_{M-1}\cdots X_{k+1}$}
  \item[(e)] evaluate fidelity \\[1mm] %(single convenient $k$ suffices)\\[1mm]
    \mbox{$f\negthickspace=\negthickspace\tfrac{1}{N^2} \Re \tr\big\{ \Lambda^{\dagger}_{M+1:k+1} X^{\phantom{ \dagger}}_{k:0} \big\}
	     \negthickspace=\negthickspace\tfrac{1}{N^2} \Re \tr\big\{\widehat{U}^\dagger_{\rm tar}X^{\phantom{ \dagger}}_{M:0}\big\}$}\\[1mm]
	and stop if $f\geq 1-\varepsilon_{\rm threshold}$ or iteration $r>r_{\rm limit}$.
    \mbox{$\frac{\partial f(X(t_k))}{\partial u_j} \approx \tfrac{-\Delta t}{N^2} \Re \tr\big\{ \Lambda^{\dagger}_{M+1:k+1}
	(i{\widehat{H}}_{u_j}+\tfrac{\partial \Gamma}{\partial u_j}) X_{k:0} \big\}$}
  \item[(g)] update amplitudes for all $k\in\mathcal T^{(r)}$ by quasi-Newton\\
     $u_j^{(r+1)}(t_k) = u_j^{(r)}(t_k) + \alpha_k\, \Hess_k^{-1}\, \frac{\partial f(X(t_k))}{\partial u_j}$ 
		or other methods (as in the text);
  \item[(h)] while $||\tfrac{\partial f_k}{\partial u_j}|| > f'_{\rm limit}$ for some $k\in\mathcal T^{(r)}$ re-iterate;\\
	else re-iterate with new set $\mathcal T^{(r+1)}$.\\
 \end{itemize}

{\em Comments:} General algorithmic scheme for synthesising quantum maps $X(T)$
 at fixed time $T$ with optimised the gate fidelity
 $f:=\tfrac{1}{N^2}\Re\tr \{{\widehat{U}}_{\rm target} X(T)\}$
 in open dissipative quantum systems. $X(t)$ denote Markovian quantum maps
 generated in step~1. 
%%%%%%%%%%%%%%%%%%%

\bigskip
%%%%%%%%%%%%%%%%%%%
\noindent
 {\bf Task 6:} {\em Optimise State Transfer between Density Operators in Open Systems}\\[2mm]
   Define boundary conditions by vectors in Liouville space
   $X_0:= \vec(\rho_0)$ and $X^\dagger_{\rm tar}:= \vec^t(\rho^\dagger_{\rm target})$;
 	fix final time $T$ and digitisation $M$ so that $T=M \Delta t$.
 \begin{itemize}
  \item[(a)] and (b) as in Task 5.

  \item[(c)] calculate forward-propagation \\
         \mbox{$ X_{k:0} := X_k X_{k-1}\cdots X_1 \vec(\rho_0)$};

  \item[(d)] calculate back-propagation \\
         \mbox{$\Lambda^\dagger_{M+1:k+1}
		:= \vec^t(\rho^\dagger_{\rm tar}) X_M  X_{M-1}\cdots X_{k+1}$}

  \item[(e)] evaluate fidelity \\[1mm]
    \mbox{$f\negthickspace=\negthickspace\tfrac{1}{N} \Re [\tr]\big\{ \Lambda^{\dagger}_{M+1:k+1} X^{\phantom{ \dagger}}_{k:0} \big\}
	     \negthickspace=\negthickspace\tfrac{1}{N} \Re [\tr]\big\{\widehat{X}^\dagger_{\rm tar}X^{\phantom{ \dagger}}_{M:0}\big\}$}\\[1mm]
	and stop if $f\geq 1-\varepsilon_{\rm threshold}$ or iteration $r>r_{\rm limit}$.

  \item[(f)] approximate gradients for all $k\in\mathcal T^{(r)}$\\[1mm]
    \mbox{$\frac{\partial f(X(t_k))}{\partial u_j} \approx \tfrac{-\Delta t}{N} \Re \tr\big\{ \Lambda^{\dagger}_{M+1:k+1}
	(i{\widehat{H}}_{u_j}+\tfrac{\partial \Gamma}{\partial u_j}) X_{k:0} \big\}$}

  \item[(g)] and (h) as in Task 5.\\
 \end{itemize}
{\em Comments:} Algorithmic scheme for
optimising state transfer between density operators in open Markovian quantum systems,
where the representation in Liouville space is required. So Task 6 can be seen as the
rank-1 version of Task 5.
%%%%%%%%%%%%%%%%%%%

%%%%
%\subsection{Further Numerical Results}
\begin{figure*}[Ht!]
{\bf  C. Further Numerical Results}\\[3mm]
\begin{center}
\includegraphics[width=0.75\textwidth]{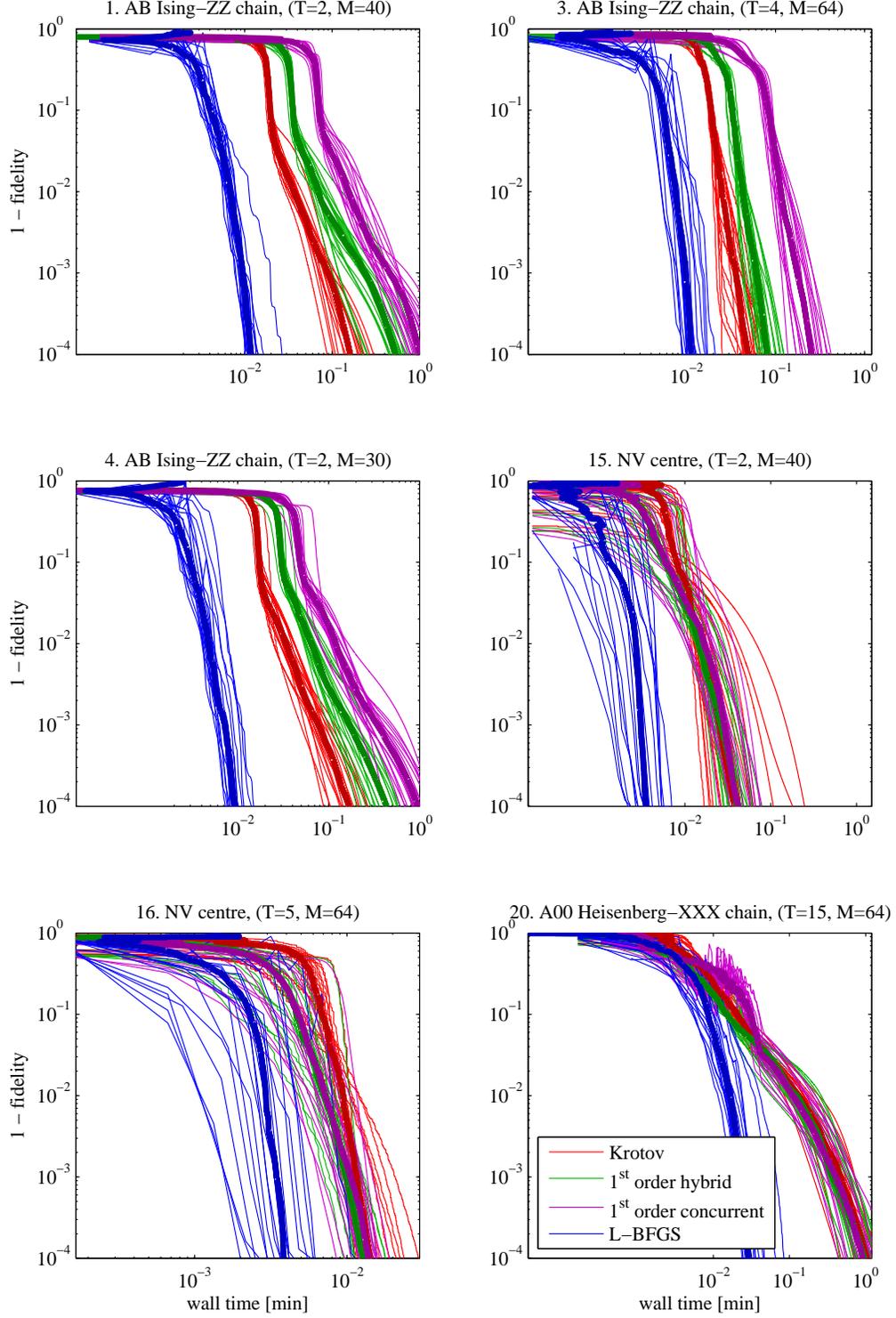}
\caption{
(Colour) Performance of a broader variety of {\em first-order} schemes compared to
the \lbfgs concurrent update.
The red traces show the plain \krotov sequential first-order update, while the first-order 
concurrent update is given in magenta and a first-order hybrid (with block size of $5$)
is given in green.
For comparison, the {\em second-order} \lbfgs concurrent update is shown in blue.
The test examples are again taken from (see Tab.~\ref{tab:test-def} and Sec.~\ref{sec:ExampleSpinChains})
with $mean(u_{ini})=0$ and $std(u_{ini})=1$ in units of $1/J$.
Note that in the (simpler) problems 1, 3, and 4 original \krotov performs fastest among all
the first-order methods, while in problems 15, 16, and 20 the variance within each method 
comes much closer to the variance among the methods so that in problem 16 the first-order
concurrent scheme outperforms the sequential one.
\label{fig:moreSpaghettis}}
\end{center}
\end{figure*}
%%%%

%%%%%%%%%%
\end{document}